\documentclass[onecolumn,amsmath,amssymb,nofootinbib,12pt]{article}
\pdfoutput=1 

\usepackage{jheppub} 

\usepackage[multiple]{footmisc}

\usepackage{tensor}
\usepackage{graphicx,subcaption}
\graphicspath{ {figures/} }	
\usepackage{amsfonts}
\usepackage{xparse}
 \usepackage{amsmath}
 \usepackage{amssymb}
 \usepackage{mathtools}
 \usepackage{tensor}
\usepackage{epsfig}
\usepackage{pdfpages}
\usepackage{bbm}
\usepackage{graphicx,epstopdf}
\usepackage[numbers]{natbib}
\usepackage[makeroom]{cancel}
\usepackage{hyperref}
\usepackage{array}
\usepackage[export]{adjustbox}
\usepackage[normalem]{ulem}
\usepackage{tcolorbox}
\usepackage{romannum}

\newcommand{\rcm}[1]{\textcolor{red}{\bf [[Rob: #1]]}}
\newcommand{\smr}[1]{\textcolor{orange}{\bf [[#1]] -- SM}}

\newcommand{\cv}{{\cal C}_\mt{V}}
\newcommand{\ca}{{\cal C}_\mt{A}}
\newcommand{\cvv}{{\cal C}_\mt{SV}}
\newcommand{\GN}{G_\mt{N}}
\newcommand{\eg}{{\it e.g.,}\ }
\newcommand{\ie}{{\it i.e.,}\ }
\newcommand{\viz}{{\it viz,}\ }
\newcommand{\mt}[1]{\textrm{\tiny #1}}
\newcommand{\reef}[1]{(\ref{#1})}

\newcommand{\Scft}{\Sigma_\mt{CFT}}

\newcommand{\tlam}{{\tilde\lambda}}

\newcommand{\be}{\begin{equation}}
\newcommand{\ee}{\end{equation}}

\newcommand{\beq}{\begin{equation}}
\newcommand{\eeq}{\end{equation}}
\newcommand{\beqa}{\begin{eqnarray}}
\newcommand{\eeqa}{\end{eqnarray}}

\newcommand{\beqs}{\begin{equation}\begin{aligned}}
\newcommand{\eeqs}{\end{aligned}\end{equation}}

\newcommand{\alB}{\alpha_{\mt B}}

\newcommand{\ket}[1]{\left| #1 \right>}

\renewcommand{\(}{\left(}
\renewcommand{\)}{\right)}
\renewcommand{\[}{\left[}
\renewcommand{\]}{\right]}

\newcommand{\veps}{\varepsilon}

\newcommand{\mC}{\mathcal{C}}

\newcommand{\mR}{\mathcal{R}}

\newcommand{\mL}{\mathcal{L}}
\newcommand{\mB}{\mathcal{B}}

\newcommand{\rmin}{r_{\pm,\mt{min}}}
\newcommand{\rf}{r_{\pm, f}}

\newcommand{\wemin}{w_{\veps,\rm min}}

\newcommand{\un}[1]{\tilde{#1}}
\newcommand{\ns}{\mathcal{N}}

\title{Complexity Equals Anything \Romannum{2}}

\author[a,b,c]{Alexandre Belin,}
\author[d]{Robert C. Myers,}
\author[e]{Shan-Ming Ruan,}
\author[a]{G\'abor S\'arosi}
\author[f]{\\and Antony J. Speranza}

\affiliation[a]{CERN, Theory Division, 1 Esplanade des Particules,\\
Geneva 23, CH-1211, Switzerland}
\affiliation[b]{Department of Theoretical Physics, University of Geneva,\\
24 quai Ernest-Ansermet, 1211 Geneva 4, Switzerland}
\affiliation[c]{Dipartimento di Fisica, Universit\`a di Milano - Bicocca \\
I-20126 Milano, Italy}
\affiliation[d]{Perimeter Institute for Theoretical Physics, \\
Waterloo, ON N2L 2Y5, Canada}
\affiliation[e]{Center for Gravitational Physics and Quantum Information,\\
Yukawa Institute for Theoretical Physics, Kyoto University,\\
Kitashirakawa Oiwakecho, Sakyo-ku, Kyoto 606-8502, Japan}
\affiliation[f]{\it Department of Physics, University of Illinois, Urbana-Champaign,\\ Urbana IL 61801, USA}

\emailAdd{alexandre.belin@unimib.it}
\emailAdd{rmyers@perimeterinstitute.ca}
\emailAdd{ruan.shanming@yukawa.kyoto-u.ac.jp}
\emailAdd{gabor.sarosi@cern.ch}
\emailAdd{asperanz@gmail.com}

\date{\today}

\abstract{We expand on our results in \cite{Belin:2021bga} to present a broad new class of gravitational observables in asymptotically Anti-de Sitter space living on general codimension-zero regions of the bulk spacetime. 
By taking distinct limits, these observables can reduce to well-studied holographic complexity proposals, \eg the volume of the maximal slice and the action or spacetime volume of the Wheeler-DeWitt patch. As with the codimension-one family found in \cite{Belin:2021bga}, these new observables display two key universal features for the thermofield double state: they grow linearly in time at late times and reproduce the switchback effect. Hence we argue that any member of this new class of observables is an equally viable candidate as a gravitational dual of complexity. Moreover, using the Peierls construction, we show that variations of the codimension-zero and codimension-one observables are encoded in the gravitational symplectic form on the semi-classical phase-space, which can then be mapped to the CFT.
}

\begin{document}

\begin{flushright}
        CERN-TH-2022-159
		\\
		YITP-22-101
		\\
	\end{flushright}

\maketitle

\section{Introduction}

Complexity quantifies how difficult it is to perform a task from a set of simple operations. For example, in quantum complexity, one is interested in constructing a unitary operator performing a particular operation by combining simple gates, which only act on a few qubits, into a quantum circuit \cite{Aaronson:2016vto,watrous}. An important aspect to keep in mind is that in complexity theory, the interest is always on robust features, such as the scaling of the complexity with the ``size" of the problem, for example, the dimensionality of the Hilbert space in quantum complexity or the number of digits if one aims at factoring a number into primes. Extracting an actual value for complexity is highly sensitive to not only the choice of the gate set, \ie allowed simple operations, but also the cost assigned to each gate. These ambiguities can be seen as a feature of complexity \cite{Brown:2021rmz}. 

Quantum complexity has recently triggered much interest in the context of black holes and holography as a new twist in the ongoing effort to connect quantum information theory to quantum gravity. The length of the wormhole for a two-sided AdS black hole grows linearly in time at late times and continues growing far beyond times at which entanglement entropies have thermalized \cite{Hartman:2013qma}. This suggests that a new quantum information measure is needed to encode the growth of the wormhole. It is believed that the holographic dual of circuit complexity, \ie holographic complexity, could capture the evolution of the black hole interior at late times \cite{Susskind:2014moa,Susskind:2014rva}. From the viewpoint of the bulk spacetime, there are three proposals for holographic complexity which have been studied extensively: complexity=volume (CV) \cite{Susskind:2014rva,Stanford:2014jda},  complexity=action (CA) \cite{Brown:2015bva,Brown:2015lvg} and complexity=spacetime volume (CV2.0) \cite{Couch:2016exn}. 

The CV conjecture \cite{Susskind:2014rva,Stanford:2014jda} proposes that the complexity is dual to the maximal volume of a hypersurface anchored on the boundary time slice $\Scft$ on which the CFT state is defined, \ie
\begin{equation}\label{eq:defineCV}
\cv(\Sigma_{\mt{CFT}}) =\,\max_{\Sigma_{\mt{CFT}}= \partial \mB}\left[\frac{\mathcal{V(B)}}{\GN \, \ell_{ \rm bulk}}\right] \,, 
\end{equation}
where $\GN$ denotes Newton's constant in the bulk gravitational theory and $\mathcal B$ corresponds to the bulk hypersurface of interest. In the CA proposal \cite{Brown:2015bva,Brown:2015lvg}, the complexity is given by evaluating the gravitational action on a region of spacetime, known as the Wheeler-DeWitt (WDW) patch. The latter can be defined as the causal development of a spacelike bulk surface anchored on the boundary time slice $\Scft$. The CA proposal is then given by 
\begin{equation}\label{eq:defineCA}
\ca(\Sigma_{\mt{CFT}}) =  \frac{I_\mt{WDW}}{\pi\, \hbar}\,. 
\end{equation}
The CV2.0 proposal generalizes and at the same time, simplifies the previous approach \cite{Couch:2016exn}. In this case, the holographic complexity is simply given by the spacetime volume of the WDW patch, namely
\begin{equation}\label{eq:defineCV2}
\cvv(\Sigma_{\mt{CFT}}) =  \frac{V_\mt{WDW}}{\GN \, \ell^2_{ \rm bulk}}\,. 
\end{equation}

It is worthwhile noting that all of these proposals for holographic complexity come with ambiguities in their definitions. For example, the definitions of both $\cv$ and $\cvv$ require the introduction of a new length scale $\ell_{\rm bulk}$ in eqs.~\eqref{eq:defineCV} and \eqref{eq:defineCV2} to make the holographic complexity dimensionless.\footnote{For simplicity, one typically chooses $\ell_{\rm bulk} = L$, \ie the curvature radius of AdS.} For the CA proposal \eqref{eq:defineCA}, a similar length scale appears in the boundary terms on the null boundaries of the WDW patch \cite{Lehner:2016vdi}. There is no natural prescription to fix these new scales and hence the precise value of the holographic complexity in any of these proposals is ambiguous. While one may worry about such ambiguities, it should be seen as a feature that connects to the fact that in complexity theory, only the scaling of the complexity matters, and the precise value of prefactors are of little or no interest. 

Two universal features which have been argued to hold for any definition of quantum complexity in a holographic setting are: At late times in the time evolution of the thermofield-double state, the complexity should grow linearly in time, and the growth rate will be proportional to the mass of the dual black hole \cite{Susskind:2014moa,Susskind:2018pmk}. The second is a universal time delay in the response of the complexity to the insertion of shock waves in the far past, known as the switchback effect \cite{Stanford:2014jda}. 

The various holographic complexity proposals have certainly drawn attention to new kinds of gravitational observables in the bulk, as well as highlighting their possible role in the AdS/CFT correspondence. This discussion was expanded in \cite{Belin:2021bga} to a broad new class of codimension-one observables, as we now briefly review  -- see  appendix \ref{revone} for further details. To begin, we may observe that the CV proposal \eqref{eq:defineCV} actually involves two independent steps. First, the maximization procedure selects a special codimension-one hypersurface in the bulk from among all possible spacelike surfaces whose boundary is fixed at $\Scft$. The second step is to evaluate a particular geometric feature, \ie the volume, of this special hypersurface. 

Given this perspective, it is natural to construct an infinite class of new (diffeomorphism-invariant) gravitational observables on codimension-one surfaces, defined in terms of two scalar functions $F_1, F_2$ \cite{Belin:2021bga}. To begin, we consider codimension-one hypersurfaces $\Sigma$ in an asymptotically AdS bulk spacetime with $d$+1 dimensions and anchored on a particular boundary time slice $\Scft$, \ie $\partial\Sigma=\Scft$. Following the previous discussion, the first step is to select a special bulk surface within this class using an extremization procedure
\begin{equation}
	\delta_{\mt{X}} \left[ \int_{\Sigma} d^d\sigma \,\sqrt{h} \,F_2(g_{\mu\nu};X^\mu)\right] =0 \,.
	\label{stepone}
\end{equation}
Here $F_2$ is a scalar functional integrated over the bulk hypersurfaces. In general, $F_2$ may depend on the bulk metric $g_{\mu\nu}$ and also the embedding functions $X^\mu(\sigma^a)$ of the hypersurfaces. For example, $F_2$ may be a scalar constructed from the background curvature and/or the extrinsic curvature of the hypersurfaces. The above extremization then allows for variations of the position of the hypersurfaces while fixing their boundary $\Scft$. This procedure selects a special codimension-one hypersurface, which we denote $\Sigma_{F_2}$. Given this hypersurface, we evaluate
\begin{equation}\label{eq:obsdef}
O_{F_1,\Sigma_{F_2}}(\Scft) =\frac{1}{\GN L} \int_{\substack{\Sigma_{F_2}  }}\!\!\!\!\! d^d\sigma \,\sqrt{h} \,F_1(g_{
	\mu\nu}; X^{\mu}) \,,
\end{equation}
where the new scalar functional $F_1$ again depends on the bulk metric and the embedding functions. 

This procedure produces a well-defined diffeomorphism-invariant observable in the bulk that characterizes some feature of the boundary state on the time slice $\Scft$. We note that in general the scalar functionals $F_1$ and $F_2$ need not coincide, \ie the choice $F_1=F_2$ is a subset of the full family of observables defined in \cite{Belin:2021bga}. The simplest choice for these functionals would be $F_1=1=F_2$, with which we recover the CV proposal \eqref{eq:defineCV}. However, the most interesting feature of this family of new observables is that infinitely many of them can exhibit the universal behaviour (\ie linear late-time growth for the thermofield-double state and the switchback effect for perturbations of this state)  that is expected of holographic complexity. Therefore it was argued in \cite{Belin:2021bga} that any of those new observables are equally viable candidates to be the gravitational dual of complexity.

In the present paper, we build on these ideas to construct yet another infinite class of gravitational observables now defined in codimension-zero regions of the bulk spacetime. Further, we will show that the new observables display the same universal features discussed above. Hence they are also good candidates for holographic complexity. Moreover, we show that particular limits of our construction reduce to the CA and CV2.0 proposals, see eqs.~\eqref{eq:defineCA} and \eqref{eq:defineCV2}.

We will also show that the variations of these observables can be captured by the gravitational symplectic form, which can then be pushed to the boundary where it is given by a symplectic form on the space of Euclidean-path integral states with sources for single-trace operators \cite{Belin:2018fxe}.\footnote{These states should be interpreted as coherent states of the quantum gravity theory \cite{Botta-Cantcheff:2015sav,Marolf:2017kvq,Belin:2018fxe,Belin:2020zjb}. See also \eg \cite{Guo:2018kzl,Bernamonti:2019zyy,Bernamonti:2020bcf} for more studies on holographic complexity of coherent states.} This is accomplished in terms of a particular conjugate variation $\delta_w$ \cite{Belin:2018bpg} which depends on the codimension-zero or codimension-one observable $W$ and gives
\begin{equation}
\delta W = \Omega(\delta, \delta_{w}) \,,
\end{equation}
where $\Omega$ is the symplectic form of general relativity.
For codimension-zero observables, their variations are obtained thanks to a covariant version of the Poisson bracket, known as the Peierls bracket \cite{Peierls:1952cb} (see also \cite{Harlow:2019yfa} for a more recent discussion on the subject).

The remainder of the paper is organized as follows: In section \ref{zero}, we describe the construction for general codimension-zero observables and the corresponding linear growth at late times.
Section \ref{symp} presents the construction of the conjugate variations $\delta w$ for either codimension-one or codimension-zero gravitational observables $W$. We will work out some examples explicitly and construct the appropriate conjugate variation for several functionals $W$. We conclude in section \ref{discuss} with a brief discussion of our results and possible future directions. Appendix \ref{revone} reviews the construction of codimension-one observables from \cite{Belin:2021bga}, as well as providing some additional details. Further, we consider adding extrinsic curvature terms in codimension-one observables in Appendix \ref{sec:appB}. Concerning the example in section \ref{zero}, we further discuss the existence and uniqueness of constant mean curvature slices in asymptotically AdS spacetime in Appendix \ref{newapp}. In Appendix \ref{app:null}, we explore the limit of the gravitational action of a codimension-zero subregion by taking its spacelike boundaries to be null hypersurfaces. In Appendix \ref{app:variation}, we derive the conjugate variations $\delta_w h^{ab}$ and $\delta_w K_{ab}$ for a general codimension-one observable $W(h^{ab}, K_{ab})$. Finally, we discuss one example of codimension-one observables using three-dimensional rotating black holes in Appendix \ref{sec:approtate}.

\section{Codimension-Zero Observables} \label{zero}
\begin{figure}[h!]
	\centering		
	\includegraphics[width=3.5in]{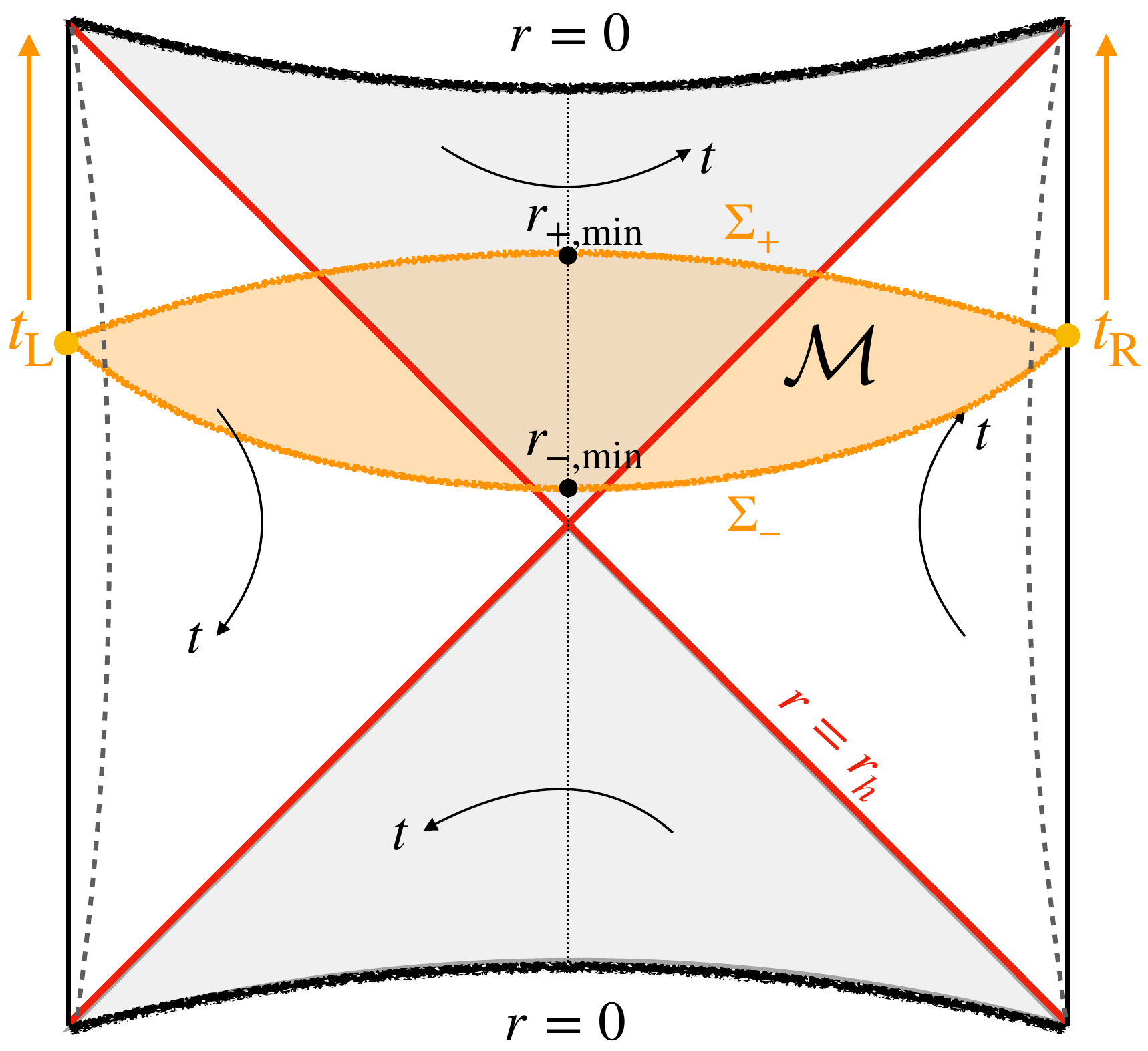}
	\caption{The codimension-zero region $\mathcal{M}$ (orange shaded patch) illustrated in an eternal AdS black hole background. 
	The future and past  boundaries of $\mathcal{M}$ are denoted by $\Sigma_\pm$, respectively, and are both anchored on the same boundary time slice $\Scft$. }
	\label{fig:spacetimevolume0}
\end{figure}

This section will build on the ideas of \cite{Belin:2021bga} to construct a broad class of gravitational observables associated with codimension-zero regions of the bulk spacetime -- see figure \ref{fig:spacetimevolume0}. Further, we will demonstrate that the new observables exhibit the universal behaviour, \ie linear late-time growth and the switchback effect, desired for holographic complexity. Our construction of observables associated with spacetime regions is, of course, inspired by the CA and CV2.0 proposals, which evaluate various geometric functionals on the WDW patch. In contrast to these proposals, we begin by identifying a particular bulk spacetime region using an extremization procedure, which mimics the first step of the generalized approach in eq.~\reef{stepone}. In parallel with the second step \reef{eq:obsdef}, the observable is then given by evaluating geometric features of this particular codimension-zero region. While this approach may seem more elaborate than that in CA and CV2.0 proposals, we show that eqs.~\eqref{eq:defineCA} and \eqref{eq:defineCV2} can be recovered as a special case of our new approach.

Let us describe our construction in more detail now. As illustrated in figure \ref{fig:spacetimevolume0}, we begin by choosing a boundary time slice $\Scft$. We then consider bulk regions $\mathcal{M}$ which are bounded by two codimension-one surfaces $\Sigma_\pm$ (\ie the future and past boundaries of $\mathcal{M}$, which we assume not to touch or cross in the bulk) anchored on $\Scft$, \ie $\partial\Sigma_\pm=\Scft$. We now define the following functional in such regions
\begin{equation}\label{eq:W1}
\begin{split}
W_{G_2,F_{2,\pm}} (\mathcal{M})&=\int_{ \Sigma_+}\!\!\!d^d\sigma \,\sqrt{h} \,F_{2,+}(g_{\mu\nu}; X^{\mu}_+) +\int_{   \Sigma_-}\!\!\!d^d\sigma \,\sqrt{h} \,F_{2,-}(g_{\mu\nu}; X^{\mu}_-) \\
	&\qquad+\frac{1}{L } \int_{\mathcal{M}}\!\!d^{d+1}x \,\sqrt{g} \ G_2(g_{
	\mu\nu})\,. 
\end{split}
\end{equation}
As in eq.~\reef{stepone}, $F_{2,\pm}$ are scalar functionals of the bulk metric $g_{\mu\nu}$ and the embedding functions $X^\mu_\pm(\sigma^a)$ of the corresponding boundaries $\Sigma_\pm$. For example, $F_{2,\pm}$ may be scalars constructed from the Riemann curvature of the bulk geometry and/or the extrinsic curvature of $\Sigma_\pm$. Similarly, $G_2$, which is integrated over the region  $\mathcal{M}$, may be a scalar constructed from the background curvature. As emphasized by our notation, $F_{2,\pm}$ and $G_2$ are three independent scalars, \ie we may choose completely different functionals on the future and past boundaries.\footnote{Let us  note that we assume that these functionals are dimensionless and so a prefactor of $1/L$ appears in the final integral in eq.~\eqref{eq:W1} for consistency of the overall dimension of $W_{G_2,F_{2,\pm}}$. This factor was chosen to be the inverse of the AdS curvature scale for simplicity, but any other choice would simply correspond to a change in the overall normalization of $G_2$.} The above functional \reef{eq:W1} is constructed to select a special bulk region by extremizing 
 \begin{equation}
 \delta_{\mt{X}_\pm}\! \left[ \,W_{G_2,F_{2,\pm}} (\mathcal{M})\,\right]=0 \,,
 \label{Xtreme1}
 \end{equation}
where as indicated, we are varying the shape of the two boundaries $\Sigma_\pm$.\footnote{If there are more than one extremal solutions, we choose the one yielding the maximum value of $W_{G_2,F_{2,\pm}}$, following the CV proposal \reef{eq:defineCV}. We return to this point in section \ref{sec:max}. For codimension-one observables, the maximization  is also discussed in Appendix \ref{sec:localmaxima}.} 

Before we move on, let us illustrate that the above extremization procedure \reef{Xtreme1} can always be recast as two independent extremizations, which are each analogous to the codimension-one case \reef{stepone}. This is actually a simple result of Stokes' theorem, which allows us to rewrite the bulk integral related to an arbitrary scalar function $G(g_{\mu\nu})$ as an integral on the boundaries, \ie $\Sigma_{\pm}$. More explicitly, it means that one can always find 
 \begin{equation}\label{eq:Stokes}
 \int_{\mathcal{M}}\!\!d^{d+1}x \,\sqrt{g} \ G_2(g_{\mu\nu})  = L \,\int_{ \Sigma_+ \cup  \Sigma_-  = \partial \mathcal{M}}\!\!\!\!d^d\sigma \,\sqrt{h} \   \widetilde{G}_2(g_{\mu\nu}; X^\mu) \,.  
 \end{equation}
Roughly speaking, this means that we can always find a primitive function $\widetilde{G}_2$ for any continuous function $G_2$ and perform Stokes' theorem.\footnote{We note that the primitive function $\widetilde{G}_2$ will generally not be a local functional of background and extrinsic curvatures on the boundaries $\Sigma_\pm$. Of course, this is in contrast with $F_{2,\pm}(g_{\mu\nu}; X^\mu)$, which are constructed as local functionals.} Another interpretation of this equality is that it is the statement that every top form (which is always closed by definition) on {non-compact} orientable manifold (such as the subregions $\mathcal{M}$ considered here) is always {exact}. 
Substituting the above expression in eq.~\reef{eq:W1}, our extremization procedure \reef{Xtreme1} reduces to two independent variations for the future and past boundaries,
\begin{equation}	\label{Xtreme2}
	\begin{split}
		&\delta_{\mt{X}_+}\! \left[  \int_{\Sigma_+} d^d\sigma \,\sqrt{h} \,\left(F_{2,+}(g_{\mu\nu};X^\mu_+)
		+\widetilde{G}_2(g_{\mu\nu}; X_+^\mu) \right)\right] =0 \,,\\
		&\delta_{\mt{X}_-}\! \left[  \int_{\Sigma_-} d^d\sigma \,\sqrt{h} \,\left(F_{2,-}(g_{\mu\nu};X^\mu_-)
		-\widetilde{G}_2(g_{\mu\nu}; X_-^\mu) \right)\right] =0 \,.	\\
	\end{split}
\end{equation}
This procedure then selects out special codimension-one surfaces as the future and past boundaries of our codimension-zero bulk region. We denote these boundaries as $\Sigma_\pm[G_2,F_{2,\pm}]$ and the corresponding bulk region as $\mathcal{M}_{G_2,F_{2,\pm}}$. Given this special codimension-zero region, we evaluate
\begin{equation}\label{eq:O1}
 \begin{split}
  &O\[G_1,F_{1,\pm},\mathcal{M}_{G_2,F_{2,\pm}}\] (\Scft)=\frac{1}{\GN L }\int_{ \Sigma_+[G_2,F_{2,+}]}\!\!\!\!\!\!\!d^d\sigma \,\sqrt{h} \,F_{1,+}(g_{\mu\nu}; X^{\mu}_+) \\
	&\qquad
+\frac{1}{\GN L }\int_{   \Sigma_-[G_2,F_{2,-}]}\!\!\!\!\!\!\!d^d\sigma \,\sqrt{h} \,F_{1,-}(g_{\mu\nu}; X^{\mu}_-) +\frac{1}{G_N L^2 } \int_{\mathcal{M}_{G_2,F_{2,\pm}}}\!\!\!\!\!\!d^{d+1}x \,\sqrt{g} \ G_1(g_{
	\mu\nu})\,.    
 \end{split}
\end{equation}
As in eq.~\reef{stepone}, $F_{1,\pm}$ are scalar functionals which may be constructed from the bulk curvature and/or the extrinsic curvature of $\Sigma_\pm$. Similarly, $G_1$, which is integrated over the region  $\mathcal{M}_{G_2,F_{2,\pm}}$, is a scalar constructed from the background curvature. Again as emphasized by our notation, $F_{1,\pm}$ and $G_1$ are three independent scalars, which can, in general, be chosen to be completely different from the functionals appearing in eq.~\reef{eq:W1}. 

Hence our codimension-zero version of the complexity=anything proposal \cite{Belin:2021bga} again follows a two-step procedure. First, we pick out the codimension-zero region, which is tied to a boundary time slice $\Scft$, by extremizing a geometric functional \reef{eq:W1}. Then we evaluate a separate geometric functional \reef{eq:O1} on this region. In all, this construction involves two independent `bulk' scalars, $G_1$ and $G_2$, and four independent `boundary' scalars, $F_{1,\pm}$ and $F_{2,\pm}$. Of course, this new proposal includes the observables originally discussed in \cite{Belin:2021bga}, \eg by setting $G_1=G_2=F_{1,-}=F_{2,-}=0$ which focuses our attention on codimension-one observables constructed on the single boundary $\Sigma_\pm[G_2=0,F_{2,+}]$. Hence it is perhaps not surprising that the analysis of \cite{Belin:2021bga} is readily extended  to show that the new codimension-zero observables exhibit the universal behaviour expected for holographic complexity. We now explicitly show these features in the following subsection.

\subsection{General Analysis}\label{sec:general}

With our discussion above, we have defined a large class of diffeomorphism-invariant observables \reef{eq:O1} which are tied to a boundary time slice of an asymptotically AdS spacetime. However, to connect these new observables to holographic complexity, we need to demonstrate that they exhibit the expected universal behaviour, \ie linear late-time growth and the switchback effect in a black hole background. 

For simplicity, we will focus our discussion here on the eternal planar black hole in $d+1$ dimensions,\footnote{For the most part, the following analysis does not depend on the details of the blackening factor $f(r)$. Hence we could easily extend the discussion to consider \eg static black holes with curved event horizons (\ie with $f(r)=k+ \frac{r^2}{L^2}\(1-\frac{r_h^{d}}{r^d}\,\)$) and/or with electromagnetic charges.} 
\begin{equation}\label{eq:BHmetric}
\begin{split}
d s^{2}&=-f(r)\, d t^{2}+\frac{d r^{2}}{f(r)}+r^{2}\, d \vec{x}^2\,,
\qquad \text{with} \quad f(r)= \frac{r^2}{L^2}\(1-\frac{r_h^{d}}{r^d}\,\)\,,
\end{split}
\end{equation}
and where $r=r_h$ is the location of the event horizon. The corresponding temperature and mass of the black hole are given by  
\begin{equation}\label{arena}
T_{\mt{BH}}=\frac{d \, r_h}{4 \pi L^2}\,, \qquad\quad M=\frac{(d-1) V_{ x}\,r_h^d}{16 \pi G_{\mathrm{N}}L^{d+1}} \,,
\end{equation}
where we have introduced $V_x$ as the regulated volume of the $(d-1)$-dimensional spatial boundary geometry, \ie $V_x=\int d^{d-1}x$.  

This two-sided black hole geometry \reef{eq:BHmetric} is dual to the thermofield double (TFD) state of two decoupled boundary CFTs on independent planar background geometries (\ie $\mathbb{R}^{d}$),
\begin{equation}\label{eq:TFD}
\ket{\psi_{\mt{TFD}}(t_{\mt{L}},t_{\mt{R}})}=\sum_{E_n} e^{-\beta E_n/2-iE_n (t_{\mt{L}}+t_{\mt{R}})/2 } \ket{n}_{\mt{L}} \otimes \ket{n}_{\mt{R}} \,.
\end{equation}
Here, we have associated the state with the time slices $t=-t_{\mt{L}}$ and $t=t_{\mt{R}}$ in the left and right boundaries, as illustrated in figure \ref{fig:spacetimevolume0}. Of course, this state is invariant under time shifts $\delta t_{\mt{L}}=-\delta t_{\mt{R}}$, which is reflected in the flow of the Killing vector $\partial_t$ in the bulk spacetime, \eg see figure \ref{fig:spacetimevolume0}. Hence, without loss of generality, we will set $t_{\mt{L}}=t_{\mt{R}}=\tau/2$ and consider the time evolution of the CFT state $\ket{\psi_{\mt{TFD}}(\tau)}$. 

Because the regions and surfaces of interest, \ie $\mathcal{M}$ and  $\Sigma_\pm$, extend from the asymptotic boundaries to the interior of the black hole, our calculations are facilitated by rewriting the metric \reef{eq:BHmetric} in terms of infalling Eddington-Finkelstein coordinates,
\begin{equation}\label{infall}
ds^2=-f(r)\,dv^2+2\,dv\, dr + \frac{r^2}{L^2}\,d \vec{x}^2\,.
\end{equation}
As usual, the infalling coordinate is defined by $v=t + r_*(r)$ with $r_*(r)= -\int^\infty_r \frac{dr'}{f(r')}$.

\subsubsection{Observables with $G_1=G_2$ and $F_{1,\pm}=F_{2,\pm}$}

To illustrate our new proposal, we begin by considering observables with $G_1=G_2$ and $F_{1,\pm}=F_{2,\pm}$. That is, the same geometric functional appears in the extremization \reef{eq:W1} and in the final observable \reef{eq:O1}. To further simplify the analysis, we focus on functionals that only depend on the background geometry (and not on the extrinsic or intrinsic geometry of the boundary surfaces, in the case of $F_{1,\pm}$). We label the corresponding observables $\mathcal{C}_{\textrm{gen}}(\tau)$ since we anchor them on the boundary time slice $t_{\mt{L}}=\tau/2=t_{\mt{R}}$, which we denote $\Scft(\tau)$. Thus eq.~\reef{eq:O1} reduces to
\begin{equation} \label{eq:O11}
    \begin{split}
      \mC_{\rm{gen}}(\tau) &=\max_{\partial\Sigma_\pm(\tau)=\Scft(\tau)} \bigg[\frac{1}{G_N L }\int_{ \Sigma_+[G_1,F_{1,+}]}\!\!\!\!\!\!\!d^d\sigma \,\sqrt{h} \,F_{1,+}(g_{\mu\nu})\\
	&+\frac{1}{G_N L }\int_{   \Sigma_-[G_1,F_{1,-}]}\!\!\!\!\!\!\!d^d\sigma \,\sqrt{h} \,F_{1,-}(g_{\mu\nu}) +\frac{1}{G_N L^2 } \int_{\mathcal{M}_{G_1,F_{1,\pm}}}\!\!\!\!\!\!d^{d+1}x \,\sqrt{g} \ G_1(g_{	\mu\nu})\bigg]\,.   
    \end{split}
\end{equation}
Due to the symmetries of the background geometry \eqref{infall}, we can parametrize the boundaries $\Sigma_\pm$ as $\(v_\pm(\sigma), r_\pm(\sigma), \vec{x}\)$, where we have introduced the `radial' coordinate $\sigma$ on these two codimension-one surfaces. The codimension-zero functional then becomes
 \begin{equation}\label{eq:defineV0}
\mC_{\rm{gen}}(\tau)=\frac{V_x}{G_N L } \sum_{\veps=+,-}\int_{\Sigma_\veps} d\sigma\, \[ \(\frac{r_\veps}{L}\)^{d-1}\sqrt{-f(r_\veps){\dot v_\veps}^2+2\dot v_\veps\,\dot r_\veps}\ a_\veps(r_\veps) - \veps \dot{v}_\veps\,b(r_\veps)\] \,,
 \end{equation}
where $\dot{v}_\pm=dv_\pm/d\sigma$, $\dot{r}_\pm=dr_\pm/d\sigma$ and the functions $a_\pm(r)$ are given by evaluating $F_{1,\pm}$ in the black hole background \reef{infall}.\footnote{To reduce the clutter in our equations here and in the following, we have dropped the subscripts from the coordinates specifying the profiles of the surfaces $\Sigma_\pm$.} Similarly, $b(r)$ is the primitive function found by evaluating $\sqrt{g}\, G_1$ in the background and integrating with respective to $r$.\footnote{Note the extra minus sign appearing in front of this term in eq.~\reef{eq:defineV0} because radial integral runs from the black hole interior to the conformal boundary, which implicitly extends from $\Sigma_+$ to $\Sigma_-$ in the infalling coordinates \reef{infall}.} That is,\footnote{Here we have assumed that $\sqrt{g}\, G_1$ is a {continuous function} of $r$ in the domain $\mathcal{M}$, which is a sufficient (but not necessary) condition for the existence of the primitive function $b(r)$. \label{footy87}}
 \begin{equation}\label{arena2}
\sqrt{g}  \, G_1(g_{\mu\nu}) = G_1(r)  \( \frac{r}{L} \)^{d-1} \equiv L\,\frac{\partial b(r)}{\partial r} \,.
 \end{equation}

Hence as expressed in eq.~\reef{eq:defineV0}, our observable reduces to two independent contributions coming from the future and past boundaries of the region $\mathcal{M}$. As observed in eq.~\reef{Xtreme2}, the next step involves extremizing each of these contributions independently. Finding the extremal boundaries can be recast as a classical mechanics problem with identifying the corresponding integrand as a Lagrangian, \ie
\begin{equation}\label{eq:Lagragianpm}
\mathcal{L}_\pm := \(\frac{r}{L}\)^{d-1}\sqrt{-f(r){\dot v}^2+2\dot v\,\dot r}\ a_\pm(r) \mp \dot{v}\,b(r)\,.
\end{equation}
Moreover, because spacetime is stationary, the effective Lagrangians $\mathcal{L}_\pm$ are $v$-independent. Hence the conjugate momenta are conserved, \ie $P^\pm_v={\partial {\cal L}_\pm}/{\partial \dot{v}}$ are constants along the entire profiles of $\Sigma_\pm$.

Now we will be interested in the time evolution of the observable \reef{eq:defineV0} with respect to the boundary time $\tau$. 
Because the boundary time evolution can be interpreted as the variation from one pair of extremal boundaries $\Sigma_{\pm}(\tau)$ to another $\Sigma_{\pm}(\tau+\delta \tau)$, where we are infinitesimally shifting the endpoints of the profiles, we can use our intuition from classical mechanics to write 
\begin{equation}\label{eq:dVdt01}
 \begin{split}
   \frac{d}{d\tau}\,\mC_{\rm{gen}}(\tau)&=\frac{V_x}{\GN L} \(  P_\tau^+(\tau)+P_\tau^-(\tau) \)\Big|_{r=\infty}\\
&= \frac{V_x}{\GN L} \(  P_v^+(\tau)+P_v^-(\tau) \)\,.
 \end{split}
\end{equation}
As usual, in the first line, only the boundary terms evaluated at the asymptotic boundary contribute since the bulk integral coming from the variation along the extremal hypersurface $\Sigma_\pm$ vanishes. The second line follows from the fact that the infalling coordinate $v$ and the usual Schwarzschild time coordinate $t$ coincide at the asymptotic boundary, and hence their conjugate momenta are also equal there. Further, as noted above, $P_v^\pm$ are conserved along the corresponding surfaces (for fixed $\tau$), so the last expression need not be evaluated at $r=\infty$. Let us add that the linear growth at late times will follow from the fact that $P_v^\pm(\tau)$ will approach constant values at large $\tau$.

Each of the contributions in eq.~\reef{eq:defineV0} is invariant under reparametrizations of the worldvolume coordinate $\sigma \to \sigma'(\sigma)$. One finds that a convenient constraint to fix the choice of this coordinate on each surface is\footnote{In general, the functions $a_\pm(r)$ may become negative in certain ranges of $r$ and in this case, we should set  $\sqrt{-f(r)\dot v^2+2  \dot v \dot r}=|a_\pm(r)| \(\frac{r}{L}\)^{d-1}$ -- see further comments after eq.~\eqref{eq:momentum02} in Appendix \ref{revone}.}
\begin{equation}\label{eq:gaugechoice}
\sqrt{-f(r)\dot v^2+2  \dot v \dot r}=a_\pm(r) \(\frac{r}{L}\)^{d-1} \,.
\end{equation}
With this constraint, the conserved momenta become
 \begin{equation}\label{eq:vmoment}
 P^\pm_v=\frac{\partial {\cal L}_\pm}{\partial \dot{v}}=\dot r -f(r)\,\dot v \mp b(r)\,.
 \end{equation}
We will choose the worldvolume coordinate $\sigma$, so it increases from the left AdS boundary to the right AdS boundary. 
Hence moving to larger (positive) $\sigma$ corresponds to moving to larger $r$ near the right quadrant of figure \ref{fig:spacetimevolume0}. With this choice, we will generally find the conserved momentum positive on the late-time surfaces.
 
Combining this pair of equations, the gauge constraint \reef{eq:gaugechoice} and the conserved momentum eq.~\eqref{eq:vmoment}, we can solve for the profile of the future and past boundaries (\ie $v_\pm(\sigma)$ and $r_\pm(\sigma)$)  with the first order equations
\begin{equation}\label{eq:dots}
\begin{split}
\dot r&=\lambda\[(P^\pm_v\pm\,b(r))^2 +f(r)\,a^2(r)\,\(\frac{r}{L}\)^{2(d-1)}\]^{1/2}\,,
\\
\dot v&=\frac{\dot{r}-P^\pm_v\mp b(r)}{f(r)}\\
&=\frac{1}{f(r)}\left(
\lambda\[(P^\pm_v\pm\,b(r))^2 +f(r)\,a^2(r)\,\(\frac{r}{L}\)^{2(d-1)}\]^{1/2}-(P^\pm_v\pm b(r))\right)\,,
\end{split}
\end{equation}
with $\lambda=\pm$. Let us note that the surfaces of interest at the right asymptotic AdS boundary (\ie at large $r$) fall inward to some minimum radius $r=r_\mt{min}$ (see figure \ref{fig:spacetimevolume0}, as well as eq.~\reef{rmin}). With our choice of $\sigma$ (see comments below eq.~\reef{eq:vmoment}), this portion of $\Sigma_\pm$ corresponds to the branch $\lambda=+$ with $\dot {r}\ge 0$ above. The branch $\lambda=-$ with $\dot {r}\le 0$ corresponds to the portion where the surfaces continue from $r=r_\mt{min}$ and climb out to the left asymptotic boundary. In addition, we note that the coordinate time $t$ along these ``extremal'' surfaces satisfies 
\begin{equation}\label{eq:dott}
 \dot{t}:= \dot{v} - \frac{\dot{r}}{f(r)} = -\frac{P^\pm_v\pm b(r) }{f(r)} \,,
\end{equation}
which applies both inside and outside of the black hole horizon.

Following the discussion in appendix \ref{revone}, the determination of $r_\pm(\sigma)$ can be cast as a simple classical mechanics problem, \viz 
\begin{equation}\label{eq:definepotential}
\begin{split}
&\dot{r}^2 + \mathcal{U}_\pm(P^\pm_v, r) =0\\
{\rm where} \qquad \mathcal{U}_\pm(P^\pm_v&, r):= U_0(r)-(P^\pm_v\pm b(r))^2 \,.
\end{split}
\end{equation}
and as in eq.~\eqref{eq:dotr02}, $U_0(r)=  -f(r)\,a^2(r)\,({r}/{L})^{2(d-1)}$. Hence we are looking for a zero-energy trajectory in a potential that is adjusted as we vary the conserved momentum \reef{eq:vmoment}.\footnote{In contrast to eq.~\eqref{eq:dotr02}, we have not separated out $P^\pm_v{}^2$ as the effective energy here. Even with the latter choice, the cross term $2 P^\pm_v\,b(r)$ would still result in the effective potential being dependent on $P^\pm_v$, hence varying this parameter would change both the effective energy and the effective potential at the same time.}
It is clear from eq.~\reef{eq:definepotential} that for large $r$, the effective potential is negative and that the trajectory reverses at 
\begin{equation}\label{rmin}
\mathcal{U}_\pm(P^\pm_v, \rmin) = 0\,. 
\end{equation}
That is, we switch from the $\lambda=+$ branch to that with $\lambda=-$ in eq.~\reef{eq:dots}. It is straightforward to show that this reversal occurs inside the horizon, \ie $\rmin\le r_h$, since this is the only region where $U_0(r)$ is positive. Further, since the trajectories begin  
at large $r$, \ie at the asymptotic AdS boundary, we always choose the largest value of $\rmin$ when there is more than one possible solution for the turning point. For a given value of $P^\pm_v$, the minimal radius may be determined by
\begin{equation}\label{eq:defrmin}
    P^\pm_v =  \mp b(\rmin) +\tlam \sqrt{U_0(\rmin)}
    \,,
\end{equation}
where $\tlam=\pm$. Alternatively, given $\rmin$, we can determine the corresponding conserved momentum with this equation. With our conventions, late times (\ie $\tau\to+\infty$) will arise with the $\tlam=+$ branch.\footnote{The $\tlam=-$ branch yields observables anchored at negative $\tau$ and in particular, $\tau\to -\infty$.}

We are interested in the time evolution of the surfaces $\Sigma_{\pm}$ with respect to the boundary time $\tau$. 
Recall that both surfaces are anchored on the boundary time slice $t_{\mt{L}}= \tau/2 = t_{\mt{R}}$. For a surface with the turning point $r=\rmin$ (and from eq.~\reef{eq:defrmin}, the corresponding conserved momentum), we can determine the boundary time by integrating eq.~\eqref{eq:dott}, \ie
\begin{equation}\label{eq:boundarytimetR02}
\tau\equiv 2\,t_{\mt{R}}= -2\int^{\infty}_{\rmin}\!\!\! dr\, \frac{P^\pm_v\pm b(r)}{f(r)\,
\sqrt{(P^\pm_v\pm b(r))^2-U_0(r)}}  \,.
\end{equation}
Note that the $1/f(r)$ factor introduces a pole in the integrand at the horizon $r=r_h$, and the integral is defined with the Cauchy principal value at this singularity. Further, the radial integral contains another singularity at $r=\rmin$ where $\mathcal{U}_\pm(P^\pm_v, r)$ vanishes. However, this singularity is integrable as long as $\partial_r\,\mathcal{U}_\pm(P^\pm_v, \rmin) \ne 0$, and hence the corresponding boundary time remains finite. However, one finds that $\tau\to\infty$ when the turning point corresponds to an extremum of the effective potential, \ie $\partial_r\, \mathcal{U}_\pm(P^\pm_v, \rmin) = 0$.

We saw in eq.~\reef{eq:defineV0} that codimension-zero observable $\mC_{\rm{gen}}(\tau)$ can be expressed as a sum of two integrals over the codimension-one boundaries $\Sigma_\pm$. Now using the gauge constraint \eqref{eq:gaugechoice}, the $\dot v$ equation \reef{eq:dots} and the symmetry of the surfaces about $r=\rmin$, this expression can be rewritten as
\begin{equation}\label{hurrah}
 \mC_{\rm{gen}}(\tau)= -\frac{2V_x}{\GN L}\sum_{\varepsilon=+,- }\int^{\infty}_{r_{\veps,\mt{min}}} \frac{ U_0(r) -\veps\, b(r)\(P^\veps_v+\veps\, b(r) \)}{f(r)\,\sqrt{-\mathcal{U}_\veps(P^\veps_v,r)}}\, dr \,.
\end{equation}
Of course, using eq.~\reef{eq:boundarytimetR02}, we have tuned the minimal radius  for each surface so that they reach the asymptotic AdS boundary on the same time slice.
Further applying eqs.~\eqref{eq:defrmin} and \eqref{eq:boundarytimetR02}, we can write the above expression as
\begin{equation}\label{eq:WGFdecmposition}
\mC_{\rm{gen}}(\tau)
 = \frac{V_x}{\GN L}\,\sum_{\varepsilon=+,-} \(P^\veps_v \, \tau+2  \int^{\infty}_{r_{\veps,\mt{min}}}\frac{\sqrt{-\mathcal{U}_\veps(P^\veps_v,r)}}{f(r)}\, dr  \) \,.
\end{equation}
Now it is straightforward to calculate the time derivative of $\mC_{\rm{gen}}$ as
\begin{equation}\label{eq:dWdtau01}
\begin{split}
 \frac{d}{d\tau}\mC_{\rm{gen}}(\tau) &= \frac{V_x}{\GN L}\,\sum_{\varepsilon=+,-}  \left( P^\veps_v + \frac{d P^\veps_v}{d\tau} \[\tau +  2\int^{\infty}_{r_{\veps,\mt{min}}}\frac{ (P^\veps_v +\veps\, b(r))}{f(r)\sqrt{-\mathcal{U}_\veps(P^\veps_v,r)}} dr \,\]\right.  \\
&\qquad\qquad\qquad\qquad \left. - 2\,\frac{d r_{\veps,\mt{min}}}{d \tau}\,\frac{\sqrt{-\mathcal{U}_\veps(P^\veps_v,r)}}{f(r)}\bigg|_{r=r_{\veps,\mt{min}}} \right)\,,\\
&= \frac{V_x}{\GN L}\,\(P_v^{+}(\tau)+  P_v^{-}(\tau)\)\,.
\end{split}
\end{equation}
where in the first line, the two terms multiplying $\frac{d P^\veps_v}{d\tau}$ precisely cancel using eq.~\eqref{eq:boundarytimetR02} and the last term in the second line vanishes because $\mathcal{U}_\veps(P^\veps_v,r)$ vanishes at $r=r_{\veps,\mt{min}}$ by eq.~\eqref{rmin}. Hence we have recovered our result in eq.~\reef{eq:dVdt01}.

\subsubsection{Linear Growth}

We would like to show that the growth rate \reef{eq:dWdtau01} becomes constant at late times and that the constant is proportional to the mass of the black hole. Hence, we examine the late-time behaviour here, \ie as $\tau \to \infty$. As we noted below eq.~\reef{eq:boundarytimetR02}, we reach this regime  when the conserved momenta are chosen such that $\partial_r\, \mathcal{U}_\pm(P^\pm_v, \rmin) = 0$. Following the notation of Appendix \ref{revone}, we denote the radius at this critical point as $\rf$ -- see eq.~\eqref{eq:definerf}. Now recall that solving eq.~\reef{eq:definepotential} requires the effective potential to be negative, and so the critical point must correspond to a local maximum of the effective potential, \ie 
\begin{equation}\label{critic}
 \mathcal{U}_\pm(P^\pm_{\infty}, \rf) = 0\,,\quad  \partial_r\,\mathcal{U}_\pm(P^\pm_{\infty}, \rf) = 0 \,, \quad  \partial^2_r\,\mathcal{U}_\pm(P^\pm_{\infty}, \rf) \le  0   \,.
\end{equation}
where we denote the conserved momentum in this late time limit as
$P^\pm_{\infty}=\lim_{\tau\to\infty}P^\pm_v(\tau)$. 

As noted below eq.~\reef{eq:dVdt01},  linear growth of the new observable at late times requires that $P_v^\pm(\tau)$ is nearly constant at large $\tau$. That is,
\begin{equation}
 \lim_{\tau \to \infty} \frac{d P^\pm_v}{d \tau} \to 0 \qquad  \text{or}  \qquad  \lim_{\tau \to \infty} \frac{d \tau}{d P^\pm_v} \to \infty\,.
\end{equation}
We now show that this asymptotic behaviour arises with some more detailed analysis. 

First, we expand the effective potential around the critical point using eq.~\reef{critic},
\begin{equation}
 \lim_{r \to \rf}\mathcal{U}_\pm(P^\pm_\infty,r) \simeq  \frac{1}{2} \,\mathcal{U}_\pm''(P^\pm_\infty,\rf)\,(r-\rf)^2 + \mathcal{O}((r-\rf)^3)\,,
\end{equation}
where $\mathcal{U}_\pm''= \partial_r^2\, \mathcal{U}_\pm(P^\pm_v, r) $.
Next, one can expand the potential in the late-time limit where $\rmin$ approaches $\rf$ to find
\begin{equation}\label{eq:PinfPv}
 \lim_{\rmin \to \rf} \mathcal{U}_\pm(P^\pm_\infty,\rmin) \simeq 2(P^\pm_v-P^\pm_\infty)(P^\pm_\infty \pm b(\rf))+\mathcal{O}((P^\pm_\infty-P^\pm_v)(\rmin-\rf)) \,.
\end{equation}
Combining the above two limits, we arrive at the asymptotic behaviour for the conserved momenta, \ie  
\begin{equation}
 \lim_{\tau \to \infty} (P^\pm_\infty - P^\pm_v) \simeq -\frac{\mathcal{U}_\pm''(P^\pm_\infty,\rf)}{4 (P^\pm_\infty \pm b(\rf))} (\rmin-\rf)^2 +\mathcal{O}((\rmin-\rf)^3) \,.
\end{equation}
Now it is straightforward to differentiate eq.~\eqref{eq:boundarytimetR02} to find
\begin{equation}\label{eq:dtRdPv}
\begin{split}
\frac{d \tau}{d P^\pm_v} &= \frac{d \rmin}{d P^\pm_v}\, \frac{2(P^\pm_v\pm b(r))}{ f(r) \sqrt{\mathcal{U}_\pm(P^\pm_v,r)}} \Bigg|_{r\to \rmin}+\int^{\infty}_{\rmin}\!\!\! dr\, \frac{2U_0(r)}{f(r) \(\mathcal{U}_\pm(P^\pm_v,r)\)^{3/2}}\,.
\end{split}
\end{equation}
Let us note that the two terms on the right-hand side above are individually divergent since from eq.~\reef{rmin}, 
$\mathcal{U}_\pm(P^\pm_v,\rmin)=0$. However, one can show that these divergences precisely cancel at any finite time $\tau$ for which $\partial_r\mathcal{U}_\pm(P^\pm_{v}, \rmin) \ne 0$. On the other hand, this cancellation fails in the late-time limit, \ie with $\rmin \to \rf$ which yields $\partial_r\mathcal{U}_\pm(P^\pm_{\infty}, \rmin) = 0$. 
After a careful analysis of the leading divergences, we obtain the following asymptotic behaviour
\begin{equation}\label{eq:dPvdtau}
\lim_{\tau \to \infty}\frac{d P^\pm_v} {d\tau} \simeq  \frac{-f(\rf) \,(-\mathcal{U}_\pm''(P^\pm_\infty,\rf)^{3/2}}{2\sqrt{2}\,(P^\pm_\infty\pm b(\rf))^2} (\rmin-\rf)^2\,,
\end{equation}
Combining eqs.~\eqref{eq:PinfPv} and \eqref{eq:dPvdtau}, we find 
\begin{equation}
\lim_{\tau \to \infty} P^\pm_v \simeq P^\pm_{\infty} - N_\pm\, e^{-\kappa_\pm \tau} \,,
\end{equation}
with $N_\pm$ are positive coefficients (independent of $\tau$) and $\kappa_\pm$ given by 
\begin{equation}
\kappa_\pm = -f(\rf) \frac{\sqrt{-2\, \mathcal{U}_\pm''(P^\pm_{\infty}, \rf)}}{P^\pm_{\infty}\pm b(\rf)}\,.
\end{equation}
Recalling that $\frac{d}{d\tau}\,\mC_{\rm{gen}}\propto (  P_v^+(\tau)+P_v^-(\tau) )$ from eq.~\reef{eq:dVdt01}, we see that at late times, the growth rate is constant up to exponentially suppressed corrections.  

Of course, the analysis and results presented above are analogous to that given for codimension-one observables in \cite{Belin:2021bga}. Further, following the discussion there, we can also show that $P^\pm_\infty$ are both proportional to the mass, where $M \propto r_h^d/L^{d}$ from eq.~\reef{arena}. Here, we recall that we have focused on the case where the functionals appearing in eqs.~\reef{eq:W1} and \reef{eq:O1} involve only background curvature invariants (and not extrinsic or intrinsic curvatures of the boundary surfaces for $F_{1,\pm}$ and $F_{2,\pm}$). 

Let us then consider the various elements comprising the effective potentials $\mathcal{U}_\pm(P^\pm_v, r)$ in eq.~\reef{eq:definepotential}. With the restriction to background curvatures, one finds that
\begin{equation}\label{arena1}
U_0=\left(\frac{r_h}{L}\right)^{2d}(w-w^2)\, a^2(w)=\left(\frac{r_h}{L}\right)^{2d}\widehat U_0(w) \,, 
\end{equation}
where we introduced the dimensionless radial coordinate $w=(r/r_h)^d$ -- compare to eq.~\reef{eq:ham3}. The fact that $a(w)$ has no explicit dependence on $r_h$ when written in terms of  $w$ is inherited from curvature invariants which have the same property for the planar black hole background \reef{infall}. We are assuming that the dimensional coefficients appearing in the functions $F_{1,\pm}$ and $F_{2,\pm}$ are independent of $r_h$ because the observable should be defined in a state independent way.\footnote{As in Appendix \ref{revone}, it is natural to absorb the dimensions in these couplings with the AdS scale.} For the primitive function $b(r)$, we have from eq.~\reef{arena2}
\beq
b(r) = \int \frac{dr}L\,\left(\frac{r}L\right)^{d-1} G_1(r)
=\frac1d \left(\frac{r_h}L\right)^{d} \int dw\, G_1(w)
= \left(\frac{r_h}L\right)^{d} \,\hat b(w)\,,
\label{arena3}
\eeq
where  $G_1(w)$ has no explicit dependence on $r_h$ when written in terms of  $w$, as above for 
$a(w)$.\footnote{Let us note here that $b(r)$ is only defined up to a constant shift, \ie the integration constant in eq.~\reef{arena3}. Eq.~\reef{eq:defrmin} shows that this shift produces a shift in the corresponding momenta, \ie with $b(r) \to b(r) + c$, the same boundary surfaces $\Sigma_\pm$ are described by $P^\pm_v\to P^\pm_v \mp c$. However, the latter shifts cancel in the sum $P^+_v(\tau)+P^-_v(\tau)$ which determines $\partial_t\mC_{\rm{gen}}(\tau)$, as shown in eq.~\reef{eq:dVdt01}.}

Combining these expressions, we may rewrite the effective potential \reef{eq:definepotential} as
\begin{equation}\label{arena5}
	\mathcal{U}_\pm(P^\pm_v, w)= \left(\frac{r_h}{L}\right)^{2d}\widehat U_0(w)-\(P^\pm_v\pm \left(\frac{r_h}L\right)^{d} \,\hat b(w)\)^2\,.
\end{equation}
Now the critical value $w_f=(r_f/r_h)^d$ can be found by combining $\mathcal{U}_\pm(P^\pm_\infty, w_f) =0=\partial_w\mathcal{U}_\pm(P^\pm_\infty, w_f)$ and takes a numerical value which depends only on the couplings appearing in the various functions (\ie $w_f$ is independent of $r_h$). Then the corresponding conserved momenta can be determined using eq.~\reef{eq:defrmin},
\begin{equation}\label{arena6}
	P^\pm_\infty = \left(\frac{r_h}L\right)^{d}\, \(\sqrt{\widehat U_0(w_f)}\mp \hat b(w_f)\)\,.
\end{equation}
As desired, we have found that $P^\pm_\infty$ are both proportional to the mass, $M \propto r_h^d/L^{d}$ as in eq.~\reef{arena}. Hence from eq.~\reef{eq:dVdt01}, the late time growth rate of this class of observables will also be proportional to the mass defined in eq.~\eqref{arena}.

\subsubsection{Switchback effect}

The analysis from the previous subsection reveals that the gravitational observable $\mC_{\rm gen}$ in the black hole background exhibits a universal linear growth at late times, namely, 
\begin{equation}
	\mC_{\rm{gen}} {(t_{\mt{L}}, t_{\mt{R}}})	\simeq \frac{V_x}{\GN L} \( P_{\infty}^+  +  P_{\infty}^-  \) |t_{\mt{R}} +t_{\mt{L}} |  + \mathcal{O}(1)\,,
\end{equation} 
where the first ``subleading'' term is actually a UV divergent constant (independent of the boundary times, $t_\mt{L}$ and $t_\mt{R}$) \cite{Carmi:2016wjl}. Of course, this universal linear growth is the first evidence that our new gravitational observables could be related to holographic complexity. In addition, holographic complexity is also expected to exhibit the so-called switchback effect \cite{Stanford:2014jda}, which describes the time dependence of complexity under perturbations. The derivation of the switchback effect for the codimension-zero observables is similar to that in \cite{Belin:2021bga} for the codimension-one case. 
In the following, we only sketch the two salient properties of the  codimension-zero observables needed to support the switchback effect. 

Let us start with a quick review: We begin by perturbing the TFD state \reef{eq:TFD} as follows
\begin{equation}\label{eq:pTFD}
	|\Psi(t_{\mt{L}},t_\mt{R})\rangle = e^{-i H_{\mt{L}} t_{\mt{L}}-i H_{\mt{R}} t_{\mt{R}}}\mathcal{O}_{\mt{L}}(t_n)...\mathcal{O}_{\mt{L}}(t_1)|\psi_{\rm TFD}(0)\rangle\,.
\end{equation}
Here $\mathcal{O}_{\mt{L}}(t_i)$ denotes a thermal-scale operator  acting on the left boundary at the boundary time $t_i$. Further, the times $t_1,\cdots,t_n$ are in an alternating ``zig-zag" order (\ie $t_{2k+1}>t_{2k}$ but $t_{2k}<t_{2k-1}$).  Further the perturbing operators are chosen to be ``small". That is, they only act on a small number of degrees of freedom, and so they only affect a small perturbation of the overall state when they are initially applied.  Assuming all of the intervals $|t_{i+1}-t_i|$ are much larger than the scrambling time $t_*$,  the complexity of this perturbed TFD state is proportional to \cite{Stanford:2014jda} 
\begin{equation}\label{eq:switchbackcomplexity}
	|t_\mt{R}+t_1|+|t_2-t_1|+\cdots +|t_\mt{L}-t_n|-2n\, t_* \,. 
\end{equation}
 In this context, $n$ is referred to as the number of switchbacks or time-folds. The contribution $-2n\, t_*$ reflects that the effect of the perturbing operators is initially small and only grows to spread through the entire system over an interval of the scrambling time $t_*$. Hence the forward and backward time evolution of the state near the time-folds cancels, and these partial cancellations result in a reduction of the complexity compared to the naively expected linear growth. 
 
 In order to demonstrate this behaviour arises for our new gravitational observables, we recall that the dual bulk geometry of the perturbed TFD state \eqref{eq:pTFD} is given by a shockwave geometry \cite{Shenker:2013yza}.  Taking a single shockwave produced by massless null matter located at $U=0$ as an explicit example, the back-reacted black hole geometry becomes  \cite{Sfetsos:1994xa,Shenker:2013yza,Stanford:2014jda}
\begin{equation}
	\begin{split}	\label{eq:shockwave}
		d s^{2}		&=-2 A(U [V+\alpha_i \Theta( U)])d U d V + B(U  [V+\alpha_i \Theta(U)]) d{\vec x}^{\,2}\,,\\
		&\ \ \ \ A(UV) = - \frac{2}{UV} \frac{f(r)}{f'(r_h)^2} \,, \qquad  B(UV) = r^2/L^2\,,
	\end{split}
\end{equation}
where $|U|=e^{-\frac{2\pi}{\beta}u}$, $|V|=e^{\frac{2\pi}{\beta}v}$ are the usual Kruskal coordinates, $\Theta( U)$ denotes the Heaviside step function, and $\alpha_i = 2 e^{-\frac{2\pi}{\beta}(t_* \pm t_i)}$ is the null shift caused by the $i$'th shockwave. 

The first property we need is that the gravitational observables \reef{eq:O11} are additive from the left side to the right side of the shockwave, \ie there are no contributions resulting from the shockwave located at $U=0$. This fact can be proven in the limit of strong shockwaves \cite{Stanford:2014jda}, by observing that all delta function contributions appearing in any scalar functional take the form of $U\delta(U),\,  U^2 \delta'(U)$ and so on \cite{Belin:2021bga}. Hence these terms naturally vanish on the shockwave, and both the bulk and boundary contributions appearing in eq.~\reef{eq:O11} are additive, as desired. However, in the following, we would like  to work with the expression \reef{eq:defineV0} where the codimension-zero observables are given by integrals along the surfaces $\Sigma_\pm$, and hence we want the additivity to extend to this form as well. Here, one needs to be careful since the terms proportional to $\dot v_\veps b(r_\varepsilon)$ are boundary terms arising from integrating the bulk contribution by parts. Since the shockwave divides the codimension-zero subregion $\mathcal{M}$ into the left and right parts, one might expect to pick up additional boundary terms there from this bulk integration. Hence we need to make sure that these potential boundary terms are the same on the two sides of the shockwave, \ie they cancel one another. This can be realized by shifting one of the primitive functions $b(r)$ by a constant to match the value on the opposite side. Recall from eq.~\reef{arena2} that the primitive functions are only defined up to an additive constant and so we are free to make such a shift. More generally, we believe this is  related to the regularization of the shockwave, \ie this shift would arise naturally if one began with a smooth background where the shockwave was replaced by a finite matter distribution -- see also footnote \ref{footy87}. 

Given the above additivity, the gravitational observable $\mC_{\rm gen}$ (which we implicitly write in the form of eq.~\eqref{eq:defineV0}, \ie as the sum of surface contributions from $\Sigma_\pm$) in the multiple shockwave geometry is simply given by the sum of each patch, \viz 
\begin{equation}\label{eq:piecewisevolume}
	\begin{split}
	\mC_{\rm{gen}} {(t_{\mt{L}}, t_{\mt{R}}})=\mC_{\rm gen}(t_{\mt{R}},V_1)&+	\mC_{\rm gen}(V_1+\alpha_1,U_2)+\cdots\\
	&+	\mC_{\rm gen}(U_{n-1}-\alpha_{n-1},V_n)+	\mC_{\rm gen}(V_n+\alpha_n,t_{\mt{L}})\,.
	\end{split}
\end{equation}
Here $\mC_{\rm gen}(.,.)$ denotes the contributions from $\Sigma_{\pm}$ with two fixed endpoints and all endpoints are located either on the left/right horizon or on the boundary. For the purpose of evaluating the gravitational observables $\mC_{\rm gen}$ in the shockwave geometry, we finally need to maximize eq.~\eqref{eq:piecewisevolume} by varying the joint points on the shockwaves.  The maximization is easily performed at late times with  strong shockwaves by noting that the contributions between the left/right horizon also present the linear growth but in terms of the ingoing/outgoing time as follows 
\begin{equation}\label{eq:linearuv}
	\mC_{\rm{gen}} {(u_{\mt{L}}, v_{\mt{R}}})	\approx  \frac{V_x}{\GN L} \( P_{\infty}^+  \,  |v_{\mt{R}}^+ +u_{\mt{L}}^+ |   + P_{\infty}^-  \,  |v_{\mt{R}}^- +u^-_{\mt{L}} |  \)+ \mathcal{O}(1)\,,
\end{equation}
where $u^{\pm}_{\mt{L}}/ v_{\mt{R}}^{\pm}$ present the coordinate value of the intersection between $\Sigma_\pm$ and left/right horizon.
In order to derive the linear growth in eq.~\eqref{eq:linearuv}, one can focus on the right part and first rewrite the value of the intersection , \ie $v_{\mt{R}}^{\varepsilon}$ in terms of 
\begin{equation}\label{eq:boundarytime}
	\begin{split}
		v_{\mt{R}}^{\varepsilon}- v_{\varepsilon,\rm{min}}&
		= \int^{r_{\rm{h}}}_{r_{\rm{min}}} dr \, \frac{\dot{v}}{\dot{r}} =\int^{r_{\rm{h}}}_{r_{\varepsilon,\rm{min}}} dr \, \frac{1}{f(r)}  \(  1 -  \frac{P^{\varepsilon}_v+ {\varepsilon}b(r)}{\sqrt{(P^{{\varepsilon}}_v + {\varepsilon} b(r))^2-U_0(r)}}  \) \,,
	\end{split}
\end{equation}
by using the definition of infalling coordinate $v$ and the extremization equations derived in eqs.~\eqref{eq:dots} for $\Sigma_\pm$. 
At late times, the extremal surface $\Sigma_\pm$ inside the horizon approaches the final slice at $r=r_{\pm,f}$.
It is then obvious that the dominant contributions to $v_{\mt{R}}^{\varepsilon}$
(\ie a logarithmic divergence) at late times originate from the integral around the regime $r \sim r_{\rm{min}} \sim r_{\pm,f}$, \ie 
\begin{equation}\label{eq:linearv}
	\lim\limits_{r_{\varepsilon,\rm{min} } \to r_{\varepsilon,f}} v_{\mt{R}}^{\varepsilon} \approx  \( \frac{ P^\veps_{\infty}+\veps\, b(r_{\varepsilon,f}) }{f(r_{\varepsilon,f})} \)  \frac{\log \(  r_{\varepsilon,\rm{min}}-r_{\varepsilon,f} \) }{\sqrt{-\frac{1}{2} \mathcal{U}_{\varepsilon}''(P_{\infty}^{\varepsilon}, r_{\varepsilon,f})}} + \mathcal{O}(1) \,.
\end{equation}
A similar formula holds for the outgoing coordinate $u_{\mt{L}}^\varepsilon$. On the other hand, the volume contribution  $\mC_{\rm{gen}} {(u_{\mt{L}}, v_{\mt{R}}})$ from the portion connecting any two points on the left and right horizon can be simplified as 
\begin{equation}
	\begin{split}
	\mC_{\rm{gen}} {(u^{\varepsilon}_{\mt{L}}, v^{\varepsilon}_{\mt{R}}})=-\frac{2V_x}{\GN L}\sum_{\varepsilon=+,- }\int^{r_{\veps,h}}_{r_{\veps,\mt{min}}} \frac{ U_0(r) -\veps\, b(r)\(P^\veps_v+\veps\, b(r) \)}{f(r)\,\sqrt{-\mathcal{U}_\veps(P^\veps_v,r)}}\, dr \,,
	\end{split}
\end{equation}
which is similar to eq.~\eqref{hurrah}. The leading contribution of $\mC_{\rm{gen}} {(u^{\varepsilon}_{\mt{L}}, v^{\varepsilon}_{\mt{R}}})$ at late times is also given by a logarithmic divergence but with a different coefficient, \viz 
\begin{equation}\label{eq:C001}
	\lim\limits_{r_{\varepsilon, \rm{min} } \to r_{\varepsilon,f}}  	\mC_{\rm{gen}} {(u^{\varepsilon}_{\mt{L}}, v^{\varepsilon}_{\mt{R}}})  \approx  \sum_{\varepsilon=+,-} \( \frac{ U_0(r_{\varepsilon,f}) -\veps\, b(r_{\varepsilon,f})\(P^\veps_{\infty}+\veps\, b(r_{\varepsilon,f})\) }{f(r_{\varepsilon,f})} \)  \frac{\log \(  r_{\varepsilon,\rm{min}}-r_{\varepsilon,f} \) }{\sqrt{-\frac{1}{2} \mathcal{U}_{\varepsilon}''(P_{\infty}^{\varepsilon}, r_{\varepsilon,f})}}   \,.
\end{equation}
Noting that the final slice is determined by 
\begin{equation}
 \mathcal{U}_\pm(P^\veps_{\infty}, r_{\varepsilon,f})  =U_0(r_{\varepsilon,f}) -\(P^\veps_{\infty}+\veps\, b(r_{\varepsilon,f})\)^2=0 \,, 
\end{equation}
we finally arrive at the desired linear growth shown in eq.~\eqref{eq:linearuv} by comparing the two expressions in eq.~\eqref{eq:linearv} and eq.~\eqref{eq:C001}. Of course, one can also consider the extremal surface with one endpoint on the horizon and one on the asymptotic boundary, for which we obtain 
\begin{equation}\label{eq:linearut}
	\mC_{\rm{gen}} {(t_{\mt{L}}, v_{\mt{R}}})	\approx  \frac{V_x}{\GN L} \( P_{\infty}^+  \,  |v_{\mt{R}}^+ +t_{\mt{L}} |   + P_{\infty}^-  \,  |v_{\mt{R}}^- +t_{\mt{L}} |  \)+ \mathcal{O}(1)\,,
\end{equation}
Besides the linear growth, we would like to highlight that another interesting and important fact for the switchback effect is that the growth rate (\ie $P^\pm_{\infty}$) is also universal for all pieces of the piecewise maximal surface $\Sigma_\pm$. Substituting the linear growth exhibited in  eqs.~\eqref{eq:linearuv} and \eqref{eq:linearut} into eq.~\eqref{eq:piecewisevolume}, and extremizing with respect to $V_1,U_2,...U_{n-1},V_n$, we find that  $V_i=-\alpha_i/2$ and $U_i=\alpha_i/2$. Finally, we can conclude that the gravitational observables $\mC_{\rm gen}$ associated with the bulk subregion whose boundaries are maximal surfaces $\Sigma_\pm$ display the desired switchback effect \eqref{eq:switchbackcomplexity}.

\subsubsection{General observables with $G_1\ne G_2$ and $F_{1,\pm}\ne F_{2,\pm}$ \label{complexsquared}}

Now we return to consider the general case with the functional \reef{eq:W1} which is extremized to define the bulk region and the functional \reef{eq:O1} evaluated on this region are defined independently. That is, the corresponding integrals involve different  scalar functions, \ie $(F_{1,\pm}, G_1)$ and $(F_{2,\pm}, G_2)$. The codimension-zero complexity observable is then reduced to
\begin{equation}\label{eq:defineW02}
O\!\[G_1,F_{1,\pm}, \mathcal{M}_{G_2,F_{2,\pm}}\]=\frac{V_x}{G_N L } \sum_{\veps=+,-}\int_{\Sigma_\veps (a_2,b_2)} \!\!\!\!\!\!\! d\sigma\, \[ \(\frac{r}{L}\)^{d-1}\sqrt{-f(r\,){\dot v}^2+2\dot v\,\dot r}\ a_1(r) - \veps \dot{v}\,b_1(r) \] \,,
 \end{equation}
 when evaluated in the planar black hole spacetime \reef{infall}. Similar to the calculations around eq.~\eqref{hurrah}, one can recast the integral along the extremal surfaces $\Sigma_\pm(a_2, b_2)$ in terms of a radial integral with
 \begin{equation}
    O\!\[G_1,F_{1,\pm}, \mathcal{M}_{G_2,F_{2,\pm}}\]=-\frac{2V_x}{\GN L} \sum_{\varepsilon=+,-} \int^{\infty}_{\rmin}\frac{dr}{f(r)} \[ \frac{  \sqrt{U_1(r)U_2(r)}- b_1(r) (b_2(r) +\veps P^\veps_v)}{\sqrt{(P^\veps_v+\varepsilon\, b_2(r))^2 -U_2(r)}} \]\,,
    \label{hurrah2}
 \end{equation}
where $U_{1,2}(r)=  -f(r)\,a_{1,2}^2(r)\,({r}/{L})^{2(d-1)}$.
 It is worth noting that the conjugate momenta $P^\pm_v$ and the corresponding minimal radii $\rmin$ are associated with $(a_2(r),b_2(r))$, \ie
 \begin{equation}
  P^\pm_v =  \mp \bar{b}_2 +\tlam\, \sqrt{\bar{U}_2} \,. 
 \end{equation}
where $\tlam=\pm$ as in eq.~\reef{eq:defrmin}, and we have introduced the `overbar' notation to indicate that the corresponding function is evaluated at $r=\rmin$, \ie $\bar{y}=y(\rmin)$. As before, our conventions are such that the $\tlam=+$ branch yields the observable for late times (\ie $\tau\to+\infty$) and {\it we adopt this sign choice throughout the following.} In the same spirit of the decomposition in eq.~\eqref{eq:WGFdecmposition}, one can re-express the above as 
\begin{equation}\label{defcon3}
\begin{split}
&O\!\[G_1,F_{1,\pm}, \mathcal{M}_{G_2,F_{2,\pm}}\] =-\frac{V_x}{\GN L}\sum_{\varepsilon=+,-} \[ \(\varepsilon \bar{b}_1 - \sqrt{\bar{U}_1} \)\tau  \right. \\
   &\qquad\left. + 2\int^{\infty}_{\rmin}\frac{dr}{f(r)} \( \frac{  \sqrt{U_1(r)U_2(r)}-( b_1(r)-  \bar{b}_1 +\veps \sqrt{\bar{U}_1} )) (  b_2(r)+\veps P^\veps_v)}{\sqrt{(P^\veps_v+\varepsilon\, b_2(r))^2 -U_2(r)}} \) \]\,.
\end{split}
\end{equation}
It is crucial to note the integrand on the minimal radius exactly vanishes since
\begin{equation}
 \[ \sqrt{U_1(r)U_2(r)}-( b_1(r)- \bar{b}_1 \pm \sqrt{\bar{U}_1} )) (  b_2(r)\pm P^\pm_v) \] \bigg|_{r=\rmin} =0 \,.
\end{equation}

Now, taking the time derivative of the observable $O\!\[G_1,F_{1,\pm}, \mathcal{M}_{G_2,F_{2,\pm}}\]$, we find\footnote{Note that there is a factor of 2 missing in the analogous equations in \cite{Belin:2021bga}, \ie eqs.~(23) and (42).}
\begin{equation}\label{hurrah77}
\begin{split}
    & \frac{d}{d\tau} \Big( O\!\[G_1,F_{1,\pm}, \mathcal{M}_{G_2,F_{2,\pm}}\] \Big)  =-\frac{V_x}{\GN L}\sum_{\varepsilon=+,-} \Bigg[ \(\varepsilon \bar{b}_1 - \sqrt{\bar{U}_1} \)\\
		&+ 2\frac{d P^\veps_v}{d \tau} \int^{\infty}_{\rmin}\frac{dr}{f(r)} \frac{ (P^\veps_v+\veps\, b_2(r))\sqrt{U_1(r)U_2(r)}+\(\veps( b_1(r)- \bar{b}_1) + \sqrt{\bar{U}_1} \)\,U_2(r)}{\((P^\veps_v-\varepsilon\, b_2(r))^2 -U_2(r)\)^{3/2}} \Bigg] \,,
\end{split}
\end{equation}
where the bulk integral term does not vanish since the surfaces $\Sigma_\pm$ were determined by extremizing with respect to the functional $W_{G_2,F_{2,\pm}}$ in eq.~\reef{eq:W1}. However, we can find that the late-time limit of the corresponding growth rate is dominated by the linear terms, \ie 
\begin{equation}\label{eq:dVdtlimit02}
\lim_{\tau \to \infty}\frac{d}{d\tau}\, \Big(O\!\[G_1,F_{1,\pm}, \mathcal{M}_{G_2,F_{2,\pm}}\] \Big) =  \frac{V_x}{\GN L} \(  P_{\infty}^+(F_1, G_1) +P_{\infty}^-(F_1, G_1) \)\,,
\end{equation}
where we have defined
\begin{equation}
P^\pm_{\infty}(F_1, G_1)\equiv \mp \bar{b}_1 + \sqrt{\bar{U}_1} \,.
\end{equation}
To produce this result \reef{eq:dVdtlimit02}, we have used the asymptotic expansion \eqref{eq:dPvdtau} for $d P^\pm_v/d\tau$ and the suppression of the divergence in the bulk integral by the vanishing of 
\begin{equation}
 \[(P^\pm_v\pm  b_2(r))\sqrt{U_1(r)U_2(r)}+\(\pm( b_1(r)-  \bar{b}_1) + \sqrt{\bar{U}_1} \)\,U_2(r) \]\bigg|_{r=\rmin}=0\,.
\end{equation}

\subsection{Spacetime volume between CMC slices}\label{sec:spacetimevolume}

In this section, we study an explicit example of the codimension-zero observable $\mC_{\rm gen}$ in eq.~\eqref{eq:O11} with $G_1=G_2$ and $F_{1,\pm}=F_{2,\pm}$. In fact, we examine the simplest example where $G_1$ and $F_{1,\pm}$ are all taken to be constants. That is, we consider the following functional:
\begin{equation}
\label{eq:CMCfunc}
\mC_{\rm gen}=\frac{1}{\GN L } \[\alpha_+ \int_{\Sigma_+}\!\!\!d^d\sigma\,\sqrt{h}+\alpha_- \int_{\Sigma_-}\!\!\!d^d\sigma\,\sqrt{h} + \frac{\alB}{L}\,\int_{\mathcal{M}}\!\!d^{d+1}x\,\sqrt{-g}\]\,,
\end{equation}
where $\alpha_\pm$ and $\alB$ are dimensionless (positive) constants. In a moment, we will consider evaluating this observable in a planar black hole background as discussed in the previous section. However, we begin with a few general remarks.  Extremizing over the shape of the boundary surfaces following eq.~\reef{Xtreme2}, one finds that the extremal boundaries $\Sigma_\pm$ are constant mean curvature (CMC) slices, \eg see \cite{marsden1980maximal,witten2017}. That is, the extremization equations can be simply expressed as\footnote{The extrinsic curvature is defined as $K_{\mu\nu} \equiv h\indices{^\alpha_\mu}\nabla_\alpha n_\nu$ where the (timelike) normal vector $n^\mu$ is chosen here to be future-directed for both spacelike hypersurfaces $\Sigma_\pm$, and 
$h\indices{^\alpha_\mu}$ is the projector onto the tangent space of the surfaces $\Sigma_\pm$.\label{footnote:defineK}} 
\begin{equation}
\label{eq:CMCextr}
K_{\Sigma_+} = -\frac{\alB}{\alpha_+L}\,,\qquad K_{\Sigma_-} =\frac{\alB}{\alpha_- L}\,,
\end{equation}
Given this geometric framework for the extremal equations, one can argue that the boundary surfaces $\Sigma_{\pm}$ always exist and are unique, as long as the cosmological constant is negative and no matter fields are excited. We leave this discussion to appendix \ref{newapp}.  

This functional \eqref{eq:CMCfunc} is also an interesting object since it can be easily related to the original holographic complexity proposals in eqs.~(\ref{eq:defineCV} -- \ref{eq:defineCV2}). For example, one can recover the CV proposal \eqref{eq:defineCV} by taking the limit $\alpha_{+} \to +\infty$ (while keeping $\alpha_{-}$ and $\alB$ fixed. That is,
\begin{equation}
 \cv=\lim_{\alpha_+ \to +\infty } \( \frac{1}{\alpha_+}\,\mC_{\rm gen} \) \,. \label{limit1}
\end{equation}
Of course, this is a simple illustration of how the codimension-one observables introduced in \cite{Belin:2021bga} are related to the new codimension-zero observables presented here. 

\begin{figure}[t]
	\centering		
	\includegraphics[width=6in]{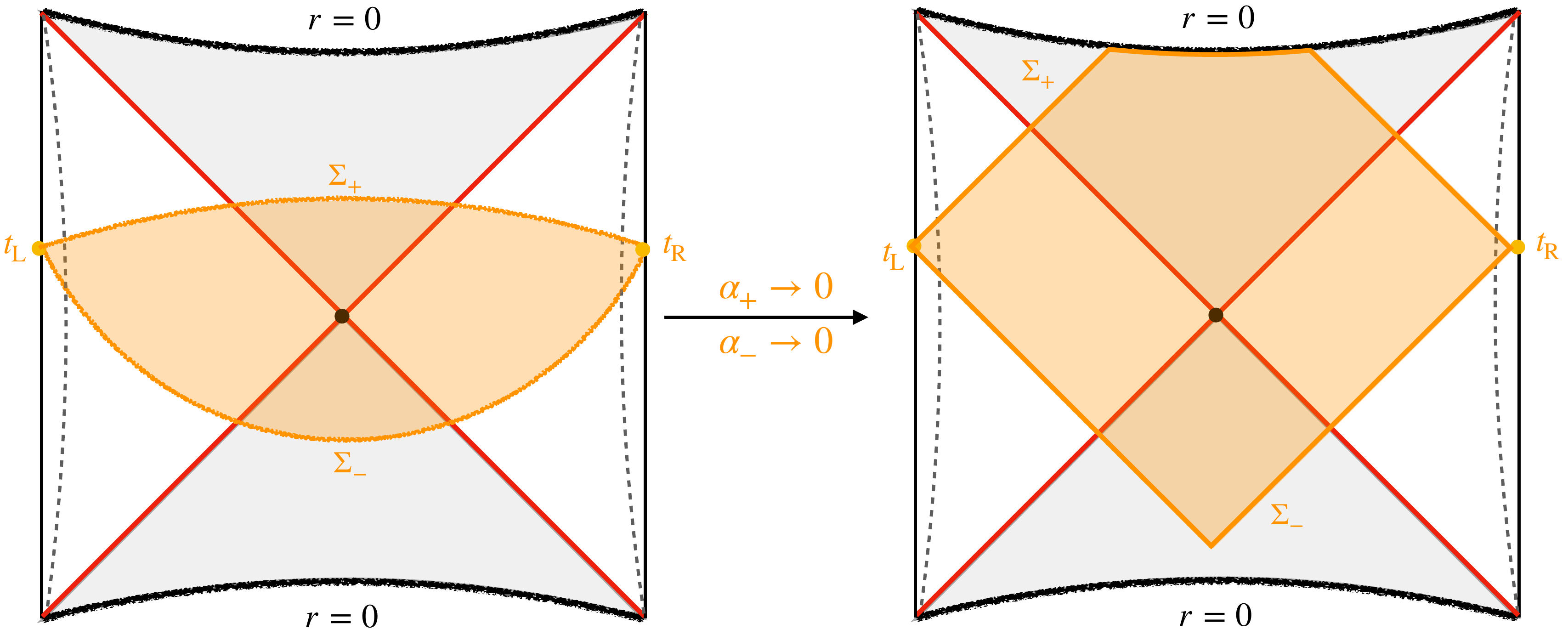}
	\caption{Left: Sketch of a codimension-zero region where boundaries $\Sigma_\pm$ are constant mean curvature slices, determined by the functional in eq.~\eqref{eq:CMCfunc}. Right: Taking the limit $\alpha_\pm \to 0$, the boundaries $\Sigma_\pm$ become null surfaces, and the region becomes the WDW patch.}
	\label{fig:spacetimevolume}
\end{figure}

Another interesting limit is to take $\alpha_\pm\to 0$ with $\alB$ fixed. From eq.~\reef{eq:CMCextr}, we see the extrinsic curvatures of the boundaries $\Sigma_\pm$ diverge in this limit, \ie $K_{\Sigma_\pm}\rightarrow \mp\infty$. This divergence indicates that these boundary surfaces are being pushed to the future and past light sheets emanating from the boundary time slice $\Scft$, as illustrated in Figure \ref{fig:spacetimevolume}. That is, the bulk codimension-zero region $\mathcal{M}$ becomes the Wheeler-DeWitt (WDW) patch in this limit! 

Hence, one recovers the CV2.0 proposal \eqref{eq:defineCV2}, \ie the spacetime volume of the WDW patch, from eq.~\reef{eq:CMCfunc} above by  taking $\alpha_\pm \to 0$ while keeping $\alB$ fixed \viz 
\begin{equation}
\cvv = \lim_{ \alpha_\pm \to0 } \(\frac1{\alB}\,\mC_{\rm gen}\)  \,. \label{limit2}
\end{equation}
We can also recover the CA proposal \eqref{eq:defineCA} with this limit and using our general proposal \reef{eq:O1}. That is, we first would extremize the functional $\mC_{\rm gen}$ in eq.~\reef{eq:CMCfunc} and take the limit $\alpha_\pm\to0$. Having obtained the WDW patch as the region $\mathcal{M}$, we then evaluate the gravitational action on this region, \ie take $G_1$ as the standard Einstein-Hilbert term (with cosmological constant contribution) and $F_{1,\pm}$, as the corresponding null boundary action terms
\cite{Parattu:2015gga, Lehner:2016vdi,Hopfmuller:2016scf, Chandrasekaran:2020wwn}. An interesting question is to examine the procedure where we extremize eq.~\reef{eq:CMCfunc} with finite $\alpha_\pm$ first, then evaluate the gravitational action on the resulting region $\mathcal{M}$ with the Gibbons-Hawking-York boundary terms on the spacelike surfaces $\Sigma_\pm$, and finally take the limit $\alpha_\pm\to0$. 
Interestingly, this null limit procedure for the action yields a finite result, although corrections to the standard null boundary terms involving the surface gravities of the null surfaces arise from this limit. The details of this limiting procedure and additional discussion of the treatment of Hayward terms in the action for codimension-two corners are given in appendix \ref{app:null}, and further discussion is given in section \ref{discuss}.

In the following discussion, we will keep $\alpha_\pm$, $\alB$ as independent constants and explicitly explore the maximization and time evolution of the functional \eqref{eq:CMCfunc} in the planar black hole background \reef{infall}.

\subsubsection{Extremal Surfaces $\Sigma_\pm$}
Now, let us explicitly evaluate the codimension-zero observable given in eq.~\eqref{eq:CMCfunc} in the black hole background given by the infalling coordinates \eqref{infall}. We begin with the bulk term in eq.~\reef{eq:CMCfunc}, which becomes 
\begin{equation}\label{eq:spacetimevolume}
\begin{split}
\frac{\alB}{L}\,\int_{\mathcal{M}}\sqrt{g} &= \frac{\alB}{L}\int d^{d-1}x \int dv \int^{r_-(v)}_{r_+(v)}dr \left(\frac{r}{L}\right)^{d-1} \\
&= \frac{\alB\,V_x}{d\,L^{d}}  \(\int_{\Sigma_- } d\sigma \, \dot v_-(\sigma)\, r_-(\sigma )^d- \int_{\Sigma_+} d\sigma\,\dot v_+(\sigma)\, r_+(\sigma)^d \)\,,
\end{split}
\end{equation}
where we have parametrized the surfaces as $r_{\pm}(\sigma),v_{\pm}(\sigma)$ and integrated by parts to get the second line. As we illustrated before, this bulk contribution splits into a sum of two independent terms associated to the  boundary surfaces $\Sigma_\pm$. The extremality conditions \reef{Xtreme2} are then equivalent to solving for two decoupled particle motions governed by the Lagrangian  $\mathcal{L}=\mathcal{L}_+({\Sigma_+})+\mathcal{L}_-({\Sigma_-})$, where
\begin{equation}
\mathcal{L}_\pm  =  \(\frac{r}{L}\)^{d-1}\sqrt{-f(r){\dot v}^2+2\dot v\,\dot r} \mp\frac{\alB}{d\,\alpha_\pm  L^{d}}\, \dot{v}\,r^d \,,
\label{noun9}\end{equation}
where we have dropped the subscripts $\pm$ on $(v(\sigma), r(\sigma))$ to reduce the clutter. Further, to simplify the following equations, we have absorbed the factors $\alpha_\pm$ as part of the overall coefficients in the observable, \eg along with ${V_x}/(\GN L)$. Compared to the analysis for the codimension-one observables (see appendix \ref{revone}), the new feature here is a `magnetic field'-like term in the Lagrangian. 

 Comparing the expressions \reef{noun9} to eq.~\reef{eq:Lagragianpm}, we see that here we have 
\begin{equation}
a_\pm(r)= 1 \,, \qquad b(r)= \frac{\alB}{d\,\alpha_\pm } \(\frac{r}{L}\)^d \,.
\end{equation}
Gauge fixing as before, with 
\begin{equation}\label{gfix2}
\sqrt{-f(r)\dot v^2+2  \dot v \dot r}=(r/L)^{d-1}\,,
\end{equation}
the conserved momentum  becomes
\begin{equation}\label{Pcon2}
P_v^{\varepsilon}=\dot r-\dot v f(r)  -  \veps\,\frac{\alB}{d\,\alpha_\veps}\, \left(\frac{r}{L}\right)^d\,,
\end{equation}
where as before, $\varepsilon = \pm$ for the future/past boundaries $\Sigma_\pm$. Combining eqs.~\reef{gfix2} and \reef{Pcon2}, we can express the equation solving for the radial profile $r(\sigma)$ as
\begin{equation}\label{simp2}
	\dot r^2 + \mathcal{U}(P_v^{\varepsilon}, r)=0\,,
\end{equation}
where 
\begin{equation}\label{simp22}
    \begin{split}
    \mathcal{U}(P_v^{\varepsilon},r)&=U_0(r)-\left(P_v^{\varepsilon}+\veps\,\frac{\alB}{d\,\alpha_\veps}\,\left(\frac{r}{L}\right)^{d}\right)^2\,,\\
{\rm with} \qquad U_0(r)&\equiv -f(r)\(\frac{r}{L}\)^{2(d-1)} =\left(\frac{r}{L}\right)^{2d} \left(\frac{r_h^{d}}{r^d}-1\right)\,.    \\
    \end{split}
\end{equation}
That is, as before, we have expressed this as a simple classical mechanics problem. However, in the present case, the potential depends on $P_v^{\varepsilon}$. This is not unexpected since canonical momentum differs from kinetic momentum due to the magnetic field. For convenience, we have combined all of the $P_v^{\varepsilon}$ terms in the potential $\mathcal{U}(P_v^{\varepsilon}, r)$ and hence the effective energy (on the right-hand side of eq.~\reef{simp2}) is zero. Recall that $U_0(r)$ is the effective potential that appears when solving for extremal surfaces for complexity=volume (and which can be recovered with the limit $\alpha_+\to\infty$ -- see eq.~\reef{limit1}).

\begin{figure}[ht!]
	\centering
	\includegraphics[width=3.05in]{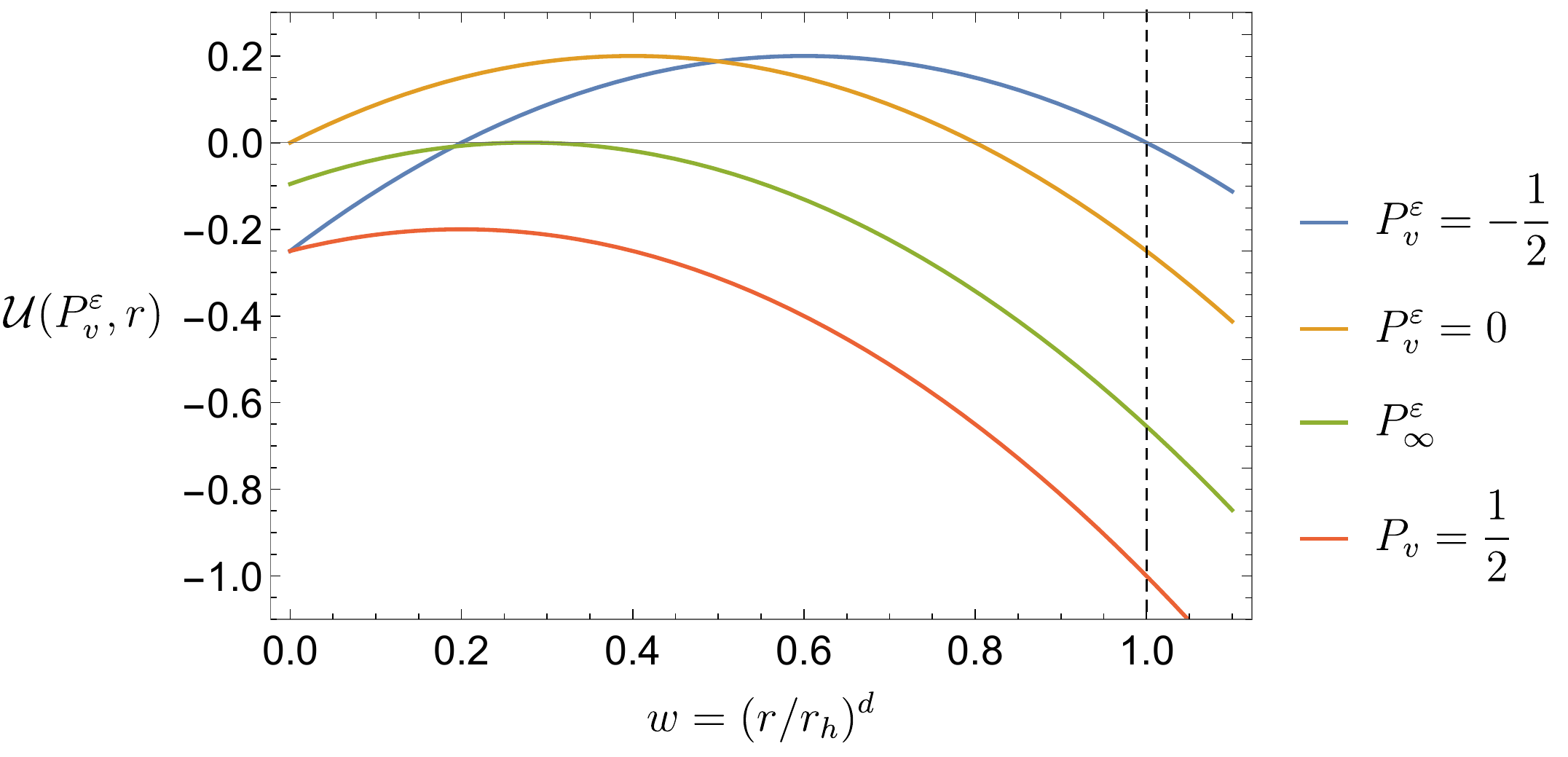}\,
	\includegraphics[width=2.95in]{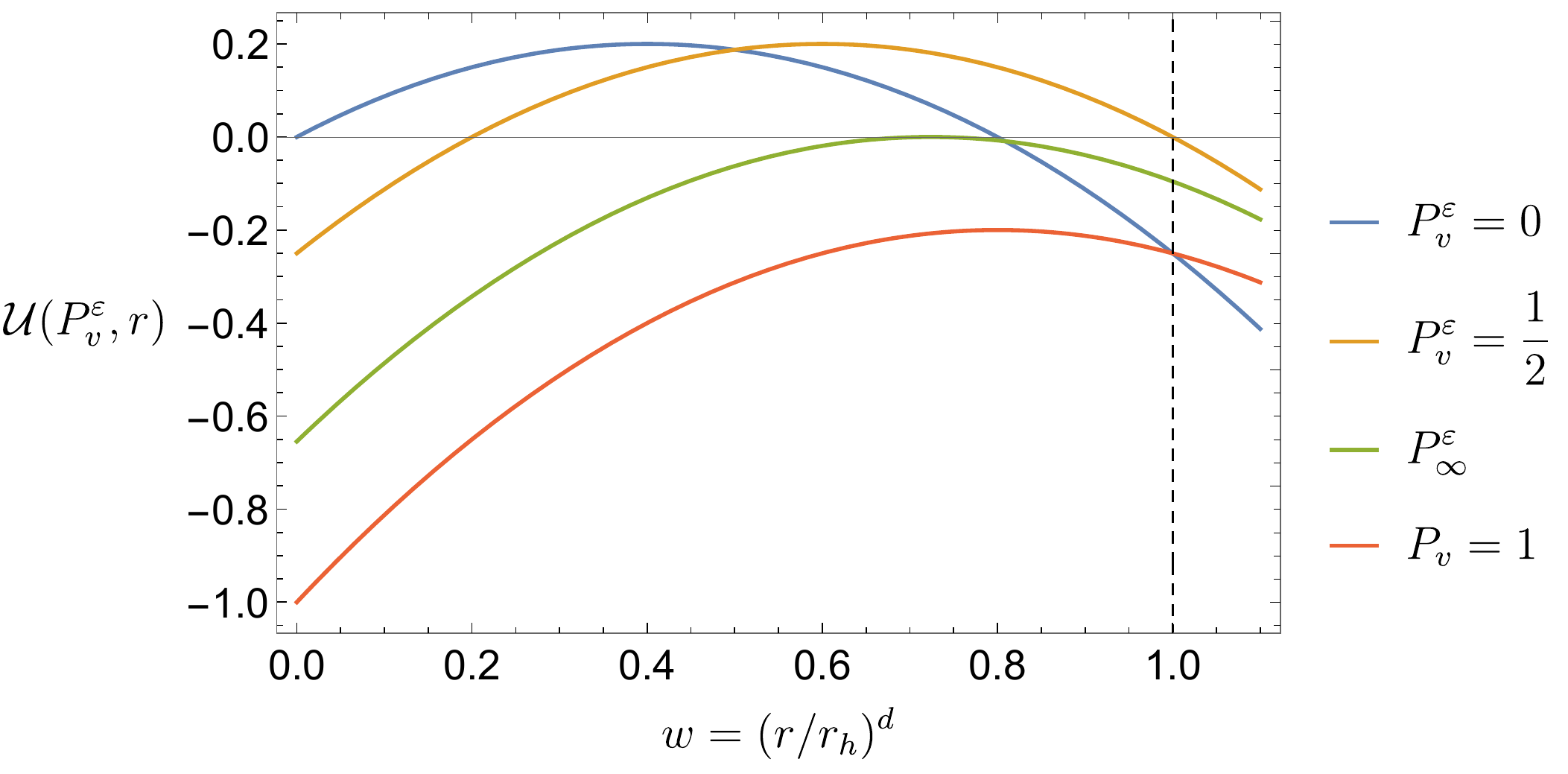}
	\caption{The effective potentials $\mathcal{U}(P_v^\varepsilon, r)$ for the future boundary $\Sigma_+$ (left) and the past boundary $\Sigma_-$ (right), respectively. Hence $\veps=+ \,(-)$ in the left (right) panel. For this figure, we have chosen $d=2$ and $\alpha_\pm=1=\alB$. }
	\label{fig:Potential01}
\end{figure}

We note that 
\begin{equation}
\begin{split}
\mathcal{U}(P_v^{\varepsilon},r=0)&= - (P_v^{\varepsilon}/\alpha_\veps)^2 < 0 \,, \\
\mathcal{U}(P_v^{\varepsilon},r\to \infty)&\simeq-\(1+\frac{\alB^2}{d^2\alpha_\veps^2}\)\(\frac{r}{L}\)^{2d}<0 \,, \\
\end{split}
\label{gamma6}
\end{equation}
In general, we will be interested in the situation where $P_v^{\varepsilon}$ is chosen such that the potential is positive over some range of $r$. The extremal surfaces of interest are then given by the zero-energy trajectories where the particle comes in from $r=\infty$, reflects off of the potential at the minimal radius $\rmin$ where $\mathcal{U}(P_v^{\pm},\rmin)=0$, and then returns to $r=\infty$. We illustrate some typical potentials in figure \ref{fig:Potential01}. 

Before proceeding, let us first show that the surfaces satisfying eqs.~\eqref{gfix2} and \eqref{simp2} with any conserved momentum $P_v^\epsilon$ are precisely the constant mean curvature slices given in eq.~\reef{eq:CMCextr}. Using the infalling coordinates \reef{infall}, the trace of the extrinsic curvature of the hypersurface $v(r)$ can be expressed as\footnote{As discussed in footnote \ref{footnote:defineK}, the normal vector $n^\mu$ is taken to be future-directed for both $\Sigma_\pm$, meaning that 
its components after lowering an index satisfy $n_\mu \propto (-1,v'(r),\vec{0})$.}
\begin{equation}
K =  \frac{   4(d-1)v'(r)  - \(2\(d-1\) f(r)  - r f'(r)\) (3- f(r)v'(r)) (v'(r))^2    - 2r v''(r)}{ 2r \( 2 v'(r) - f(r) v'(r)^2) \)^{3/2}} \,,
\end{equation}
where $v'\equiv\partial v(r)/\partial r$.
Using our intrinsic parametrization of the boundary surfaces, \ie $(v(\sigma), r(\sigma))$, we can rewrite the above result as
\begin{equation}
\begin{split}
K &=    \frac{ 4(d-1) \dot{v}\dot{r}^2 - \left( 2(d-1)f(r) +r f'(r)  \right) \(3\dot{r}- f(r) \dot{v}\)\dot{v}^2      - 2 r\( \dot{r}\ddot{v}-\dot{v}\ddot{r}\) }{2 r (2 \dot{v} \dot{r}-f(r) \dot{v}^2)^{3/2}  }  \,. 
\end{split}
\end{equation}
After imposing our gauge-fixing condition \eqref{gfix2}, this expression reduces to 
\begin{equation}
 K =\frac{1}{ \sqrt{-f(r) (r/L)^{2(d-1)}+\dot{r}^2 }} \( -\(\frac{L}{r} \)^{d-1}r \ddot{r} +  \frac12 \(2(d-1) f(r) +rf'(r) \)  \)\,,
\end{equation}
where we have substituted $\dot{v}(\sigma)$ and $\ddot{v}(\sigma)$ (derived by taking the derivative of eq.~\eqref{gfix2}) in terms of $\dot{r}(\sigma)$ and $\ddot{r}(\sigma)$. We note that this expression does not distinguish between the future and past boundary surfaces, however, this distinction arises in the final step where we eliminate $\dot r$ and $\ddot r$ using the radial equation of motion \reef{simp2}. The final result with any conserved momentum $P_v^\veps$ is then
\begin{equation}\label{eq:CMCexample}
K_{\Sigma_{\varepsilon}} (P_v^\varepsilon) = -\veps\, \frac{ \alB }{\alpha_\varepsilon L}\,.
\end{equation}
Hence we have precisely recovered eq.~\reef{eq:CMCextr} and confirmed that the extremal boundaries are constant mean curvature slices.
While we are generally interested in positive coefficients, \ie $\alB,\,\alpha_\pm>0$, we note that the above expression is invariant under $ \alB/\alpha_\veps \to - \alB/\alpha_\veps , \ \varepsilon \to - \varepsilon$. That is, the two surfaces $\Sigma_\pm$ are interchanged if, \eg we flip the sign of $\alB$.

\begin{figure}[ht!]
	\centering
	\includegraphics[width=2.7in]{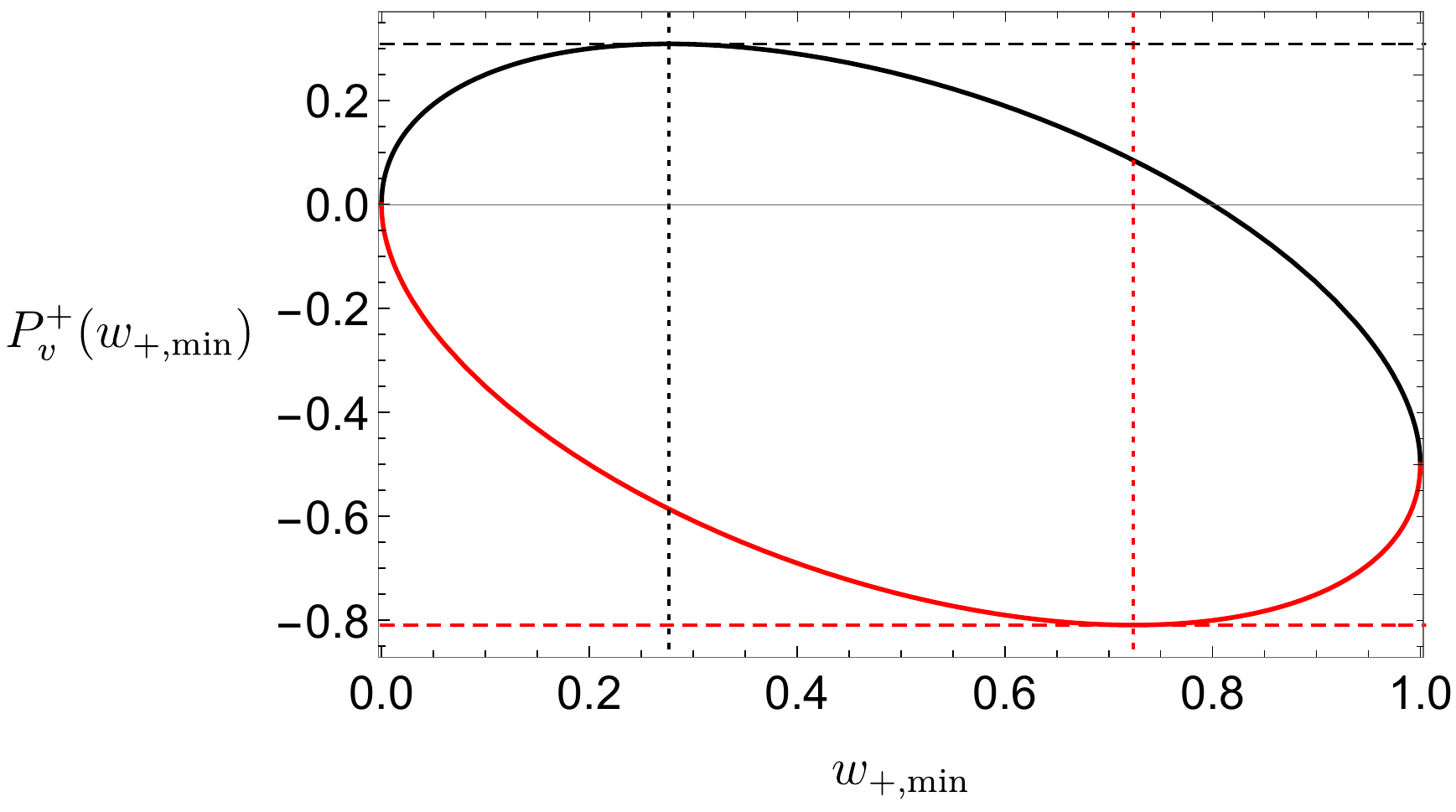}
	\includegraphics[width=3.25in]{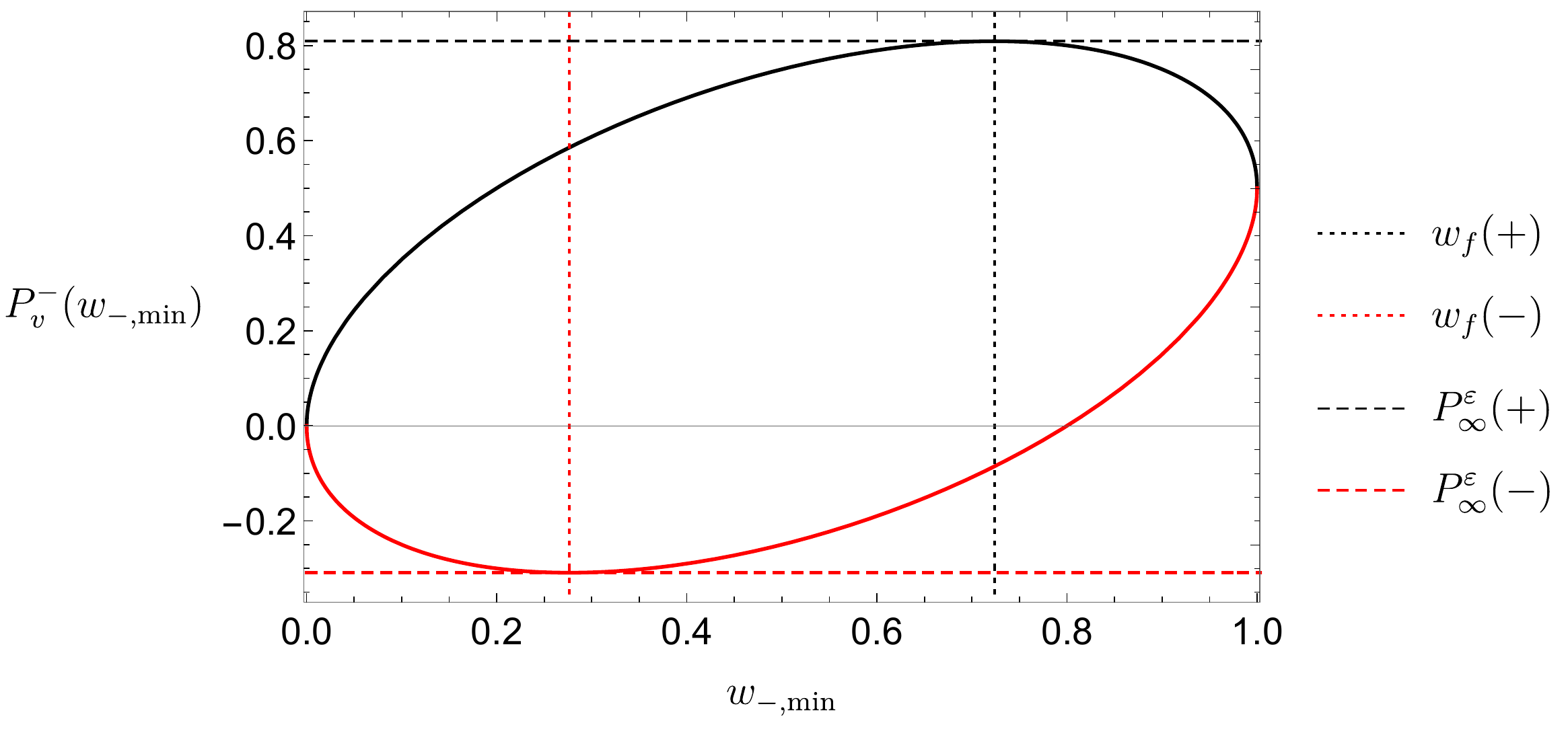}
	\caption{Illustration of the relation between the conserved momentum $P_v^{\varepsilon}$ and the minimal radius $\wemin$ in eq.~\reef{ellipse}, for $\alB=1=\alpha_\pm$, $x=1$ and $d=2$.
	The solutions $P_v^{\varepsilon}(+)$ in eq.~\reef{turn44} correspond to the black arcs, while $P_v^{\varepsilon}(-)$ correspond to the red arcs. The critical momenta $P_\infty^\veps(\pm)$ in eq.~\reef{limits22} correspond to the maximum and minimum vertical extrema of the ellipses marked by the horizontal dashed lines. The corresponding turning points $w_f(\pm)$ given in eq.~\reef{limits33} are marked by the vertical dashed lines.}
	\label{fig:Pvwmin}
\end{figure}

Turning to the time evolution of the extremal surfaces in the planar black hole background \reef{infall}, we introduce a dimensionless radial coordinate and mass parameter,
\begin{equation}\label{simp33}
w \equiv \(\frac{r}{r_h} \)^d \,,\qquad x \equiv \(\frac{r_h}{L} \)^d \propto M \,.
\end{equation}
Recall the extremal surfaces are described by the classical mechanics problem in eqs.~\reef{simp2} and \reef{simp22} and the trajectories of interest come in from $r=\infty$, reflect of the potential at $\mathcal{U}(P_v^{\varepsilon}, r=\rmin)=0$ and return to $r=\infty$.
Hence using eq.~\reef{simp22}, the turning points $\rmin$ for the surfaces $\Sigma_\pm$ are determined by 
\begin{equation}\label{eq:definermin}
 -f(\rmin) \(\frac{\rmin}{L}\)^d = \(P_v^{\pm} \pm \frac{\alB}{ d\,\alpha_\pm }\(\frac{\rmin}{L}\)^d   \)^2 \,.
\end{equation}
Using the dimensionless variables in eq.~\reef{simp33}, the latter is simply recast as the following quadratic equation
\begin{equation}\label{ellipse}
	0=\left(1+\frac{\alB^2}{d^2\alpha_\veps^2}\right)x^2\,\wemin^2-\left( x-\veps\,\frac{2\alB}{d\,\alpha_\veps}\,P_v^\veps\right)x\,\wemin+P_v^{\veps}{}^2\,.
\end{equation}
This corresponds to an ellipse in the ($P^\veps_v,\wemin$)-plane, as illustrated in figure \ref{fig:Pvwmin}. Solving eq.~\reef{ellipse} for the turning points yields
\begin{equation}
\begin{split}
\wemin=  \frac{d\,\alpha_\veps }{ 2x ( d^2 \alpha_\veps^2 +  \alB^2)}\[ (d\,\alpha_\veps x -2\veps \alB  {P_v^\veps}) +\hat{\lambda} \sqrt{d^2\alpha_\veps^2\(x^2-4 P_v^\veps{}^2\)- 4\veps d\,\alpha_{\varepsilon}\alB \, P_v^\veps x  } \]\,,
\end{split}
\label{turn22}
\end{equation}
where $\hat\lambda=\pm$.\footnote{A simple check of this result is to consider the limit $\alpha_+ \to \infty$ as in eq.~\reef{limit1}, which leads to the result associated with the extremal volume surface, \ie 
$\lim\limits_{\alpha_+ \to \infty} w_{+,\mathrm{min}}=  \frac{1}{2} \big( 1 +\hat\lambda \sqrt{1 - {4P^{+\,2}_v}/{x^2}} \big)$.} Alternatively, for a given minimal radius, we can determine the conserved momentum $P_v^\epsilon$ as 
\begin{equation}
 P_v^{\varepsilon} (\pm) = x \(  -\frac{\varepsilon \alB}{ d \alpha_{\varepsilon}}\, w_{\varepsilon,\rm{min}} \pm \sqrt{ w_{\varepsilon,\rm{min}} \( 1 - w_{\varepsilon,\rm{min}}  \) }    \)   \,.
 \label{turn44}
\end{equation}
Here the plus/minus signs correspond to moving the boundary time slice towards positive or negative times. Note that for $P_v^{\varepsilon}$ to be real, we must have $0\le\wemin\le1$. That is, the turning point lies between the singularity (\ie $w=0$) and the horizon (\ie $w=1$). Finally, we remark that in this expression, the conserved momentum $P_v^{\varepsilon}$ is proportional to the dimensionless mass parameter $x \propto M$.

Recall that we are generally interested in the case where there are two real solutions for eq.~\reef{turn22}, \eg see figure \ref{fig:Potential01}, and further the larger root is the correct turning point for the trajectories of interest. That is, we must choose $\hat\lambda=+$ in the above expression.\footnote{The choice $\hat\lambda=-$ would correspond to a trajectory that begins at the singularity $r=0$, reflects off the potential at the corresponding $\wemin(\hat\lambda=-)$ (which actually corresponds to the maximum radius, rather than the minimum), and returns to $r=0$. Hence these extremal surfaces are anchored to the singularity, rather than the asymptotic boundary.} As illustrated in figure \ref{fig:Pvwmin}, we must also choose the momentum $P^\veps_v$ such that the argument of the square-root above is positive, \ie
\beq
 P^\veps_\infty(-)\le P^\veps_v\le P^\veps_\infty(+)
\label{limits}
\eeq
where
\beq\label{limits22}
\begin{split}
P^\veps_\infty(+)&\equiv
\frac{x}2\(\sqrt{1+\frac{\alB^2}{d^2\alpha^2_\veps}}-\veps\,\frac{\alB}{d\,\alpha_\veps}\)\,,\qquad
 P^\veps_\infty(-)\equiv -\frac{x}2\(\sqrt{1+\frac{\alB^2}{d^2\alpha^2_\veps}}+\veps\,\frac{\alB}{d\,\alpha_\veps}\)\,.
\end{split}
\eeq
The limiting values $P^\veps_\infty(\pm)$ correspond to the critical values of the momenta where eq.~\reef{turn22} yields a single solution. The corresponding turning points $w_f=(r_f/r_h)^d$ are
\begin{equation}\label{limits33}
w_{f}(\pm) =  
\frac{1}{2}\(1 \mp \frac{\veps}{\sqrt{d^2\alpha_\veps^2/\alB^2 +1}}\)\,.
\end{equation}
Tuning the momenta to the critical values \reef{limits22} produces the critical potentials with
\begin{equation}
\mathcal{U}(P_v^\varepsilon,w)\big|_{w=w_f}=0\,, \quad \partial_w \mathcal{U}(P_v^\varepsilon,w) \big|_{w=w_f, P_v^\varepsilon= P_\infty^\varepsilon}=0\,,
\end{equation}
as in eq.~\eqref{critic}, for which the boundary time diverges. We will find below (\eg see eq.~\reef{eq:twolimits}) that $P^\veps_v\to P^\veps_\infty(+)$ corresponds to $\tau\to+\infty$ and $P^\veps_v\to P^\veps_\infty(-)$, to $\tau\to-\infty$.  

\subsubsection{Time Evolution of Extremal Surfaces}
\begin{figure}[ht!]
	\centering
	\includegraphics[width=2.53in]{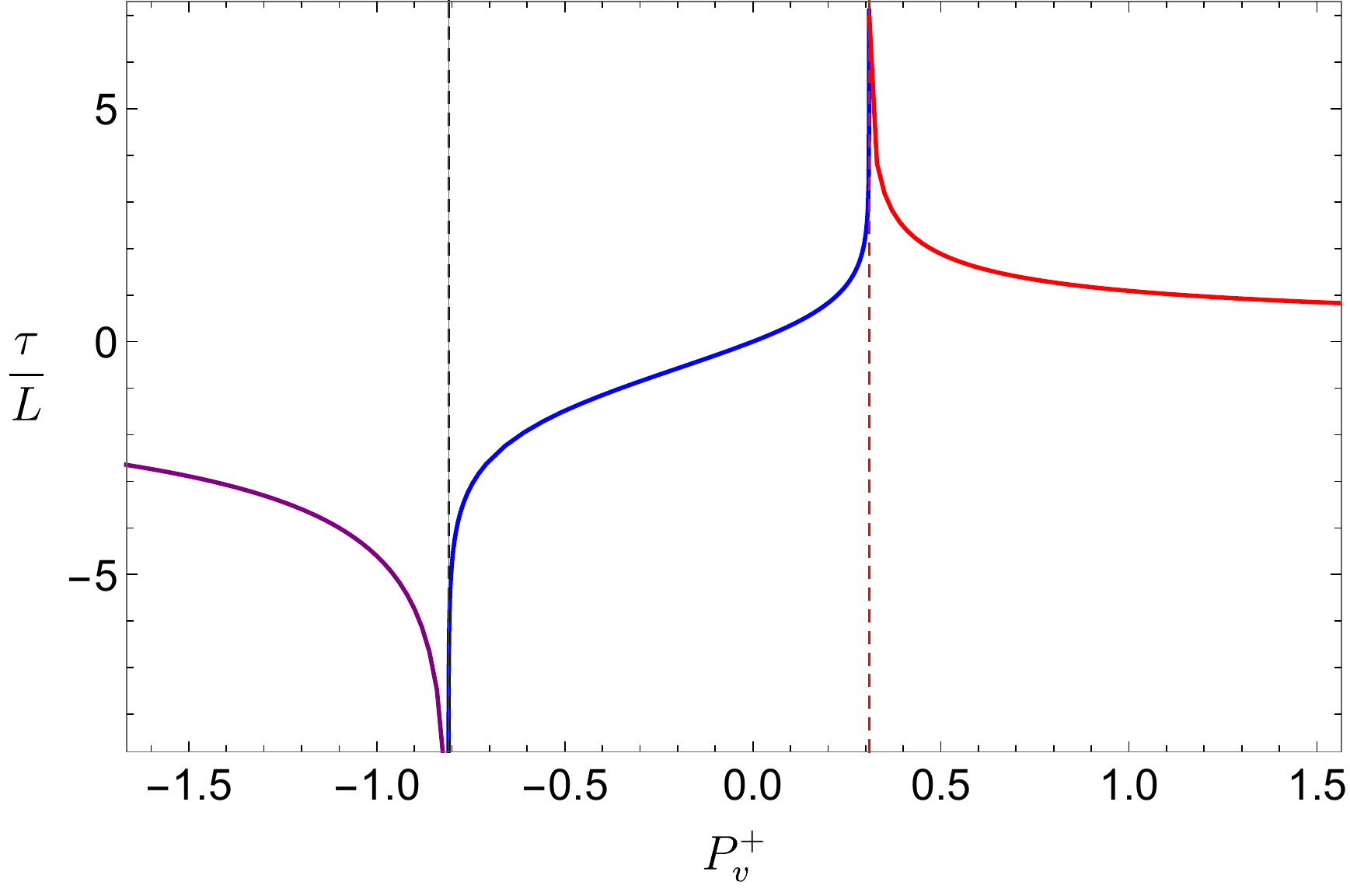}
	\includegraphics[width=3.47in]{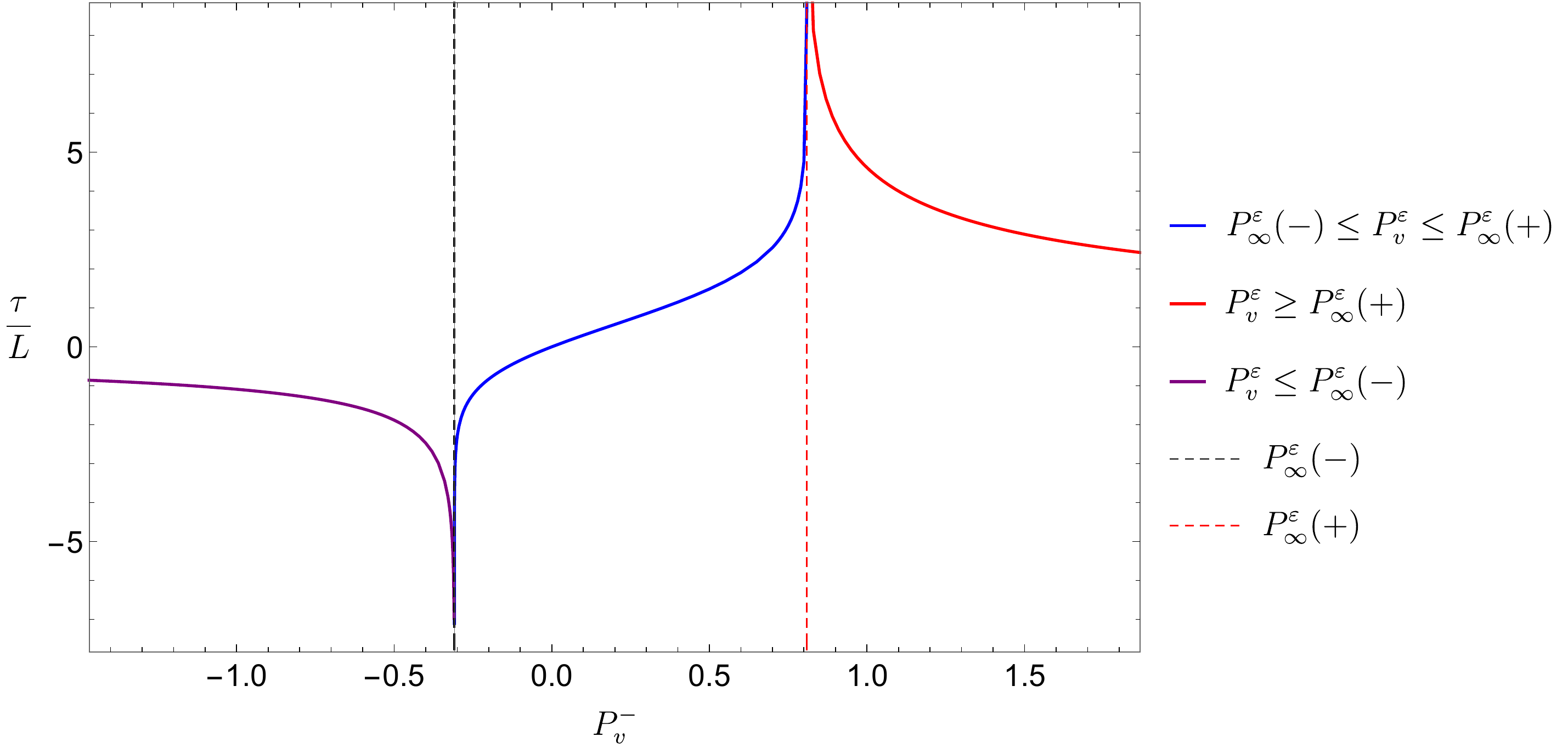}
	\caption{The relation between the boundary time $\tau$ and the conserved momentum $P_v^\varepsilon$. We set $\alB=1=\alpha_\pm$, $x=1$ and $d=2$ in both panels and $\varepsilon=+,-$ for the left and right plot, respectively.}
	\label{fig:tR}
\end{figure}

\subsubsection*{Regime A:  $P_{\infty}^\epsilon(-)\le P_v^\epsilon\le P_{\infty}^\epsilon(+)$} 

With regards to the time evolution of the boundary time slice $\Sigma_\tau$, we have shown in eq.~\eqref{eq:dWdtau01} that the growth rate of the codimension-zero observable $\mC_{\rm gen} (\tau) $ is 
\begin{equation}\label{eq:dCdtauzero}
\frac{d}{d\tau}\mC_{\rm{gen}} = \frac{V_x}{\GN L}\,\( \alpha_+ \, P_v^{+}(\tau)+  \alpha_-\, P_v^{-}(\tau)\)\,.
\end{equation}
Recall that $\alpha_\pm$ were absorbed as part of the prefactors in the previous subsection -- see discussion below eq.~\reef{noun9}. Hence we would like to determine the conserved momentum $P_v^\varepsilon(\tau)$ as a function of the boundary time. From the definition of the infalling coordinate \reef{infall} and the equations of motion \reef{eq:dots}, we can obtain the evolution of time coordinate  $t(\sigma)$ along the extremal surfaces as follows 
\begin{equation}\label{eq:tdot}
\dot{t}(\sigma) \equiv   \dot{v} - \frac{\dot{r}}{f(r)} = -\frac{1}{f(r)} \(  P_v^\varepsilon + \frac{\varepsilon\,\alB}{d\, \alpha_\veps} \( \frac{r}{L} \)^d   \)\,.
\end{equation}
Now, if we fix the value of conserved momentum $P_v^\varepsilon$, we can integrate the above expression using $\dot r =\sqrt{-\mathcal{U}(P_v^\varepsilon,r)}$ to determine the corresponding boundary time $t_{\mt{R}}$ as
\begin{equation}
t_{\mt{R}}- t(\rmin) = - \int_{\rmin}^\infty \frac{dr}{ f(r)\sqrt{-\mathcal{U}(P_v^\varepsilon,r)}} \,  \(P_v^\varepsilon + \frac{\varepsilon\alB}{d\, \alpha_\veps} \( \frac{r}{L} \)^d   \) \,,
\label{horse2}
\end{equation}
where $t(\rmin)$ denotes the value of the time at the minimal radius. However, with our symmetric setup (\ie $t_{\mt{R}}=t_{\mt{L}}$), the latter is simply given by $t(\rmin)=0$. An example of $\tau=2t_\mt{R}$ as a function of $P^\veps_v$ is explicitly plotted in figure \ref{fig:tR}.  

There are three potentially singular points in the integral: $r=\infty$, $r=r_h$, and $r=\rmin$. We consider each of these in turn. First, near the asymptotic boundary, the leading behaviour of the integrand is given by 
\begin{equation}
  \(P_v^\varepsilon + \frac{\varepsilon\alB}{d\, \alpha_\veps} \( \frac{r}{L} \)^d   \) \frac{1}{ f(r)\sqrt{-\mathcal{U}(P_v^\varepsilon,r)}} \sim \frac{1}{f(r)} \sim \frac{1}{r^2} \,,  
\end{equation}
which yields a finite integral as $r \to \infty$. Second, one may worry about the horizon $r=r_h$ where $f(r)$ vanishes and so the integrand diverges as 
\begin{equation}
     \(P_v^\varepsilon + \frac{\varepsilon\alB}{d\, \alpha_\veps} \( \frac{r}{L} \)^d   \) \frac{1}{ f(r)\sqrt{-\mathcal{U}(P_v^\varepsilon,r)}} \sim    \(P_v^\varepsilon + \frac{\varepsilon\alB}{d\, \alpha_\veps} \( \frac{r_h}{L} \)^d   \) \frac{1}{f'(r_h)(r-r_h)}\,.
\end{equation}
However, as noted below eq.~\ref{eq:boundarytimetR02}, one can evaluate the integral using the Cauchy principal value, \ie 
\begin{equation}
\lim_{\epsilon \to 0} \( \int_{\rmin}^{r_h - \epsilon} +  \int_{r_h + \epsilon}^{\infty} \) \,  \(   \frac{P_v^\varepsilon + \frac{\varepsilon\alB}{d\, \alpha_\veps} \( \frac{r}{L} \)^d}{ f(r)\sqrt{-\mathcal{U}(P_v^\veps,r)}} \) \, dr\,,
\end{equation}
which is finite due to the cancellation of divergences from the two sides of the horizon.

Finally, $\mathcal{U}(P_v^\varepsilon,r)$ vanishes at $r=\rmin$. However, as noted previously, the resulting divergence is integrable as long as $\partial_r\mathcal{U}(P_v^\varepsilon,r)|_{r=\rmin}\ne0$, and the corresponding boundary time is finite. However, in the special case where the momentum is tuned so that the effective potential has a double zero at the turning point, the boundary time diverges. As discussed above and shown figure \ref{fig:Pvwmin}, there are two critical values of the momentum, $P^\veps_\infty(\pm)$, for each sign of $\veps$.
As shown in figure \ref{fig:tR} then, one finds
\begin{equation}\label{eq:twolimits}
\begin{split}
 P_v^\varepsilon &\to P_{\infty}^\varepsilon(+) \,, \qquad   \tau \to +\infty\,,\\
 P_v^\varepsilon &\to P_{\infty}^\varepsilon(-) \,, \qquad   \tau \to -\infty\,.
\end{split}
\end{equation}

Combining the late-time limit for the growth rate in eq.~\eqref{eq:dCdtauzero} with the critical momenta in \reef{limits22}, we find  
\begin{equation}\label{eq:dCdtau}
\begin{split}
    \lim_{\tau \to \infty} \frac{d\mC_{\rm{gen}}}{d\tau} &= \frac{V_x}{\GN L}\,\( \alpha_+ \, P_{\infty}^{+}(+)+  \alpha_- P_{\infty}^{-}(+)\) \\ 
    &= \frac{8 \pi M }{d(d-1)} \(  \sqrt{\alB^2 +d^2\, \alpha_+^2} + \sqrt{\alB^2+d^2\, \alpha_-^2}\)  \,,
\end{split}
\end{equation}
which is the expected linear growth with a rate proportional to the mass of the black hole.\footnote{We note that for $\tau\to-\infty$, the result is identical except for an overall sign.} It is interesting to consider the two limits pointed out in eqs.~\reef{limit1} and \reef{limit2}. First, we take $\alpha_+ \to +\infty$ (while keeping $\alB$ and $\alpha_-$ fixed) to recover the standard CV result. Combining eqs.~\reef{limit1} and \eqref{eq:dCdtau} then yields
\begin{equation}
\begin{split}
 \lim_{\tau \to \infty} \frac{d\mC_{\rm V}}{d\tau}  =  \lim_{\alpha_+ \to +\infty}  \lim_{\tau \to \infty}  \(\frac{1}{\alpha_+}\,\frac{d\mC_{\rm{gen}}}{d\tau} \)&= \frac{8 \pi M }{d-1}  \,,
\end{split}
\end{equation}
which is the expected growth rate for the CV conjecture, \eg \cite{Stanford:2014jda,Carmi:2017jqz}. Alternatively, following eq.~\reef{limit2}, we take $\alpha_\pm \to 0$ (while keeping $\alB$  fixed) to recover the CV2.0 proposal. The corresponding late-time linear growth is then given by 
\begin{equation}
\begin{split}
 \lim_{\tau \to \infty} \frac{d\mC_{\rm SV}}{d\tau}  =  \lim_{\alpha_\pm \to 0}  \lim_{\tau \to \infty} \(\frac{1}{\alB}\,\frac{d\mC_{\rm{gen}}}{d\tau} \)&= \frac{16 \pi M }{d(d-1)}  \,,
\end{split}
\end{equation}
which again reproduces the expected growth rate for the CV2.0 proposal, \eg \cite{Couch:2016exn} (where the bulk pressure is identified as $P=- \frac{\Lambda}{8\pi \GN}$).

\subsubsection*{Regime B:  $P_v^\epsilon 
\le P_{\infty}^\epsilon(-)$ or $P_v^\epsilon \ge  P_{\infty}^\epsilon(+)$ }

Similar to the codimension-one case discussed in Appendix \ref{revone}, we can also obtain multiple extremal surfaces for $\Sigma_{\pm}$. 
Besides the extremal surface joining the two sides at the minimal radius, there is also another type of extremal surface which moves from $r=\infty$ into the singularity at $r=0$ and where we allow it to reflect outward to $r=\infty$.\footnote{This type of extremal surfaces are similar to those discussed in \cite{Jorstad:2022mls} for de Sitter spacetime. However, the volume of the portion along the singularity in the AdS black hole is zero.}. These solutions of the extremality equations correspond to taking either $P_v^{\varepsilon} \ge P_{\infty}^{\varepsilon}(+)$ or $P_v \le P_{\infty}^{\varepsilon}(-)$. In other words, we are outside the range in eq.~\reef{limits}, thus the  effective potential never reaches zero, and there is no turning point before the singularity -- see figure \ref{fig:Potential01}. 
The complete connected but {\it non-smooth} extremal surfaces $\Sigma_\pm$ are obtained by joining two symmetric left and right portions at $r=0, t=0$ on the singularity.\footnote{Such solutions could be considered with the usual CV proposal but are traditionally discarded. However, within the framework of generalized codimension-one observables constructed in \cite{Belin:2021bga}, these non-smooth surfaces arise as a limit of a class of smooth surfaces when the higher curvature couplings are taken to zero. But we would like to emphasize that the extremal surfaces with the maximum volume are still the traditional smooth surfaces, \ie the non-smooth surfaces are subdominant saddle points.}  

Let us consider the asymptotic values of the boundary time with the limit $|P_v^{\varepsilon}| \to \infty$. Applying this limit to eq.~\eqref{horse2}, we find 
\begin{equation}
\begin{split}
 t_{\mt{R}} \( P^\veps_v \to \pm\infty\)  &= \mp \int^\infty_0 \frac{dr}{f(r)}
 = \mp \frac{L^2}{r_h}\, \frac{\pi}{d} \,\cot\(\frac{\pi}{d}\) \,.\\
\end{split}
\end{equation}
for $d>1$ \footnote{Similar to the codimension-one extremal surface we explored here, the corresponding codimension-$d$ extremal surface, \ie geodesics bouncing oﬀ at the singularity, was studied before in \cite{Fidkowski:2003nf} and was found to start from the same boundary time.}. In line with the previous discussion, we have set $t(\rmin=0)=0$ in this expression. As shown in figure \ref{fig:tR}, as the conserved momentum $P_v^\veps$ decreases from infinity to the critical value $P_\infty^\varepsilon (+)$, the corresponding boundary time $\tau$ increases from the above asymptotic value to $+\infty$.
Similarly, as $P_v^\veps$ increases from minus infinity to $P_\infty^\varepsilon (-)$, the boundary time increases from the above asymptotic value to $-\infty$.


\subsubsection{Maximization}\label{sec:max}

Although the surfaces above are not smooth, we may still consider them as candidate saddle points which compete with the smooth extremal surfaces considered previously for $P_{\infty}^\epsilon(-)\le P_v^\epsilon\le P_{\infty}^\epsilon(+)$. Here, we examine the question of which of these surfaces maximize the value of the observable $\mC_{\rm gen}$ given in eq.~\eqref{eq:CMCfunc}, and we find that the smooth extremal surfaces 
always give the maximal value.

Let us rewrite the codimension-zero observable $\mC_{\rm gen}$ as  
\begin{equation}\label{eq:volumepmintegral}
\begin{split}
    \mC_{\rm gen} (\tau) 
    &=\frac{2V_x}{\GN L} \sum_{\varepsilon=+,-} \int^{\infty}_{r_{\veps,\rm min}}  \,\frac{\alpha_\veps\, dr}{\sqrt{-\mathcal{U}( P_v^{\varepsilon}, r)} } \[ \(\frac{r}{L}\)^{2(d-1)} + \frac{\varepsilon\,\alB}{d\,\alpha_\veps}  \(\frac{r}{L} \)^d  \frac{1}{f(r)}\( P_v^\varepsilon + \frac{\varepsilon\,\alB}{d\,\alpha_\veps}  \(\frac{r}{L} \)^d  \)  \]  \\
\end{split},
\end{equation}
by using the extremality equations \reef{simp2} and \reef{simp22} in eq.~\eqref{hurrah}. Furthermore, it will be useful to recast this expression as a time integral, {
\begin{equation}
    \begin{split}
        \mC_{\rm gen} (\tau) &= -\frac{2V_x}{\GN L} \sum_{\varepsilon=+,-} \int_{0}^{t_{\mt{R}}} \frac{ \alpha_{\varepsilon}\, dt }{P_v^{\varepsilon} + \frac{\varepsilon\,\alB}{d\,\alpha_\veps} \( \frac{r}{L} \)^d    }\\
        &\qquad\qquad\qquad\times\ \[ f(r)\(\frac{r}{L}\)^{2(d-1)} + \frac{\varepsilon\,\alB}{d\,\alpha_\veps}  \(\frac{r}{L} \)^d \( P_v^{\varepsilon} + \frac{\varepsilon\,\alB}{d\,\alpha_\veps}  \(\frac{r}{L} \)^d \)  \] \,. 
    \end{split}
    \label{ggg}
\end{equation}
If we consider the smooth extremal surfaces with $P_v^\varepsilon \le P_\infty^\varepsilon(+)$, then in the late-time limit $\tau =2 t_{\mt{R}} \to \infty$ (with $P_v^\varepsilon \to P_\infty^\varepsilon(+)$), both extremal surfaces $\Sigma_\pm$ approach a constant radius surface at $r=r_{\pm,f}$. Then  after substituting eq.~\eqref{eq:definermin} into  eq.~\eqref{ggg}, $\mC_{\rm gen}$ reduces to 
\begin{equation}\label{eq:CVlimit01}
    \mC_{\rm gen} (\tau \to \infty ) =\frac{V_x}{\GN L} \sum_{\varepsilon=+,-} \alpha_{\varepsilon} \,  P_{\infty}^{\varepsilon}(+)\, \int_{0}^{\tau \to \infty}   d\tau \,.
\end{equation}

On the other hand, we have found another class of non-smooth extremal surfaces with $P_v^\varepsilon \ge P_\infty^\varepsilon(+)$. In this case, late boundary time $\tau \to \infty$ is achieved by approaching $P_{\infty}^\varepsilon$ from the above, but the corresponding extremal surface is not a constant radius surface. This class of extremal surfaces starts from the asymptotic boundary and extends to the singularity at $r=0$. Using the inequality $P_v^\varepsilon \ge P_{\infty}^\varepsilon(+)$, the following always holds along these surfaces
\begin{equation}
\mathcal{U}(P_v^\varepsilon, r)= -f(r) \(\frac{r}{L}\)^d - \(P_{\infty}^{\varepsilon} + \frac{\varepsilon \alB}{ d \alpha_{\varepsilon}}\(\frac{r}{L}\)^d   \)^2 \le 0\,.
\end{equation}
As noted above, in the late time limit, we approach the critical momentum, which we denote as $ P_{\infty}^{\varepsilon} (+) \leftarrow P_v^{\varepsilon}$,\footnote{At the infinite time, we have four possible configurations due to two extremal surfaces.}, we can decompose the integral \reef{ggg} as 
\begin{equation}
\begin{split}
 \lim_{ P_{\infty}^{\varepsilon} (+) \leftarrow P_v^{\varepsilon}  }    \mC_{\rm gen} &= \frac{V_x}{\GN L} \sum_{\varepsilon=\pm}  \int_{0}^{\tau} \frac{ \alpha_{\varepsilon}\, d\tau }{P_{\infty}^{\varepsilon}  + \frac{\varepsilon\,\alB}{d\,\alpha_\veps} \( \frac{r}{L} \)^d    } \[ P_{\infty}^{\varepsilon}  \(P_{\infty}^{\varepsilon}  + \frac{\varepsilon\,\alB}{d\,\alpha_\veps}  \( \frac{r}{L} \)^d   \) +\mathcal{U}(P_{\infty}^\epsilon, r) \] \,.
\end{split}
\end{equation}
where the first term is nothing but the result in eq.~\eqref{eq:CVlimit01}. Firstly, we note that all spacelike hypersurface anchoring on the infinite boundary time stay inside the horizon. Noting the various solutions of $P_{\infty}^{\varepsilon}$ given in eq.~\eqref{limits22}, we will always have 
\begin{equation}
P_{\infty}^{\varepsilon}(+)  + \frac{\varepsilon\,\alB}{d\,\alpha_\veps}  \( \frac{r}{L} \)^d  > 0 
\end{equation}
for $r\le r_h$. Secondly, we obviously have $\mathcal{U}(P_{\infty}^\epsilon, r) \le 0$ for this type of extremal surface. Assuming $\alpha_\varepsilon \ge 0$ for both surfaces $\Sigma_\pm$, we can conclude that 
\begin{equation}
 \lim_{ P_v^\varepsilon \to P_{\infty}^\varepsilon(+)}    \mC_{\rm gen}(\tau \to \infty)  \ge  \lim_{ P_{\infty}^{\varepsilon} (+) \leftarrow P_v^{\varepsilon} }  \mC_{\rm gen}(\tau \to \infty) \,.
\end{equation}
In other words, the late-time extremal surface yielding the maximal value of the observable is the smooth extremal surface  at $r=r_f$, \ie the  surface approached by taking the limit $P_v^\varepsilon \to P_{\infty}^\varepsilon(+)$ from below.
Using the same argument as for the maximal volume associated with the effective potential (see Appendix \ref{sec:localmaxima} for more details), one can further show that $\mC_{\rm gen}$ is maximized at any boundary time $\tau$ by the surfaces with the smaller momentum, \viz
\begin{equation}
 \mC_{\rm gen}(\tau )  \big|_{P_v^\varepsilon \le P_{\infty}^\varepsilon(+)} >  \mC_{\rm gen}(\tau ) \big|_{P_v^\varepsilon  \ge P_{\infty}^\varepsilon(+)}  \,.
\end{equation}
A similar proof also applies for negative boundary times with  
\begin{equation}
 \mC_{\rm gen}(\tau )  \big|_{P_v^\varepsilon \ge P_{\infty}^\varepsilon(-)} >  \mC_{\rm gen}(\tau ) \big|_{P_v^\varepsilon  \le P_{\infty}^\varepsilon(-)}  \,.
\end{equation}


\section{Gravitational Observables and the Symplectic Form} \label{symp}

In this section we discuss gauge invariant functionals on the phase space of asymptotically AdS gravity theories. 
An important concept will be the Hamiltonian vector field conjugate to such observables. In classical mechanics, a Hamiltonian vector field $X_f$ conjugate to a function $f$ on phase space is obtained by pulling up the indices of $\mathrm{d}f$ using the inverse 
of the symplectic form $\Omega$:
\beq
\label{eq:babyconjugate}
(X_f)^a=\Omega^{ab}\partial_b f.
\eeq
Such vector fields on phase space are to be thought of as linearized variations of initial data. This is more natural in field theory, so we will refer to the Hamiltonian vector field as the conjugate variation from now on.

In gravitational field theories, the variation conjugate to an observable can be found using a construction due to Peierls \cite{Peierls:1952cb}, which will be briefly reviewed below in section \ref{sec:Peierls} (see \cite{Kirklin:2019xug,Harlow:2019yfa,Harlow:2021dfp,Goeller:2022rsx} for recent uses and a more in-depth review). The Peierls construction is very general, but a disadvantage is that finding the conjugate variation requires solving equations of motion for an auxiliary system. It will therefore be advantageous to directly adapt the naive formula \eqref{eq:babyconjugate} in gravity, which does not require solving differential equations. When using the canonical formalism, in section \ref{sec:canonical}, we first analyze the conjugate variation for functionals defined on spacelike codimension-one slices on which the functional is extremal\footnote{We will only need that the equations of motion are second order, so the discussion in this section applies also to Lovelock gravity.}. The canonical construction is equivalent with the one of Peierls, and indeed we will show in section \ref{sec:peierls_extremality} that observables are required to be extremized in the Peierls construction. 

There are certain gauge invariant observables that do not seem to be in this class of extremized functionals. As explained in section \ref{complexsquared}, we could fix a gauge invariant slice by extremizing a functional, and then integrate a different functional on this slice. An example in this class would be to integrate any covariant density on the maximal volume slice. We will close this section by explaining how these observables are in fact related to extremized functionals, and show how to find their conjugate variations.

\subsection{Canonical formalism for codimension one}
\label{sec:canonical}

 In the following, we will focus on pure GR for simplicity. However, it is straightforward to extend the discussion by including matter. According to the canonical decomposition, the initial data are the induced metric $h_{ij}$ and its conjugate momenta $\pi^{ij}$ on a Cauchy slice $\Sigma$. The momenta are related to the extrinsic curvature of the Cauchy slice $\Sigma$ via 
 \begin{equation}
    \pi^{ij}\equiv \sqrt{h}(K^{ij}-h^{ij}K)\,.
 \end{equation}
The symplectic form between two variations $\delta_1, \delta_2$ is defined by \cite{Lee:1990nz}
\begin{equation}\label{eq:canonicalsymplectic}
\Omega(\delta_1,\delta_2)=\int_{\Sigma}\[\delta_1\pi^{ij}\delta_2 h_{ij}-\delta_1 h_{ij}\delta_2\pi^{ij} \]\,.
\end{equation}
This quantity is equivalent with the covariant phase space symplectic form up to boundary terms localized on $\partial \Sigma$. In particular, this expression is conserved on-shell, in the sense that it is independent of the slice $\Sigma$ provided that the anchoring $\partial\Sigma$ is fixed.

Let us first consider a functional $W(h_{ij},\pi^{ij})$ associated with a density functional $w(h_{ij},\pi^{ij})$ which depends on the initial data on the Cauchy slice $\Sigma$, \viz
\begin{equation}
W(h_{ij},\pi^{ij}) := \int_{\Sigma} w(h_{ij},\pi^{ij}) \,.
\end{equation}
We can construct the conjugate variation $\delta_{w}$ by naively applying the formula \eqref{eq:babyconjugate}
\begin{equation}\label{eq:gennewYork}
\delta_{w} \pi^{ij}=-\frac{\delta w}{\delta h_{ij}}, \quad \delta_{w} h_{ij} = +\frac{\delta w}{\delta \pi^{ij}}\,.
\end{equation}
Using \eqref{eq:canonicalsymplectic}, it is obvious that
\begin{equation}\label{eq:deltaWdeltaO}
\Omega(\delta,\delta_{w})=\delta W \equiv \int_\Sigma \delta w \,.
\end{equation}
However, for $\delta_{w}$ to be a legitimate initial data variation, it must preserve the constraints of GR. Denoting by $\mathcal{H}_\perp$ the Hamiltonian constraint, and by $\mathcal{H}^k$ the momentum constraint, we must require
\begin{equation}\label{eq:Hamiltonian_variations}
\begin{split}
0=\delta_{w} \mathcal{H}_{\perp} &= \frac{\delta \mathcal{H}_{\perp}}{\delta h_{ij}}\delta_{w} h_{ij} + \frac{\delta \mathcal{H}_{\perp}}{\delta \pi^{ij}}\delta_{w} \pi^{ij}=  \frac{\delta \mathcal{H}_{\perp}}{\delta h_{ij}}\frac{\delta w}{\delta \pi^{ij}}-\frac{\delta \mathcal{H}_{\perp}}{\delta \pi^{ij}}\frac{\delta w}{\delta h_{ij}}\equiv  \lbrace \mathcal{H}_{\perp},w\rbrace_{\rm P.B.}, \\
0=\delta_{w} \mathcal{H}^{k} &=\frac{\delta \mathcal{H}^k}{\delta h_{ij}}\delta_{w} h_{ij} + \frac{\delta \mathcal{H}^k}{\delta \pi^{ij}}\delta_{w} \pi^{ij}=\frac{\delta \mathcal{H}^k}{\delta h_{ij}}\frac{\delta w}{\delta \pi^{ij}}- \frac{\delta \mathcal{H}^k}{\delta \pi^{ij}}\frac{\delta w}{\delta h_{ij}}\equiv \lbrace \mathcal{H}^{k},w\rbrace_{\rm P.B.}\,,
\end{split}
\end{equation}
where ${\rm P.B.}$ denotes the Poisson bracket.\footnote{Given two functions $f$ and $g$ defined on the phase space constructed by the canonical coordinates $(q^a, p_a)$, the Poisson bracket can be defined by the following equivalent expressions: 
\begin{equation}
 \{f, g\}_{\rm P.B.} \equiv \frac{\partial f}{\partial q^a}\frac{\partial g}{\partial p_a}- \frac{\partial f}{\partial p_a}\frac{\partial g}{\partial q^a}\equiv  \Omega (X_g, X_f)= X_g f=-X_f g \,.  
\end{equation}
with the symplectic form taken as $\Omega \equiv dp_{a} \wedge dq^a$. 
This is equivalent to defining the Poisson bracket as $\{f, g\}_{\rm P.B.} \equiv \Omega^{-1} (\mathrm{d}f, \mathrm{d}g)=X_g f $ by using the inverse of the symplectic form $\Omega^{-1}$ ($\mathrm{d}$ is the exterior derivative acting on the phase space).
This latter definition is the one used for the Poisson bracket employed throughout
this section.} The second equation just entails diffeomorphism invariance in the surface, and it is automatic if we have constructed $W$ by integrating the volume density $\sqrt{h}$ times a functional that transforms as a scalar under diffeomorphism fixing the slice $\Sigma$. The first equation is more interesting. Taking the Hamiltonian density $\mathcal{H}_{\rm ADM}= N\mathcal{H}_{\perp} + N^a\mathcal{H}_a $, Hamilton's equations read
\begin{equation}
\label{eq:Hamiltonseq}
\frac{d\pi^{ij}}{dt}=-\frac{\delta \mathcal{H}_{\rm ADM}}{\delta h_{ij}} \,, \quad  \quad \frac{dh_{ij}}{dt}= \frac{\delta \mathcal{H}_{\rm{ADM}}}{\delta \pi^{ij}}  \,,
\end{equation}
where $t$ is the time in an ADM $d+1$ decomposition.\footnote{We will use this notation throughout the section, but we should emphasize that $t$ is the ADM bulk time, and \textit{not} the CFT time.}  We have that $\delta_w \mathcal{H}_{\rm ADM} = N \delta_w \mathcal{H}_\perp$, because $\mathcal{H}_\perp = \mathcal{H}_a =  0$ in the background and we have already argued for $\delta_w \mathcal{H}_a = 0$ from the diffeomorphism invariance of the density $w$. Using this, it follows that the condition that the Hamiltonian constraint is satisfied by this initial data deformation translates to
\beq
\delta_{w} \mathcal{H}_{\perp}=\frac{1}{N} \lbrace \mathcal{H}_{\rm ADM}, w\rbrace_{\rm P.B.} =  \frac{1}{N} \frac{dW}{dt} =0 \,.
\eeq
Since this must hold for an arbitrary choice of lapse, it then implies that $W$ must be extremal under any shape variation of the Cauchy slice $\Sigma$.\footnote{The Hamiltonian constraint will generate time evolution that is trivial at the boundary.} 

We learned therefore that for extremal $W$, \eqref{eq:gennewYork} corresponds to an allowed deformation of initial data. Equivalently, the initial data $(h_{ij}$, $\pi^{ij})$ defines a complete time dependent solution $g_{ab}$ of Einstein's equations, while \eqref{eq:gennewYork} defines a particular linearized solution $\delta_{w} g_{ab}$ around this. These two ways of thinking about $\delta_{w}$ are equivalent in globally hyperbolic spacetimes. In the next subsection, we review the Peierls construction, which obtains the spacetime variation $\delta_{w} g_{ab}$ directly by solving an auxiliary problem. Before doing this, we make some further comments about the conjugate variations given in eq.~\eqref{eq:gennewYork}.

Let us discuss the simplest example, where $W$ is the volume of the Cauchy slice, \ie $W=\int_\Sigma \sqrt{h}$. Gauge invariance requires us to extremize $W$ with respect to $\Sigma$. On such a $\Sigma$, the conjugate variation \eqref{eq:gennewYork} reads as
\begin{equation}
\label{eq:oldnewYork}
\delta_{w} \pi^{ij}= - \frac{1}{2}\sqrt{h}h^{ij}\,, \qquad \delta_{w} h_{ij} =0\,.
\end{equation}
This initial data variation was studied in detail in \cite{Belin:2018bpg} and was called new York transformation, because of its relation to evolution in York time.

It is useful to have the form of the general variation \eqref{eq:gennewYork} in terms of the extrinsic curvature $K_{ij}$ of the slice $\Sigma$ in GR. The final result is given by 
\begin{equation}\label{eq:genYorkK}
\begin{split}
\delta_{w} h_{ij}&=A_{ij}-\frac{{A^k}_k }{d-1}h_{ij} \,,\\
\delta_{w} K_{ij}&=B_{ij}-\frac{{A^k}_k}{2(d-1)} K_{ij}-\frac{h_{ij}}{2(d-1)}\(2 {B^k}_k-K_{kl}A^{kl} \)\,,
\end{split}
\end{equation}
where we have defined 
\begin{equation}
A^{ij}=\frac{1}{\sqrt{h}}\frac{\delta w}{\delta K_{ij}} \,, \qquad  B_{ij}=\frac{1}{\sqrt{h}}\frac{\delta w}{\delta h^{ij}} \,.
\end{equation}
The derivations for those conjugate variations $\delta_{w} h_{ij}, \delta_{w} K_{ij}$ are presented in Appendix \ref{app:variation} with more detail.

\subsection{Holographic dual of the conjugate variation}
In the AdS/CFT context, the bulk symplectic form $\Omega(\delta_1,\delta_2)$ has a remarkably simple expression in the boundary CFT \cite{Belin:2018fxe}:
\begin{equation}\label{eq:bndysympl}
\Omega(\delta_1,\delta_2) = i \int_{S^+}dx^+\int_{S^-}dx^- \langle \mathcal{O}^i(x^+)^\dagger \mathcal{O}^j(x^-)\rangle_{\lambda,\rm{con}}[\delta_1 \lambda_i(x^+)\delta_2 \lambda_j(x^-)-\delta_2 \lambda_i(x^+)\delta_1 \lambda_j(x^-)].
\end{equation}
In this expression, $S=S^+ \cup S^-$ is the Euclidean space that the CFT is defined on, $\langle \mathcal{O}^i(x^+)^\dagger \mathcal{O}^j(x^-)\rangle_{\lambda,\rm{con}}$ denotes the connected correlator of two single trace operators $\mathcal{O}^i$ on $S$ with (potentially complex) sources $\lambda$ turned on, and $\delta_{1,2}\lambda$ are deformations of these sources. The sources are such that if we use them to prepare a state $|\lambda\rangle$ by doing a path integral over $S^-$, then the path integral over $S^+$ prepares the conjugate state $\langle \lambda|$, \ie 
\begin{equation}
\begin{split}
    |\lambda\rangle =T \exp \({-\int_{S^-} d x^- \lambda(x^-) \mathcal{O}(x^-) } \)|0\rangle\,, \quad \langle\lambda| =\langle 0| T \exp \({-\int_{S^+} d x^+ \lambda^\ast(x^+) \mathcal{O}(x^+) } \) \,. 
\end{split}
\end{equation}
In other words, the manifold $S$ with the sources $\lambda$ must have a time reflection plus complex conjugation symmetry. 

After the normalization of the CFT states $|\lambda\rangle$, equation \eqref{eq:bndysympl} is nothing but the Berry curvature two-form. This Berry curvature agrees with the bulk symplectic form in the bulk geometry dual to the state $|\lambda\rangle$ \cite{Belin:2018fxe,Belin:2018bpg,Kirklin:2019ror}. The corresponding dual spacetime can be obtained by finding the (generally complex) Euclidean bulk geometry where the bulk fields have boundary conditions set by the $\lambda$ sources, and then Wick rotating over an appropriate bulk slice to a real Lorentzian spacetime. The variations $\delta_{1,2}\phi$ entering the bulk symplectic form are linearized solutions around this background whose Euclidean boundary conditions are set by $\delta_{1,2}\lambda$.

Supposing we have an extremal functional $W$ in the bulk and have found the conjugate variation $\delta_{w} \phi$ around a solution that is dual to a state $|\lambda\rangle$, one can find a source deformation $\delta_{w} \lambda$, that gives rise to the linearized solution $\delta_{w} \phi$.\footnote{The inverse problem to this is not well posed, see \cite{Belin:2020zjb}.} This source deformation $\delta_{w} \lambda$ is thus an intrinsically boundary object and satisfies $\delta W = \Omega(\delta, \delta_{w})$, where $\Omega$ is the boundary Berry curvature two form defined in eq.~\eqref{eq:bndysympl}, and $\delta$ is an arbitrary variation of the state. 

Note that $\delta_{w}\lambda$ may depend on the state $|\lambda\rangle$ in a complicated way. In certain cases, it could have a natural meaning and display some form of universality. For example,  when $W$ is defined as the area of a space-like codimension-two surface, $\delta_{w} \lambda$ inserts a small conical deficit angle, corresponding to changing the R\'enyi index \cite{Dong:2017xht}. When $W$ is the volume of a codimension-one slice, $\delta_{w}\lambda$ was explored in \cite{Belin:2018bpg}, but it is still unknown how universal the variation should be in general states. We will comment on this further in the section \ref{discuss}. 

Instead of studying the meaning of $\delta_{w}\lambda$ for a given functional $W$, our main point here is that the boundary object $\delta_{w}\lambda$ exists for each diff-invariant functional $W$.\footnote{A subtlety here is that as we have seen before, the slice or region extremizing $W$ may not exist, or may not be unique. When the region does not exist, there is also no $\delta_{w}\lambda$ that could be defined. When it is not unique, there is a distinct $\delta_{w}\lambda$ corresponding to all the different extrema.} The recipe to construct it is to find the conjugate variation in the bulk either using eq.~\eqref{eq:gennewYork} from the canonical formalism or the Peierls construction, then Wick rotate it to Euclidean, and finally follow it to the Euclidean boundary. In section \ref{sec:conjugate_spacetimevolume}, we will provide another example of such a calculation for the case where the functional is the spacetime volume discussed in section \ref{sec:spacetimevolume}.  

\subsection{Peierls construction}\label{sec:Peierls}
To construct the variation $\delta_{w}$ conjugate to an arbitrary diffeomorphism invariant functional $W$ (codimension-zero or codimension-one), we first briefly review the so-called Peierls construction. It is a covariant way of ``pulling the indices up" of $\delta W$ with the symplectic form. The Peierls construction is equivalent to the simple formula \eqref{eq:gennewYork} for codimension-one functionals, but it applies to a more general class of functionals. For example, we can apply it to the functionals defined as integrals between two non-intersecting Cauchy slices $\Sigma_-$ and $\Sigma_+$, as shown in Figure \ref{fig:peierls}.

Let us consider a diffeomorphism invariant gravitational theory described by a given action $S (\phi)$ where we have denotes all dynamical fields as $\phi$. Then, Peierls' prescription for generating $\delta_{w}\phi$ is given by the following \cite{Peierls:1952cb}: 
\begin{enumerate}
\item Deform the gravitational action by adding the functional $W$ with a small coefficient $\varrho$, \viz  $S\mapsto S-\varrho W$
\item Pick a background solution $\phi_0$ which solves the equations of motion from the undeformed action $S$. Then, one can construct linearized solutions $\phi_{\mt{R}/\mt{A}}=\phi_0 +\varrho \delta_{\mt{R}/\mt{A}} \phi$ associated with the deformed action. The subscript $R/A$ is referred to as the retarded/advanced solution. Obviously, we have that $\delta_{\mt R} \phi=0$ in the past of $\Sigma_-$ while $\delta_{\mt A} \phi=0$ in the future of $\Sigma_+$. Physically speaking, we obtain $\phi_{\mt R}$ by starting from the unperturbed solution $\phi_0$ and following how it changes under time evolution due to the perturbation $\varrho W$. $\phi_{\mt A}$ is the time reversed picture of this, where we look for the precursor solution that ``lands" on $\phi_0$ after the perturbation has been switched off. See Figure~ \ref{fig:peierls} for an illustration.

\item The conjugate variation dual to the functional $W$ is then given by 
\begin{equation}
\delta_{w} \phi = \delta_{\mt{R}} \phi-\delta_{\mt{A}} \phi\,.
\end{equation}
This is a solution of the EOM \textit{without} the deformation, since acting on either $\delta_{\mt{R/A}}\phi$ with the undeformed linearized equation of motion gives the same inhomogeneous source term $\varrho \frac{\delta W}{\delta \phi}|_{\phi_0}$, so it cancels in the difference. Following this Peierls construction, the corresponding conjugate variation $\delta_w \phi$ satisfies our target, \ie $\delta W=\Omega(\delta, \delta_{w})$.
\end{enumerate}

\begin{figure}[ht!]
	\centering
	\includegraphics[width=3.8in]{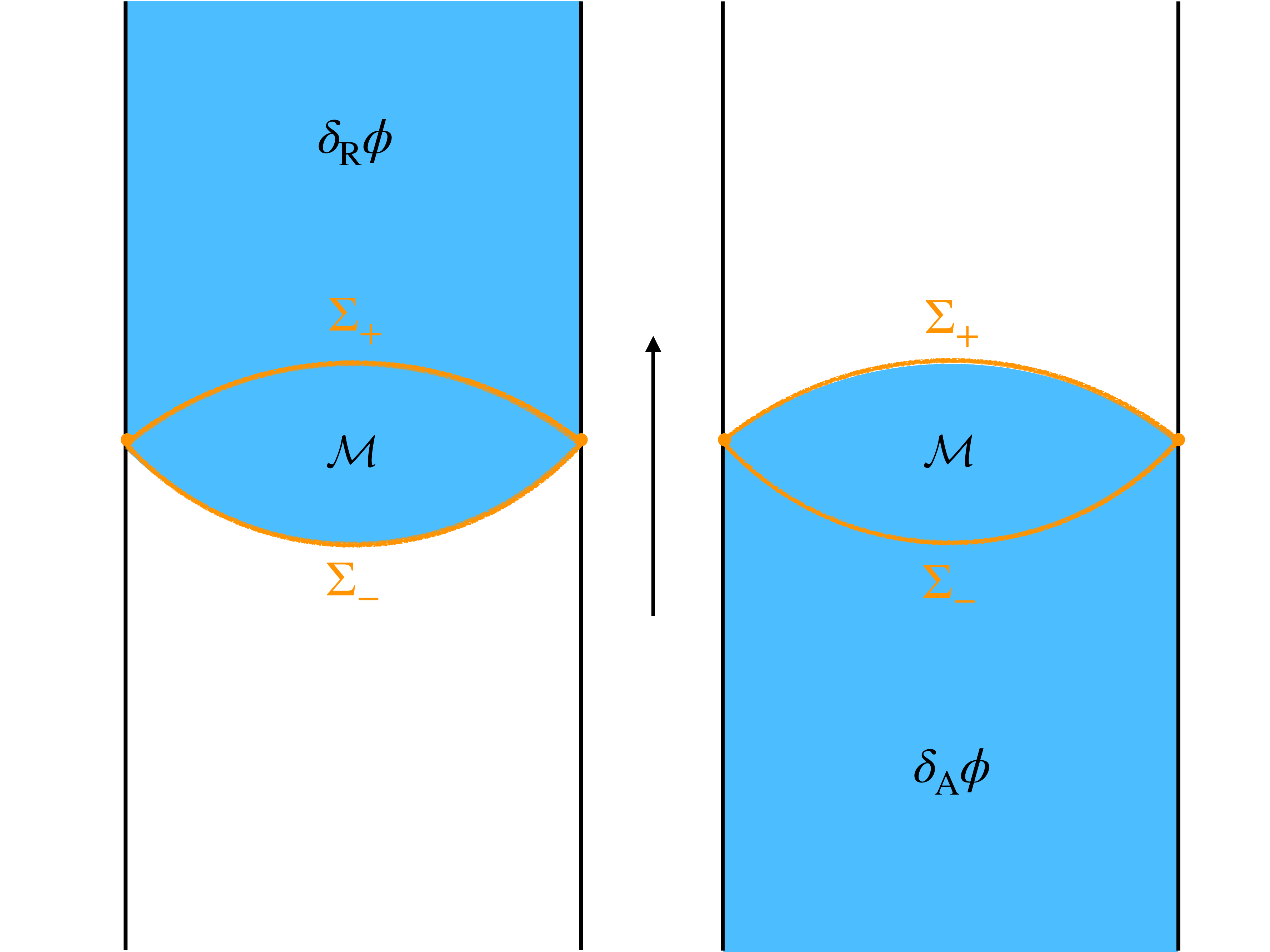}
	\caption{Illustration of the retarded and advanced solutions in the Peierls construction. The action is deformed by the functional $W$ in the bulk region $\mathcal{M}$, which lies between the two surfaces denoted $\Sigma_{\pm}$ (\ie the two orange curves). The blue shading shows the support of the retarded solution $\delta_{\mt{R}} \phi$ in the left panel, and that of the advanced solution $\delta_{\mt{A}} \phi$ in the right panel.}
	\label{fig:peierls}
\end{figure}

A modern proof of these statements can be found in \cite{Harlow:2019yfa}, which we will not repeat here. 
In the rest of this paper, we will restrict to the case where $\Sigma_{\pm}$ hit the boundary on the same boundary Cauchy slice, that is, $\partial \Sigma_+ = \partial \Sigma_-$.\footnote{An even more general class of observables would be to have $\partial \Sigma_+$ be in the future of $\partial \Sigma_-$, hence having a time interval in the boundary, where possible boundary deformations could be turned on as well. We will briefly comment on the possible uses of such observables in the discussion section.}

\subsubsection{Extremality in the Peierls construction}
\label{sec:peierls_extremality}

In the above construction, it seems like we have obtained the conjugate variation to an arbitrary functional $W$, while in the canonical discussion of section \ref{sec:canonical}, we have seen that for $\delta_{w}$ to exist, the functional must be extremized. In this section, we discuss why this extremization is automatically enforced in the covariant Peierls picture. In particular, we emphasize that even if $W$ is defined in a codimension-zero region between two non-intersecting Cauchy slices $\Sigma_\pm$, one has to extremize $W$ with respect to the shapes of $\Sigma_{\pm}$.

The basic point is that the equations of motion for the deformed gravitational action, \ie 
\beq
\label{eq:Peierlsdeform}
S_\varrho=S-\varrho W \,,
\eeq
enforce this extremization. This may seem at first a bit surprising since we didn't include the embedding coordinates of the Cauchy slices as dynamical variables.  However, one can consider the deformed action \eqref{eq:Peierlsdeform} as having two branes at $\Sigma_{\pm}$ acting as interfaces between regions with $W$ being turned on or off, and it is a well-known fact that the junction conditions and 
Einstein equation fully determine the locations of a brane \cite{Freivogel:2005qh}. 

More formally, the requirement that the deformed action $S_\varrho$ is diffeomorphism invariant means that
\begin{equation}
\delta_\xi S_\varrho = \int \left[\frac{\delta S_\varrho}{\delta \phi}\mathcal{L}_\xi \phi  + \frac{\delta S_\varrho}{\delta X^I}\mathcal{L}_\xi X^I \right]=0 \,, 
\end{equation}
where $\phi$ denotes all dynamical fields including the metric, $X^I$ are embedding fields specifying the location of the surfaces $\Sigma_\pm$, and $\xi$ is an arbitrary diffeomorphism. Evaluate this variation around a solution of the equations of motion with $\frac{\delta S_\varrho}{\delta \phi}=0$, one can obtain
\begin{equation}
\frac{\delta S_\varrho}{\delta X^I}\partial_\mu X^I =0\,.
\end{equation}
When the gradients $dX^I$ are linearly independent, this implies that the embedding fields satisfy their equations of motion $\frac{\delta S_\varrho}{\delta X^I}=0$, which reduces to the extremization condition for the surfaces $\Sigma_{\pm}$ in the limit $\varrho \to 0$. We can specify the location of two codimension-one surfaces with two scalar fields, so the $dX^I$, $I=\pm$ will in general be linearly independent for spacetimes that are at least two dimensional. This argument was made in \cite{Jacobson:2015mra}, where it was used for a rather different
purpose.v The argument only relies on the diffeomorphism invariance of the original action, and hence should apply for any higher curvature theory (Lovelock or otherwise) with any matter content.

\subsubsection{Examples of conjugate variations}
It is educational to see how the machinery of the Peierls construction reviewed in section \ref{sec:Peierls} reproduces the conjugate variation for a particular functional $W$. In this subsection, we consider two explicit examples. 

\subsubsection*{A. Re-deriving the new York transformation for the volume}
First of all, let us take the codimension-one functional $W=\int_{\Sigma}\sqrt{h}$ which is used in the CV conjecture \eqref{eq:defineCV}. Adding this to the usual Einstein-Hilbert action, one get the deformed action $S-\varrho W$. This new term $W$ acts as a stress tensor in the equations of motion, namely 
\begin{equation}
T^{\mu \nu} = + \varrho \frac{2}{\sqrt{g}} \frac{\delta w}{\delta g_{\mu \nu}}\,.
\end{equation}
Using the ADM \cite{Arnowitt:1959ah} variables $(h_{ij}$, $K_{ij})$, and working in Gaussian normal coordinates around $\Sigma$ with $N=1$, $N^i=0$, one gets 
\begin{equation}
T^{00}=T^{0i}=0 \,, \qquad T^{ij}=\varrho h^{ij} \delta(t)\,.
\end{equation}
It is clear that the variation $\varrho W$ results in a delta function on the slice $\Sigma$ (corresponding to ADM time $t=0$) since $W$ is an integral over this time slice only. Let us think about the equations of motion in ADM formulation. The fact that $T^{00}=T^{0i}=0$ means that the Hamiltonian and momentum constraints are unchanged. There is also no change in $\dot{h}_{ij}$ since the stress tensor does not enter here. The only change happens in the equation for $\dot{K}_{ij}$:
\begin{equation}
    \dot{K}_{ij} =(\text{r.h.s. in Einstein gravity}) + \frac{1}{2}\varrho h_{ij} \delta(t) \,.
\end{equation}
We may now write the retarded solution as
\begin{equation}
K^{\mt{R}}_{ij} = K^0_{ij} + \varrho\Theta(t) \delta_{\mt{R}} K_{ij}, \quad h^{\mt{R}}_{ij} = h^0_{ij} + \varrho\Theta(t)\delta_{\mt{R}} h_{ij} \,,
\end{equation}
where $h^0$, $K^0$ denote the background solution. Putting this in the ADM equations for $\dot{h}_{ij}$ and $\dot{K}_{ij}$ gives
\beqs
\varrho\delta(t) \delta_{\mt{R}} h_{ij} + \varrho \Theta(t) \frac{d}{dt} \delta_{\mt{R}} h_{ij} &= -2\varrho \Theta(t) \delta_{\mt{R}} K_{ij},\\ 
\varrho \delta(t) \delta_{\mt{R}} K_{ij} + \varrho \Theta(t) \frac{d}{dt} \delta_{\mt{R}} K_{ij} &= \varrho \Theta(t) (\text{linearized r.h.s. for } \delta_{\mt{R}} K_{ij}) + \frac{1}{2}\varrho \delta(t) h_{ij}.
\eeqs
Matching the coefficients of $\Theta(t)$ terms tells us that $\delta_{\mt{R}} h_{ij}$ and $\delta_{\mt{R}} K_{ij}$ solve the \textit{undeformed} linearized EOM around $(h^0_{ij},K^0_{ij})$, for $t>0$. Matching the coefficients of $\delta(t)$ gives:
\beq\label{eqn:retsoln}
\delta_{\mt{R}} h_{ij}(t=0)=0\,, \quad \delta_{\mt{R}} K_{ij}(t=0)=\frac{1}{2} h_{ij}(t=0).
\eeq
We can construct the advanced solution in the same way by taking 
\beq
K^{\mt{A}}_{ij} = K^0_{ij} + \varrho\Theta(-t) \delta_{\mt{A}} K_{ij}, \quad h^{\mt{A}}_{ij} = h^0_{ij} + \varrho\Theta(-t) \delta_{\mt{A}} h_{ij}.
\eeq
The difference then is that on the l.h.s of the EOMs the delta functions appear with an extra sign due to $\Theta(-t)$. This results in initial data
\beq
\delta_{\mt{A}} h_{ij}(t=0)=0, \quad \delta_{\mt{A}} K_{ij}(t=0)=-\frac{1}{2} h_{ij}(t=0).
\eeq
According to Peierls' recipe, the new York transformation is
\beqs
\delta_w h_{ij}&=\Theta(t)\delta_{\mt{R}} h_{ij}-\Theta(-t) \delta_{\mt{A}} h_{ij} \,,\\
\delta_w K_{ij}&=\Theta(t)\delta_{\mt{R}} K_{ij}-\Theta(-t) \delta_{\mt{A}} K_{ij}\,. 
\eeqs
It is clear that the delta functions now cancel from $\frac{d}{dt}\delta_w h_{ij}$ and $\frac{d}{dt}\delta_w K_{ij}$, so these are linearized solutions to the undeformed EOMs. They correspond to initial data\footnote{There is a $\Theta(0)$ formally in these expressions, but the correct prescription is to thicken the functional to be between $\Sigma_{\pm}$ with $\Sigma_-$ at $t=-\epsilon$ and $\Sigma_+$ at $t=\epsilon$. For this thickened functional, the retarded and advanced solutions now have support that overlaps in the region $-\epsilon<t<\epsilon$. Therefore, this sends $\Theta(t)\mapsto \Theta(t+\epsilon)$ and $\Theta(-t)\mapsto \Theta(-t+\epsilon)$ in the formulas, so we must formally put $\Theta(0)=1$. }
\begin{equation}
\delta_w h_{ij}(t=0)=0, \quad \delta_w K_{ij}(t=0)=h_{ij}(t=0)\,,
\end{equation}
so we indeed recover the canonical results \eqref{eq:oldnewYork} from the Peierls construction. 

\subsubsection*{B. Spacetime volume in vacuum AdS}
\label{sec:conjugate_spacetimevolume}


As a second example, we define the functional $W$ as that given in eq.~\eqref{eq:CMCfunc}. 
Following the Peierls recipe, we consider the bulk part of the deformed action
\begin{equation}\label{eq:exampledeformed}
\frac{1}{16 \pi \GN} \int \sqrt{-g}(\mathcal{R}+\frac{d(d-1)}{L^2}) - \varrho W \,.
\end{equation}
Between $\Sigma_-$ and $\Sigma_+$ the solution must look like
\begin{equation}
ds^2=-d\tau^2+ L_{\rm def}^2\cos^2 \frac{\tau}{L_{\rm def}} d\Sigma^2\,,
\end{equation}
with 
\begin{equation}
L_{\rm def}=L \sqrt{\frac{1}{1- \frac{16 \pi \varrho \alB }{d(d-1)}}}=  L \(1 + \frac{8\pi \varrho \alB }{d(d-1)} \)  + O(\varrho^2)\,.
\end{equation}
We consider the retarded ansatz
\beqs
ds^2_{\mt{R}} &= -d\tau^2 + [\Theta(-\tau_--\tau)L^2 \cos^2 \frac{\tau}{L}  + \Theta(\tau+\tau_-)\Theta(\tau_+-\tau) L_{\rm def}^2 \cos^2 \frac{\tau-\xi_1^{\mt{R}}}{L_{\rm def}} \\ & + \Theta(\tau-\tau_+)L^2 \cos^2 \frac{\tau-\xi_2^{\mt{R}}}{L} ]d\Sigma^2.
\eeqs
Here, we have chosen the location of the surfaces $\Sigma_\pm$ to be at $\tau = \pm \tau_{\pm}$, while $\xi^{\mt{R}}_{1,2}$ are time shifts that are further parameters of the ansatz. We determine these four parameters by solving the equations of motions derived from the deformed action \eqref{eq:exampledeformed}. The ansatz above is a piecewise solution to eqs.~\eqref{eq:exampledeformed} and \eqref{eq:CMCfunc}. Taking account of the contributions coming from the codimension-one boundary terms in eq.~\eqref{eq:CMCfunc}, we merely need to enforce the Israel junction conditions, namely
\begin{equation}
  \begin{split}
      \Delta h_{ij} &=0 \,, \\
     \Delta K_{ij}-h_{ij} \Delta K& = -\frac{8 \pi \varrho \, \alpha_+}{L}  h_{ij} \,,
  \end{split}
\end{equation}
where $\Delta h_{ij}$ and $\Delta K_{ij}$ are the jump in the induced metric and extrinsic curvature across the surface $\Sigma_+$. Matching the induced metric across the interfaces results in
\begin{equation}
\begin{split}
L^2 \cos^2 \frac{\tau_-}{L}&=L_{\rm def}^2 \cos^2 \frac{\tau_-+\xi_1^{\mt{R}}}{L_{\rm def}} \,, \\
L_{\rm def}^2 \cos^2 \frac{\tau_+-\xi_1^{\mt{R}}}{L_{\rm def}}&=L^2 \cos^2 \frac{\tau_+-\xi_2^{\mt{R}}}{L}\,, \\
\end{split}
\end{equation}
while the junction conditions for the extrinsic curvatures impose
\begin{equation}
\begin{split}
\frac{1}{L}\tan \frac{\tau_-}{L}- \frac{1}{L_{\rm def}}\tan \frac{(\tau_-+\xi_1^{\mt{R}})}{L_{\rm def}} &= -\frac{8\pi \varrho \alpha_-}{(d-1)L} \,,\\
\frac{1}{L} \tan \frac{(\tau_+-\xi_2^{\mt{R}})}{L}- \frac{1}{L_{\rm def}} \tan \frac{(\tau_+-\xi_1^{\mt{R}})}{L_{\rm def}} &= -\frac{8\pi \varrho \alpha_+}{(d-1)L}\,.\\
\end{split}
\end{equation}
Performing the expansion $L_{\rm def}=L + \varrho \delta L$ with $\delta L = \frac{8\pi\alB }{d(d-1)} L$, these equations then reduce to
\begin{equation}
    \begin{split}
    \xi_1^{\mt{R}} &= \varrho \, \delta L \left( \frac{\tau_-}{L} +\cot \frac{\tau_-}{L}\right) \,,\\
\xi_2^{\mt{R}} &= \varrho \, \delta L \left( \frac{\tau_-+\tau_+}{L}+\cot \frac{\tau_-}{L}+\cot \frac{\tau_+}{L}\right)\,,\\
\tan \frac{\tau_-}{L} &= \frac{(d-1)\delta L}{8\pi \alpha_- L} =\frac{\alB}{d\alpha_-}\,,\\
\tan \frac{\tau_+}{L} &= \frac{(d-1)\delta L}{8\pi \alpha_+ L}=\frac{\alB}{d\alpha_+}\,.\\
    \end{split}
\end{equation}
Noting that the last two equations give not only $\tau_{\pm}$ but also the extremality conditions. Indeed, on constant $\tau$ slices, we have $K=-\frac{d}{L} \tan \frac{\tau}{L}$. Evaluated on the two hypersurface $\Sigma_\pm$ located at $\tau = \tau_+$ and $\tau=-\tau_-$, the two constraint equations are nothing but the same extremality conditions shown in eq.~\eqref{eq:CMCextr}}.

For the advanced solution, we may consider the ansatz
\beqs
ds^2_{\mt{A}} &= -d\tau^2 + [\Theta(-\tau_--\tau)L^2\cos^2 \frac{\tau-\xi^{\mt{A}}_2}{L}  + \Theta(\tau+\tau_-)\Theta(\tau_+-\tau) L_{\rm def} \cos^2 \frac{\tau-\xi_1^{\mt{A}}}{L} \\& + \Theta(\tau-\tau_+)L^2\cos^2 \frac{\tau}{L} ]d\Sigma^2\,,
\eeqs
which leads to the solution to the junction conditions, $\xi_1^{\mt{A}}=-\xi_1^{\mt{R}}|_{\tau_-\mapsto \tau_+}$ and $\xi_2^{\mt{A}}=-\xi_2^{\mt{R}}$. Taking the difference between the retarded and advanced solutions, we find a continuous solution which is a simple shift in York time
\beq
\delta_{w} : \tau \mapsto \tau + 2\delta L \left( \frac{\tau_-+\tau_+}{L}+\cot \frac{\tau_-}{L}+\cot \frac{\tau_+}{L}\right),
\eeq
where $\tau_{\pm}$ are understood to be functions of $\alpha_{\pm}$.
Since this is a diffeomorphism, it gives rise to a boundary term in the symplectic form. As a result, all variations of $W$ around the vacuum solution are given by boundary terms, just like for codimension-one functionals \cite{Belin:2018bpg}. Moreover, because $\delta_{w}$ is only a shift in York time, the variation is the same as that for the volume, and the entire $\alpha_{\pm}$ dependence is encoded in a prefactor of the symplectic form (\ie as a prefactor in the conjugate variation). These comments carry over to the boundary dual of the conjugate variation, that is, the boundary background metric $\gamma_{ij}$ is deformed as $\delta_w \gamma_{ij} \propto \text{sign}(t_{\mt{E}})\gamma_{ij}$ \cite{Belin:2018bpg}, where $t_{\mt{E}}$ is Euclidean time on the boundary, and the $\alpha_{\pm}$ dependence is purely in the prefactor of the deformation.

\subsection{Complexity = Anything$^2$} \label{sec:c=a^2}
As discussed in the introduction, not all diffeomorphism invariant observables take the form of a single extremized functional. Another class of quantities consists of specifying the slice by extremizing a first functional, and evaluating a different functional on it. Since both the slice specification and the functional are diffeomorphism invariant, one would expect these quantities to be accessible through the CFT. This seems to suggest that the space of functionals is in fact ``squared", where the slice and quantity to integrate can be evaluated separately. We will now explain how these quantities fit in the general framework of Peierls and in particular how to construct conjugate variations by using the canonical formalism of section \ref{sec:canonical}. 

Considering the general codimension-one functional defined in eq.~\eqref{eq:obsdef}, \viz 
\begin{equation}
O_{F_1,\Sigma_{F_2}}(\Scft) =\frac{1}{\GN L} \int_{\substack{\Sigma_{F_2}}}\!\!\!\!\! d^d\sigma \,\sqrt{h} \,F_1(g_{
	\mu\nu}; X^{\mu}) \,,
\end{equation}
we can relate it to an auxiliary one with
\begin{equation}
    F_\epsilon = F_2+\epsilon F_1\,.
\end{equation}
For simplicity, we will focus on the case where $F_2=1$, \ie we integrate the density $F_1$ on the maximal volume slice. In this case, the auxiliary density is $F_\epsilon = 1 + \epsilon F_1$ and our auxiliary observable is given by 
\be \label{newobs}
O_\epsilon = \int_{\Sigma_\epsilon } d^d\sigma \sqrt{h}(1 + \epsilon F_1) \,,
\ee
where $\Sigma_\epsilon$ denotes the hypersurface which extremizes the functional $O_\epsilon$. If we think of this problem perturbatively in $\epsilon$, we have
\be
\Sigma_\epsilon =\Sigma_V + \epsilon \Sigma^{(1)}+\mathcal{O}(\epsilon^2) \,,
\ee
where $\Sigma_V$ is referred to as the maximal volume slice.
In terms of this expansion, one finds 
\begin{equation}
O_\epsilon=\int_{\Sigma_V} d^d\sigma \sqrt{h}(1 + \epsilon F_1)+\mathcal{O}(\epsilon^2) \,.
\end{equation}
Because the surface $\Sigma_V$ extremizes the volume, there is no leading order contribution from $\Sigma^{(1)}$. As a consequence, one can then relate $O_{F_1,\Sigma_{V}}$ to a family of extremized observables $O_\epsilon$ by\footnote{In fact, $\frac{d}{d\epsilon}  O_\epsilon$ at finite $\epsilon$ is just $F_1$ evaluated on
the surface $\Sigma_\epsilon$, since the contribution 
due to changing the location of the surface is a second order 
effect due to extremization of $O_\epsilon$.} 
\begin{equation}
O_{F_1,\Sigma_{V}}=\frac{d}{d\epsilon}  O_\epsilon \bigg|_{\epsilon=0} \,.
\end{equation}
Supposing $\delta_{O_\epsilon}$ is the transformation that generates variations of $O_\epsilon$ through the symplectic form, we have
\begin{equation}
 \delta O_{F_1,\Sigma_{V}} =  \Big[\frac{d}{d\epsilon} \int_{\Sigma_\epsilon} \omega(\delta, \delta_{O_\epsilon}) \Big]_{\epsilon=0} =\int_{\Sigma'} \omega\left(\delta, \left[ \frac{d}{d\epsilon}\delta_{O_\epsilon}\right]_{\epsilon=0} \right)\,,   
\end{equation}
where we have used the fact that the symplectic form is on-shell conserved to evaluate it on an $\epsilon$ independent slice $\Sigma'$, and then used linearity in the slots to move in with the $\epsilon$ derivative. We learn that the conjugate variation to the functional $O_{F_1,\Sigma_{V}}$ is derived as 
\beq
\label{eq:anythingsqconj}
\delta_{O_{F_1,\Sigma_{V}}}=\left[\frac{d}{d\epsilon}\delta_{O_\epsilon}\right]_{\epsilon=0}\,.
\eeq
Note that this indeed solves the linearized equations of motion, since it is a difference of two nearby solutions, \ie  $\delta_{O_\epsilon}$ for two different values of $\epsilon$.

Finally, let us make contact with the canonical discussion of section \ref{sec:canonical}. In order to do this, it is important to stress that while the first order change in the auxiliary extremal surface $\Sigma_\epsilon$ does not appear in the value of the observable $O_{F_1,\Sigma_{V}}$, its conjugate variation $\delta_{O_{F_1,\Sigma_{V}}}$ \textit{does} depend on it. For example, applying eq.~\eqref{eq:gennewYork} for $O_{F_1,\Sigma_{V}}$ on the maximal volume slice, we would end up with an initial data variation that would violate the Hamiltonian constraint! The resolution is that the initial data variation coming from the prescription above eq.~\eqref{eq:anythingsqconj} has extra terms involving shape variations compared to eq.~\eqref{eq:gennewYork}. To see this explicitly in the canonical formalism, we pick ADM coordinates so that the maximal volume slice is at $t=0$, while the extremal slice for $O_\epsilon$ is at $t=\epsilon$. We can apply
\eqref{eq:gennewYork} to $O_\epsilon$:
\beqa
\delta_{O_\epsilon}\pi^{ij}&= -\frac{\delta O_\epsilon}{\delta h_{ij}}\big|_{t=\epsilon} =-
\frac{\delta \text{Vol}}{\delta h_{ij}}\big|_{t=0} - \epsilon \left( \frac{\delta O_{F_1,\Sigma_{V}}}{\delta h_{ij}}\big|_{t=0} + \frac{\delta \frac{d}{dt}\text{Vol}}{\delta h_{ij}}\big|_{t=0}\right)+ \cdots \,,\\
\delta_{O_\epsilon}h^{ij}&=\frac{\delta O_\epsilon}{\delta \pi^{ij}}\big|_{t=\epsilon} =
\frac{\delta \text{Vol}}{\delta \pi^{ij}}\big|_{t=0} + \epsilon \left( \frac{\delta O_{F_1,\Sigma_{V}}}{\delta \pi^{ij}}\big|_{t=0} + \frac{\delta \frac{d}{dt}\text{Vol}}{\delta \pi^{ij}}\big|_{t=0}\right) + \cdots\,,
\eeqa
where $\rm{Vol} = \int \sqrt{h}$ is the volume of a slice at fixed ADM time $t$.
The important point is that the variation of $\frac{d}{dt}\text{Vol}$ does not vanish at $t=0$, even though $\frac{d}{dt}\text{Vol}$ itself does. Indeed, $\frac{d}{dt}\text{Vol}=-\int N \sqrt{h}K=\frac{1}{d-1}\int N \pi^i_i$, therefore, using eq.~\eqref{eq:anythingsqconj} we obtain
\begin{equation}\label{eq:deltaO}
\begin{split}
    \delta_{O_{F_1,\Sigma_{V}}} \pi^{ij} &= -\frac{\delta O_{F_1,\Sigma_{V}}}{\delta h_{ij}} - \frac{N}{d-1}\pi^{ij}\\
\delta_{O_{F_1,\Sigma_{V}}} h_{ij} &= \frac{\delta O_{F_1,\Sigma_{V}}}{\delta \pi^{ij}} + \frac{N}{d-1}h_{ij}\,,
\end{split}
\end{equation}
which has extra terms compared to naively applying eq.~\eqref{eq:gennewYork} to $O_{F_1,\Sigma_{V}}$. Note that $N$ here is a \textit{fixed} lapse, selected by the condition that the extremal surface for $O_\epsilon$ is at $t=\epsilon$ to leading order in $\epsilon$. This agrees with the condition that the above initial data variation solves the Hamiltonian constraint\footnote{In this equation, by $\frac{d}{dt} O_{F_1,\Sigma_{V}}$ we mean the \textit{density} $\frac{\delta O_{F_1,\Sigma_{V}}}{\delta \pi^{ij}}\dot \pi^{ij}+\frac{\delta O_{F_1,\Sigma_{V}}}{\delta h_{ij}}\dot h_{ij}$, and not its integral.}
\beq
0=\delta_{O_{F_1,\Sigma_{V}}} \mathcal{H}_\perp = - \frac{1}{N}\frac{d}{dt} O_{F_1,\Sigma_{V}} - \frac{\dot{\pi}}{d-1} \quad \Rightarrow \quad N=\frac{(1-d)}{\dot \pi}\frac{d O_{F_1,\Sigma_{V}}}{d t},
\eeq
where $\pi={\pi^i}_i$ and we have used Hamilton's equation \eqref{eq:Hamiltonseq}. 
When we plug this $\delta_{O_{F_1,\Sigma_{V}}}$ \eqref{eq:deltaO} into the canonical symplectic form defined in eq.~\eqref{eq:canonicalsymplectic}, we obtain
\beqa
\Omega(\delta, \delta_{O_{F_1,\Sigma_{V}}}) = \int_{\Sigma_V}\left[\frac{\delta O_{F_1,\Sigma_{V}}}{\delta \pi^{ij}}\delta \pi^{ij}+\frac{\delta O_{F_1,\Sigma_{V}}}{\delta h_{ij}}\delta h_{ij}\right]-\int_{\Sigma_V} \frac{1}{\dot \pi}\frac{d O_{F_1,\Sigma_{V}}}{dt} \delta \pi\,.
\label{eq:canonicalO}
\eeqa
This is indeed the total variation of the gauge invariant observable $O_{F_1,\Sigma_{V}}$ as we will now explain. $O_{F_1,\Sigma_{V}}$ is defined by integrating a density $F_1$ over the maximal volume slice $\Sigma_V$. When we vary the state, in the new geometry, the maximal volume slice moves, and since $O_{F_1,\Sigma_{V}}$ is not extremal on it, it picks up a first order change from moving the slice. This is the second term in  $\Omega(\delta, \delta_{O_{F_1,\Sigma_{V}}})$ shown in eq.~\eqref{eq:canonicalO}, where $\delta \pi/{\dot \pi}$ measures how much ADM time $t$ has changed due to the variation in York time $\delta \pi$, and this is multiplied by how much the density changes under the change of ADM time. The first term in eq.~\eqref{eq:canonicalO} is just the change in the density on the maximal volume slice of the original geometry. 

Alternatively, we may understand the need for the second term as follows. We collect the terms proportional to $\delta \pi$ in eq.~\eqref{eq:canonicalO}, \ie 
\beq
\int_{\Sigma_V}\left[\frac{\delta O_{F_1,\Sigma_{V}}}{\delta \pi}-\frac{1}{\dot \pi}\frac{d O_{F_1,\Sigma_{V}}}{dt}\right]\delta \pi=-\int_{\Sigma_V}\left[\frac{\delta O_{F_1,\Sigma_{V}}}{\delta {\bar\pi}^{ij}}{\dot{\bar\pi}}^{ij}+\frac{\delta O_{F_1,\Sigma_{V}}}{\delta h_{ij}}{\dot h}_{ij}\right]\frac{\delta \pi}{\dot \pi}.
\eeq
In this formula, the role of the second term is therefore to cancel the variation $\frac{\delta O_{F_1,\Sigma_{V}}}{\delta \pi}$, which should indeed not enter the physical variation of $O_{F_1,\Sigma_{V}}$, since in all states the observable is defined on the maximal volume slice $\pi=0$.

A gauge adapted to the observable $O_{F_1,\Sigma_{V}}$ is to specify states by their initial data on their maximal volume slices $K=0$. One can always do this by specifying the traceless part of $\pi^{ij}$ and the conformal metric, and then the scale is fixed by the Hamiltonian constraint \cite{York:1972sj}. In this gauge we clearly have $\delta K=0$ (and hence $\delta \pi=0$ since $K=0$) for all initial data variations. Therefore, the extra term in eq.~\eqref{eq:canonicalO} is manifestly absent in this gauge.

\section{Discussion}\label{discuss}

In this paper, we examined a generalization of the recent {\it ``complexity equals anything"} proposal \cite{Belin:2021bga} for holographic complexity by considering a new infinite class of gravitational observables living on codimension-zero subregions of the bulk spacetime. We showed that these new gravitational observables exhibit two universal features, late-time linear growth and the switchback effect in an eternal black hole spacetime, which are expected properties for holographic complexity. In section \ref{sec:spacetimevolume}, we investigated the explicit example in eq.~\reef{eq:CMCfunc} where the volumes of a codimension-zero subregion and of the codimension-one boundary surfaces were added with distinct coefficients. By taking distinct limits, we reproduced the well-studied CV and CV2.0 proposals -- see eqs.~\reef{limit1} and \reef{limit2}. Further, we showed how one could straightforwardly recover the CA proposal \reef{eq:defineCA} using this example as well. In particular, with the limit where the boundary coefficients vanish, \ie $\alpha_\pm\to0$, while the bulk coefficient $\alB$ was fixed, the extremal bulk region $\mathcal{M}$ becomes the WDW patch. One can then evaluate the gravitational action on this extremal region to recover the CA proposal, \ie $G_1$ is chosen to be the Einstein-Hilbert term plus cosmological constant and $F_{1,\pm}$, the boundary action terms. We conclude with some open questions.

\subsection*{Two different limits towards the WDW patch}
In considering the CA proposal from the observable \reef{eq:CMCfunc}, one may also take an alternate approach where one first extremizes with finite $\alpha_\pm$ and evaluates the gravitational action on the resulting extremal region $\mathcal{M}$ defined between two CMC boundaries.\footnote{A similar observable was constructed for a three-dimensional bulk in \cite{Chandra:2022pgl}.}\footnote{Let us add a short comment about this observable with finite $\alpha_\pm$: A simple question is whether the gravitational action between two CMC slices always yields a positive quantity, which would be required for the observable to have an interpretation as the holographic dual of complexity in the boundary theory. It turns out that this is determined by the leading divergences coming from contributions near the asymptotic boundaries. Using the Fefferman-Graham expansion, it is straightforward to show that the latter always combine to yield a positive result. In fact, the leading terms (proportional to $1/\delta^{d-1}$) from the bulk Einstein and boundary GHY terms cancel. Hence the leading contribution comes entirely from the Hayward joint terms where the CMC surfaces meet at the asymptotic boundary, and this is positive.} Then, we could consider the resulting observable in the limit $\alpha_\pm\to0$ where $\mathcal{M}$ becomes the WDW patch. Whether this approach yields the same result as the CA proposal is examined in detail in appendix \ref{app:null}, and here we comment on the salient points. 

Of course, the limit of the bulk term yields the desired contribution for the WDW patch. Next, considering the Gibbons-Hawking-York (GHY) boundary terms \cite{PhysRevLett.28.1082,PhysRevD.15.2752} on $\Sigma_\pm$, one would find that the null limit is ambiguous in general since it depends on the choice of foliation approaching the null surface. However, here we are taking the limit through a family of CMC slices, which fixes this ambiguity in the choice of foliation. Although the extrinsic curvature diverges as the foliation becomes null, the induced volume form on the surface goes to zero in the limit so that the GHY contribution remains finite. It reduces to the expected null boundary term \cite{Parattu:2015gga,Lehner:2016vdi,Hopfmuller:2016scf,Chandrasekaran:2020wwn} plus an integral of various surface gravity quantities along the null surface. The latter quantities depend on components of the metric transverse to the null surface, and hence the null limit action requires a variational principle that fixes these additional metric components as well. As mentioned in Appendix \ref{app:null}, fixing the surface gravities in the variational principle is likely achievable as a gauge condition. Hence, we expect it not  to substantially alter the Dirichlet variational principle.

This leaves us to consider the Hayward joint term \cite{Hayward:1993my,1994PhRvD..50.4914B} where $\Sigma_\pm$ meet at their codimension-two boundaries at the cutoff surface. In this case, we find that the joint term diverges in the null limit, as was also recently noted in \cite{Chandra:2022pgl}. However, this divergence can be removed by adding a constant counterterm, which does not affect the finite $\alpha_\pm$ variational principle. Note that this counterterm is necessary even before taking the cutoff surface to infinity since it is associated with the null limit instead of divergences arising near the asymptotic boundary. As with the GHY term, the resulting renormalized joint term agrees with the standard one for corners involving null surfaces examined in \cite{Lehner:2016vdi} up to corrections involving the surface gravities. Again, this implies a modified variational principle that requires the surface gravities to be fixed, but, as before, this choice should be accessible as a gauge condition.  

Hence, we conclude that the null limit $\alpha_\pm\to0$ then results in a finite action that agrees with standard expressions for the action of regions 
with null boundaries and joints associated with a Dirichlet-variational principle \cite{Lehner:2016vdi}, up to finite corrections involving the surface gravities of the null surface. It would be interesting to further examine the null limit variational principle and to understand any implications arising from the surface gravity corrections to the action. In particular, it would be useful to verify that the associated variational principle agrees with the standard Dirichlet principle with an additional gauge condition, as we conjecture here. Furthermore, it would be interesting to understand if the surface gravity corrections have additional implications for interpreting the null limit action as a measure of holographic complexity.

\subsection*{Complexity ambiguities and MERA\footnote{Multi-Scale Entanglement Renormalization Ansatz.}}
 
As we have seen, there is a great deal of freedom in constructing these new observables (and those in \cite{Belin:2021bga}). Yet, they all exhibit the same universal behaviour expected of holographic complexity (\ie linear late-time growth and the switchback effect). Our proposal is partly motivated by the ambiguities arising in prescribing a microscopic definition of complexity in the boundary theory. A simple example would be that different kinds of gates in the quantum circuits could be weighted differently in evaluating the complexity. To pursue this example further, let us consider the MERA representation of the bulk AdS spacetime \cite{Swingle:2009bg,Swingle:2012wq}. In particular, the MERA representation of a thermofield double state involves three distinct classes of tensors \cite{evenbly2015tensor}. The isometries and disentanglers comprising the asymptotic regions on either side, and the infrared tensors at the neck joining the two sides of the tensor network. The latter is certainly a distinct resource as they are not unitary gates, so presumably, they must assign a different weight in evaluating the complexity of the tensor network. This situation might be smoothly emulated in holographic complexity by extending the usual extremal codimension-one surfaces of the CV proposal by introducing a term proportional to the square of the Weyl tensor -- see appendix \ref{revone}. As a result, the contributions in the infrared of the dual black hole geometry are given a different weight than those from the asymptotic regions, similar to the previous MERA construction. In any event, it will be interesting to make a more explicit connection between the ambiguities in our geometric bulk constructions and the complexity models for the boundary theory. Perhaps an interesting tool in this line of investigation will be the first law techniques developed in \cite{Bernamonti:2019zyy,Bernamonti:2020bcf}.

\subsection*{The Peierls construction}

An important upshot of our investigation into new holographic complexity measures is the development of a broad class of diffeomorphism-invariant observables and an examination of their properties as functions 
on the bulk classical phase space.  In particular, in section \ref{symp}, we explicitly constructed the conjugate phase space variations $\delta_w$ associated with each observable using the Peierls bracket construction
\cite{Peierls:1952cb,Harlow:2019yfa,Harlow:2021dfp}. In addition to conceptually clarifying properties of these observables in the bulk phase space, the computations performed in this work provide several 
examples of using the Peierls construction for practical computations, adding to the tools recently developed in \cite{Harlow:2021dfp}.

One advantage of using the Peierls construction is the way in which the diffeomorphism invariance of the observable is enforced by the procedure. As explained in section \ref{sec:peierls_extremality}, functionals which superficially appear to not be diffeomorphism-invariant due to dependence on choices of embedding surfaces are instead extremized during the Peierls construction.  Since their location is dynamically determined by the extremization condition, the resulting observable is diffeomorphism-invariant after all; the Peierls construction effectively forces their diffeomorphism-invariance upon us. This observation could be particularly relevant to computing the action of observables such as the HRT area operator on the classical phase space, as was recently investigated in \cite{Kaplan:2022orm}. It would also be interesting to 
relate this result to the extension of the Peierls bracket to
gauge-dependent observables proposed in \cite{Marolf:1993af}.

We further showed in section \ref{sec:c=a^2} that the Peierls construction could also be used to compute the action of $(\text{anything})^2$ observables, which are evaluated on a surface obtained by extremizing a different functional. Such relational observables could have many applications in investigations of gravitational observable algebras in classical GR and holography. For example, operators dressed to the RT surface were investigated in \cite{Almheiri:2017fbd} using geodesics launched from the RT surface. The $(\text{anything})^2$ construction in which the reference observable is the area of the RT surface may provide a more convenient class of dressed observables with similar properties.  

Note that for these observables to be globally well-defined, solutions to the extremality equation should exist, and for the observable to be fully diffeomorphism-invariant, the solutions should be unique. Existence and uniqueness of solutions to the functional $\mC_{\rm gen}$ defined in equation (\ref{eq:CMCfunc}) were examined in appendix \ref{newapp}, and were argued to hold. However, for more general functionals involving integrals of curvature invariants, neither property is guaranteed. Generically, these observables may have multiple extrema (see, for example, appendix \ref{revone}), suggesting possible problems with their interpretation of the gravitational phase space and the applicability of the Peierls construction to them. In solutions around a given background, one can resolve this ambiguity by specifying the solution to be close to a reference one in the background, but the existence of multiple extrema likely hints at nonperturbative effects when examining the behaviour of the observable on the full gravitational phase space. It would be interesting to further investigate these effects in the future.  

\subsection*{More general codimension-one observables} 
Our new zoo of gravitational observables based on codimension-zero spacetime regions also naturally connects to  the codimension-one observables introduced in \cite{Belin:2021bga}, as illustrated in eq.~\reef{limit1}. The latter is essentially a specific example of setting $G_2=0=F_{2,-}$ while keeping $F_{2,+}$ finite. One can further extend the family of codimension-one observables by first using finite $G_2$ and $F_{2,\pm}$ to determine the extremal boundary surfaces and then choosing $G_1=0=F_{1,-}$ to evaluate an observable on $\Sigma_+$ with some finite $F_{1,+}$. A simple example would be extremizing the functional in eq.~\reef{eq:CMCfunc} with general coefficients but then evaluating the functional with $\alpha_+=1$ and $\alpha_-=0=\alpha_{\mt B}$ on the resulting geometry. The resulting observable is then the volume of a codimension-one surface with constant mean curvature, \ie $K=1/(\alpha_+ L)$.

\subsection*{More general backgrounds} 

Our analysis focused on the eternal planar black hole background \reef{eq:BHmetric}. Hence in our discussion of the new gravitational observables, we assumed that the bulk and boundary functionals in eqs.~\reef{eq:W1} and \eqref{eq:O1} were geometric, \ie they were scalar functionals of the background metric and the embedding functions. More generally, one can also consider the role of matter terms in this new class of holographic complexity. For example, in a charged black hole background, extending our construction with scalar terms involving the gauge field, \eg $F_{\mu\nu}F^{\mu\nu}$, $F^{\mu\nu}F^{\rho\sigma}C_{\mu\nu\rho\sigma}$ or $n^\mu F_{\mu\nu}F^{\nu}{}_{\!\rho}\, n^\rho$, would produce appreciable and potentially interesting new effects. Let us recall that this kind of contribution already appears in the CA proposal \reef{eq:defineCA}, \eg see \cite{Brown:2018bms,Goto:2018iay}. 

Of course, another interesting family of backgrounds in which to investigate the complexity=anything proposal would be rotating black holes. In this case, one may expect that the growth rate should be proportional to the entropy times the temperature \cite{Susskind:2014moa}.\footnote{The entropy gives a measure of the relevant number of degrees of freedom while the temperature gives the appropriate dimensionful scale for the time rate of change.} This result was confirmed for the CA and CV proposals for the three-dimensional rotating BTZ background by \cite{Brown:2015lvg} and \cite{Couch:2018phr}, respectively. In higher dimensions, this result for the growth rate can also be recovered in  the large black hole limit (or alternatively, the small spin limit) \cite{Couch:2018phr,Cai:2016xho,Auzzi:2018zdu,Frassino:2019fgr,AlBalushi:2020rqe,AlBalushi:2020ely,Bernamonti:2021jyu}. In appendix \ref{sec:approtate}, we examine the growth rate associated with the functional in eq.~\reef{eq:CMCfunc} with general coefficients in the rotating BTZ background. We find that it is indeed given by the product of the entropy times the temperature, up to an overall factor depending on $\alB$ and $\alpha_\pm$. It would, of course, be interesting to extend this analysis to rotating black holes in higher dimensions and more general gravitational observables.

\subsection*{Observables that probe boundary time bands}

Another interesting extension of these new holographic observables may be as follows: In our construction, the surfaces $\Sigma_{\pm}$ are tied to the same time slice on the asymptotic boundary. An extension would then be to allow for a separation between where these two surfaces reach the AdS boundary. These observables may be interesting for a number of reasons. For example, consider this functional in a state dual to some bulk geometry and deform the state by adding a particle in the bulk. If backreaction can be neglected, but the particle is still heavy enough to travel on a time-like geodesic, the difference between these two actions will be the action of the particle itself, which is just the length of its worldline between the surfaces $\Sigma_-$ and $\Sigma_+$. Now, if the geometry we're probing is geodesically complete (\eg vacuum AdS), this length will grow forever if we fix $\Sigma_-$ and move $\Sigma_+$ to the future by increasing $t_0$. However, when the geometry is a black hole with a spacelike singularity, time-like geodesics have a finite length and increasing $t_0$ will lead to saturation. Hence this new functional provides a probe of whether or not the bulk spacetime is geodesically complete.
Let us add that this extended family of gravitational observables may also have a relation to recent discussions of holographic complexity, which consider the optimal quantum circuit to evolve one boundary state into another, \eg see \cite{Chagnet:2021uvi,Caputa:2018kdj,Erdmenger:2020sup,Erdmenger:2021wzc,Flory:2020eot,Flory:2020dja,Chandra:2021kdv,Chandra:2022pgl}

\subsection*{Codimension-2 generalized observables}
 
In \cite{Belin:2021bga}, the CV proposal was extended to construct a general family of gravitational observables based on codimension-one surfaces. We are extending the CA and CV2.0 proposals to a broad family of observables based on codimension-zero spacetime regions. Indeed, a similar extension could be considered for the extremal codimension-two surfaces, which appear in holographic entanglement entropy. That is, we could construct new gravitational observables involving extremizing and evaluating nontrivial scalar functionals on codimension-two bulk surfaces anchored to some spacelike curve on the asymptotic boundary. The question then becomes what the interpretation of these new observables would be in the boundary theory. In contrast to holographic complexity, there is a derivation of the HRT prescription for holographic entanglement entropy \cite{Lewkowycz:2013nqa,Dong:2016hjy}, which singles out extremal codimension-two surfaces. In the case where the replica number is not precisely one, the extremal surfaces (\ie ``cosmic branes" \cite{Dong:2016fnf}) are dual to twist operators in the boundary theory. Hence one suggestion of the generalized observables is that they are dual to new operators in the boundary theory resulting from fusing the usual twist operators with other (codimension-two) operators. This would be similar to the discussion in \cite{Belin:2013uta,Belin:2014mva} where twist operators fused with Wilson lines in the boundary are dual to extremal surfaces carrying a flux of the dual gauge field.\footnote{There seems to be an important distinction with the charged Renyi entropies: in the limit $n\to1$, even though the conical defect disappears, the magnetic flux remains. Therefore, even at $n=1$, the bulk geometry is deformed by the generalized twist operator. If we wish to find a generalized twist operator which produces the integral of some general functional on a codimension-two surface, it must be such that it does not deform the spacetime at $n=1$, but only affects the first derivative. This would be like making the chemical potential $\mu\sim n-1$ in the case of the charged Renyi entropies.} In any event, pursuing this direction in the case of $d=2$ boundary theories where the twist operators are local operators would be interesting.

\subsection*{A preferred observable?}

An important open question remains whether certain observables are preferred over others. As we have argued throughout this paper, the late-time growth and the switchback effect are universally reproduced by all our observables, so they cannot be used to select an observable over any other. Nevertheless, it is possible that higher curvature terms in the observables need to be small for the observable to have a simple boundary interpretation. If we think about effective actions (\ie theory space rather than observable space), higher curvature terms are suppressed by the scale of new physics (typically the string or Planck scale) and are controlled in the CFT by the gap to higher spin operators \cite{Heemskerk:2009pn,Afkhami-Jeddi:2016ntf, Meltzer:2017rtf, Belin:2019mnx, Kologlu:2019bco}. Is there such an organizational principle for observables? An interesting example is the codimension-two surfaces appearing in the entanglement entropy. If we add higher curvature terms to the bulk action, the RT prescription must be modified to add extrinsic curvature terms \cite{Camps:2013zua,Dong:2013qoa}. This does not mean the area of extremal surfaces is not a good observable, it obviously still is, but it no longer has a simple translation in the boundary theory as an entanglement entropy. In this context, it seems that higher curvature terms in the bulk action and higher curvature terms in the observable go hand in hand. Is this a general principle? 

This touches on the question of higher curvature corrections in holographic complexity. We emphasize that in this paper, we have discussed observables in a bulk gravitational theory given by Einstein gravity. However, the CA proposal \reef{eq:defineCA} in a theory where the bulk action is modified by  higher curvature interactions would naturally produce an observable with new contributions involving powers of the background curvature similar to those discussed here, \eg see the discussion in \cite{Alishahiha:2017hwg}. Moreover, in \cite{Alishahiha:2015rta,Bueno:2016gnv,Hernandez:2020nem}, it was argued that adding higher curvature perturbations to Einstein gravity would naturally entail adding higher curvature corrections to the CV proposal \reef{eq:defineCV} with results similar to the codimension-one observables introduced in \cite{Belin:2021bga}. It would be very interesting to understand better how theory space synergizes with observable space, and we hope to return to this question in the future.

\section*{Acknowledgments}

We would like to thank Michal Heller, Ted Jacobson and Tomonori Ugajin for useful comments and discussion. The authors are also grateful for being able to meet in person at the {\it Qubits on the Horizon 2} meeting, which was sponsored by the ``It from Qubit'' and ``Scanning New Horizons'' collaborations. The work of AB is supported by
the NCCR 51NF40-141869 The Mathematics of Physics
(SwissMAP). Research at Perimeter Institute is supported in part by the Government of Canada through the Department of Innovation, Science and Economic Development Canada and by the Province of Ontario through the Ministry of Colleges and Universities. RCM is supported in part by a Discovery Grant from the NSERC of Canada, and by the BMO Financial Group.  RCM and SMR are supported by the Simons Foundation through the ``It from Qubit'' collaboration. SMR is also supported by MEXT-JSPS Grant-in-Aid for Transformative Research Areas (A) ``Extreme Universe'', No. 21H05187. AJS is supported by the Air Force Office of Scientific Research under award number FA9550-19-1-036.

\appendix
\section{Codimension-One Observables} \label{revone}

As a comparison with the analysis for codimension-zero observables in section \ref{zero}, we focus on the simpler codimension-one observables. In particular, we discuss an explicit example by adding a Weyl squared term in the generalized volume functional, \ie 
\begin{equation}\label{eq:generalziedC2}
	\mathcal{C}_{\rm gen}= \frac{1}{\GN L} \int d^d\sigma \,\sqrt{h_{ij}}\left(1 + \lambda \,L^4\,C^2 \right)\,, \quad \quad F_1=1 + \lambda \,L^4\,C^2 \,,
\end{equation}
where $ h_{ij}$ is the induced metric on the chosen hypersurface and $C^2\equiv C_{\mu\nu\rho\sigma}\,C^{\mu\nu\rho\sigma}$ denotes the square of the Weyl tensor with respect to the bulk spacetime, \viz
\begin{equation}
	C^2\equiv \mathcal{R}_{\mu\nu\rho\sigma} \mathcal{R}^{\mu\nu\rho\sigma}-\frac{4}{D-2} \mathcal{R}_{\mu\nu} \mathcal{R}^{\mu\nu}+\frac{2}{(D-1)(D-2)} \mathcal{R}^{2} \,.
\end{equation}
This is also the example briefly discussed in the appendix of \cite{Belin:2021bga}. We provide more detailed analyses of the time evolution of extremal surfaces here and stress the maximization problem among competing extremal surfaces. 

Let us take the black hole background defined in eq.~\eqref{eq:BHmetric} for more explicit results in the following. Assuming the hypersurface is parametrized by $(v(\sigma), r(\sigma))$ in the infalling coordinates, one can then rewrite the generalized volume as
\begin{equation}\label{eq:defineV}
\mC_{\rm{gen}}=\frac{V_x }{\GN L} \int_\Sigma d\sigma\,\(\frac{r}{L}\)^{d-1}\!\sqrt{-f(r\,){\dot v}^2+2\dot v\,\dot r}\ a(r)\,,
\end{equation}
where the factor $a(r)$ associated with the Weyl squared term reads 
\begin{equation}\label{eq:CCar}
	a(r) = 1 + \tilde{\lambda}\,\frac{L^4\omega^{2(d-2)}}{r^{2d}} \,, 
\end{equation}
with $\tilde{\lambda} =d(d-1)^2(d-2)\,\lambda$. Of course, the more general analysis discussed in section \ref{sec:spacetimevolume} applies to this particular example after choosing a vanishing bulk term, \ie $G_1=0$ with $b(r)=0$. As defined in eq.~\eqref{eq:definepotential}, the corresponding effective potential without the momentum part is then given by 
\begin{equation}\label{eq:ham3}
	\begin{split}
		U_0(r)&\equiv  -f(r)\,a^2(r)\,\(\frac{r}{L}\)^{2(d-1)}=\(\frac{r_h}{L}\)^{2d}\left(w-w^2\right)\left(
		1 + \frac{\tilde{\lambda}}{w^2}\right)^2 \,.
	\end{split}
\end{equation}
with using the dimensionless radial coordinate $w= (r/r_h)^d$. In the following, we will show that the time evolution of the extremal surfaces related to the codimension-one measure $\mC_{\rm gen}$ is determined by the profiles of the effective potential $U_0(r)$. In this example with a Weyl squared term, the profiles of $U_0(r)$ are then determined by the dimensionless parameter $\tilde{\lambda}$. We show some characteristic plots for the effective potential $U_0(r)$ with various $\tilde{\lambda}$. 

\begin{figure}[ht!]
	\centering
	\includegraphics[width=3in]{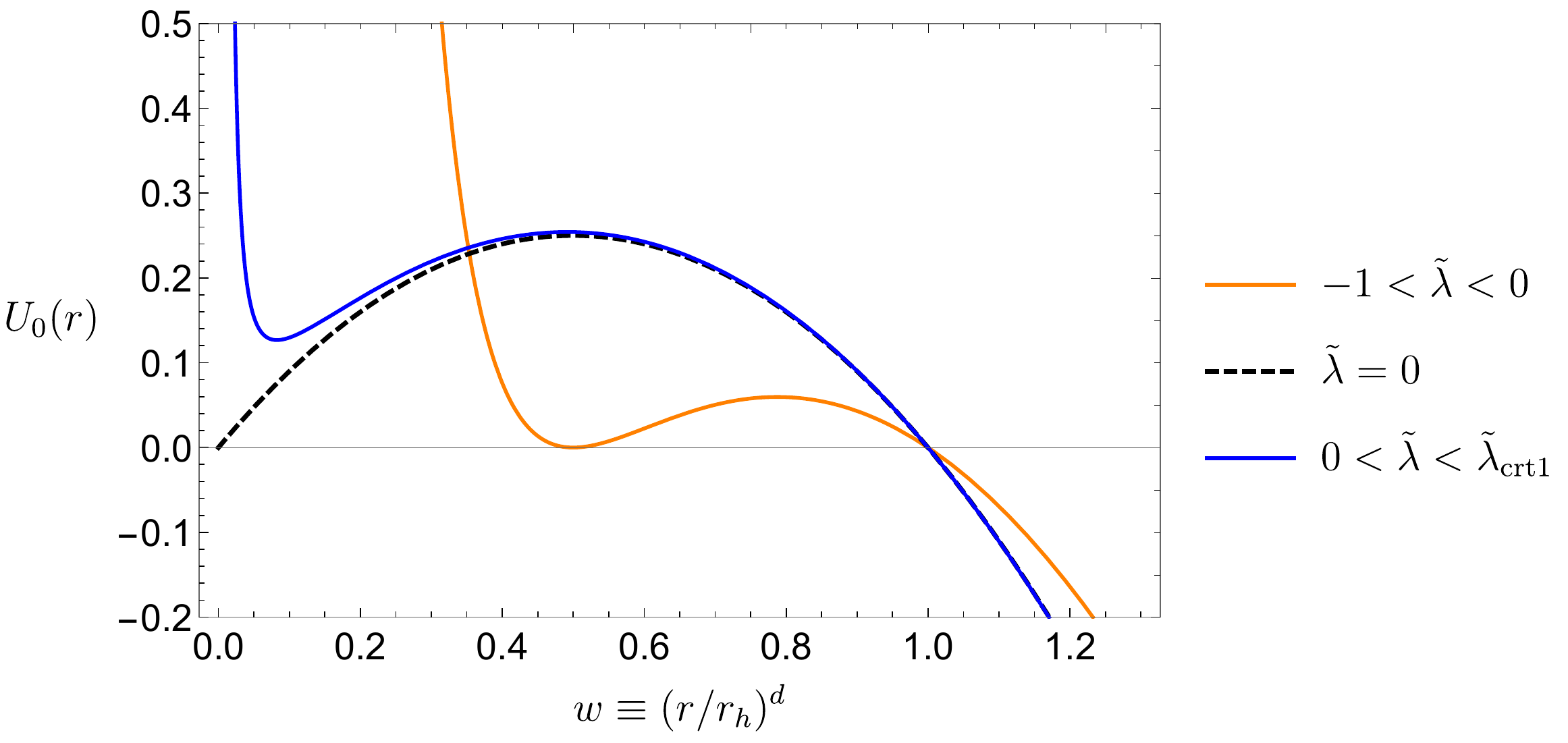}
	\includegraphics[width=3in]{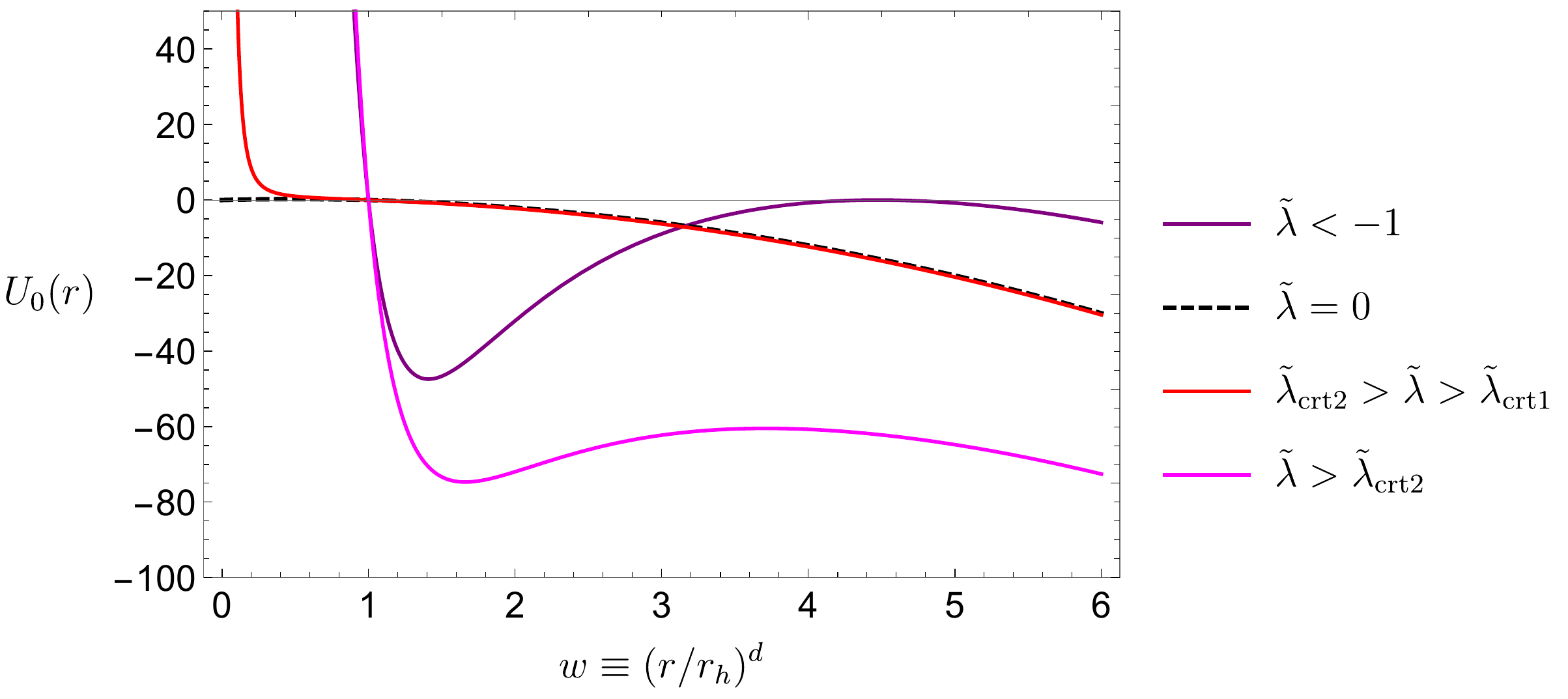}
	\caption{Left: The effective potentials $U_0$ with $-1<\tilde{\lambda} < \tilde{\lambda}_{\rm{crt}1}$. Right: The effective potentials with $\tilde{\lambda} <-1$ or $\tilde{\lambda} > \tilde{\lambda}_{\rm{crt}1}$. The late-time linear growth of $\mC_{\rm gen}$ is associated with the existence of the local maximum located inside the black hole interior (\eg taking $-1<\tilde{\lambda} < \tilde{\lambda}_{\rm{crt}1}$).}
	\label{fig:potentialC2}
\end{figure}

In order to discuss the time evolution of extremal surfaces with various patterns of effective potential $U_0(r)$, it is convenient to distinguish $U_0(r)$ by their extremal points, which are determined by $\partial_r U_0(r)=0 $. By using the explicit form derived in eq.~\eqref{eq:ham3}, one can find that the positions of the extrema of $U_0(r)$ satisfy
\begin{equation}
	\tilde{\lambda}= - w^2\,,\qquad \text{or} \qquad 
	\tilde{\lambda} = \frac{w^2(2w-1)}{2w-3} \,.
\end{equation}
However, the number of solutions depends on the value of $\tilde{\lambda}$ as shown in Figure \ref{fig:potentialC2}. Obviously, the solution of the first extremal equation $\tilde{\lambda}= - w^2$ only exist when $\tilde{\lambda} <0$. For the second extremal equation, we can get two solutions for $\tilde{\lambda} < \tilde{\lambda}_{\mathrm{crt}1}= \frac{1}{8} (47-13\sqrt{13}) $ and $\tilde{\lambda} > \tilde{\lambda}_{\mathrm{crt}2}= \frac{1}{8} (47+13\sqrt{13})$. In the middle regime $\tilde{\lambda}_{\mathrm{crt}1}<\tilde{\lambda} <\tilde{\lambda}_{\mathrm{crt}2}$, there is no extremal point. Moreover, we are interested in the extremal point behind the black hole horizon. With taking 
\begin{equation}
 -1<\tilde{\lambda} < \tilde{\lambda}_{\mathrm{crt}1}= \frac{1}{8} (47-13\sqrt{13}) \,,
\end{equation}
we can have at most two extremal points inside the black hole interior. Furthermore, it is straightforward to show that one of two extremal points corresponds to the local maximum and another is the local minimum. For later use, let us refer to the local maximum as the final slice at $r=r_f$, which plays a vital role in producing the linear growth of the codimension-one observable $\mC_{\rm gen}$.

\subsection{Extremal Surfaces and Time Evolution}
Similar to the discussion in section \ref{zero}, we can explore the time evolution of the extremal surfaces by recasting it as a classical mechanics problem with an effective action $\mC_{\rm gen}$ defined in eq.~\eqref{eq:defineV}. First of all, we choose the same gauge condition as eq.~\eqref{eq:gaugechoice}, \ie 
\begin{equation}\label{eq:gaugechoice02}
\sqrt{-f(r)\dot v^2+2  \dot v \dot r}=a_\pm(r) \(\frac{r}{L}\)^{d-1} \,.
\end{equation}
due to the reparametrization invariance of $\sigma$ for the hypersurface $(v(\sigma), r(\sigma))$. Secondly, the conserved momentum $P_v$ is derived as 
\begin{equation}\label{eq:momentum02}
P_v
=\frac{a(r)\,(r/L)^{d-1}\left(\dot r -f(r)\,\dot v\right)}{\sqrt{-f(r\,){\dot v}^2+2\dot v\,\dot r}}=\dot r -f(r)\,\dot v\,,
\end{equation}
which can be also reduced from eq.~\eqref{eq:vmoment} with setting $b(r)=0$.
With boundary conditions, the extremal surface with respect to the generalized volume defined in eq.~\eqref{eq:generalziedC2} can be obtained by solving eqs.~\eqref{eq:gaugechoice02} and \eqref{eq:momentum02}. Our analysis of the time evolution of the extremal surface could apply to all regimes of $\tilde{\lambda}$. However, we note that $a_\pm (r)$ could be also negative in some regions of $r$ when $\tilde{\lambda}<0$. Correspondingly, we should take $\sqrt{-f(r)\dot v^2+2  \dot v \dot r}= |a_\pm(r)| \(\frac{r}{L}\)^{d-1}$ in order to make the hypersurface parameter $\sigma$ always real. However, the extra sign can be absorbed into the sign of $P_v$ defined in eq.~\eqref{eq:momentum02}, resulting in the same extremization equations. So we will ignore this sign problem in the following. 

As shown in eqs.~\eqref{eq:dots}, the extremal equations can be decomposed into two first derivative equations:
\begin{equation}\label{eq:dotr02}
	\dot{r}^2 = P_v^2 - U_0(r)\,, 
\end{equation}
and 
\begin{equation}\label{eq:dtdr}
	\dot{t}\equiv \dot{v} - \frac{\dot{r}}{f(r)}= \frac{-P_v \, \dot{r}}{f(r)\sqrt{P_v^2+f(r)\,a^2(r)\,(r/L)^{2(d-1)}}}   \,.
\end{equation}
for $\dot{r}, \dot{v}$, respectively. Since the conserved momentum $P_v$ and the effective potential $U_0(r)$ are decoupled in this case, we can separate the corresponding two terms in the radial equation in eq.~\eqref{eq:dotr02}. With a given conserved momentum, one can get an extremal surface by solving eq.~\eqref{eq:dotr02}. For completeness, we further note that the complete equation of motions with second derivatives read
\begin{equation}\label{eq:EoMfull}
	\begin{split}
		2\ddot{r}&= \partial_r \(  f(r)a^2(r) (r/L)^{2(d-1)} \)=- \partial_r {U}_0(r)\,,\\
		2\ddot{v}&=  \partial_r \( a^2(r) (r/L)^{2(d-1)} \) - \dot{v}^2 \partial_r f(r)\,.\\
	\end{split}
\end{equation}
As we introduced before, the final slice at $r=r_f$ corresponds to the local maximum of $U_0(r)$, \ie satisfying 
\begin{equation}\label{eq:definerf}
U_0'(r_f) = \partial_r U_0(r_f) =0 \,, \qquad U_0''(r_f) < 0 \,.
\end{equation}
From the extremality equation, we can find that the final slice (if it exists) is also an extremal surface with respect to the codimension-one functional $\mC_{\rm gen}$.

\begin{figure}[ht!]
	\centering
	\includegraphics[width=3in]{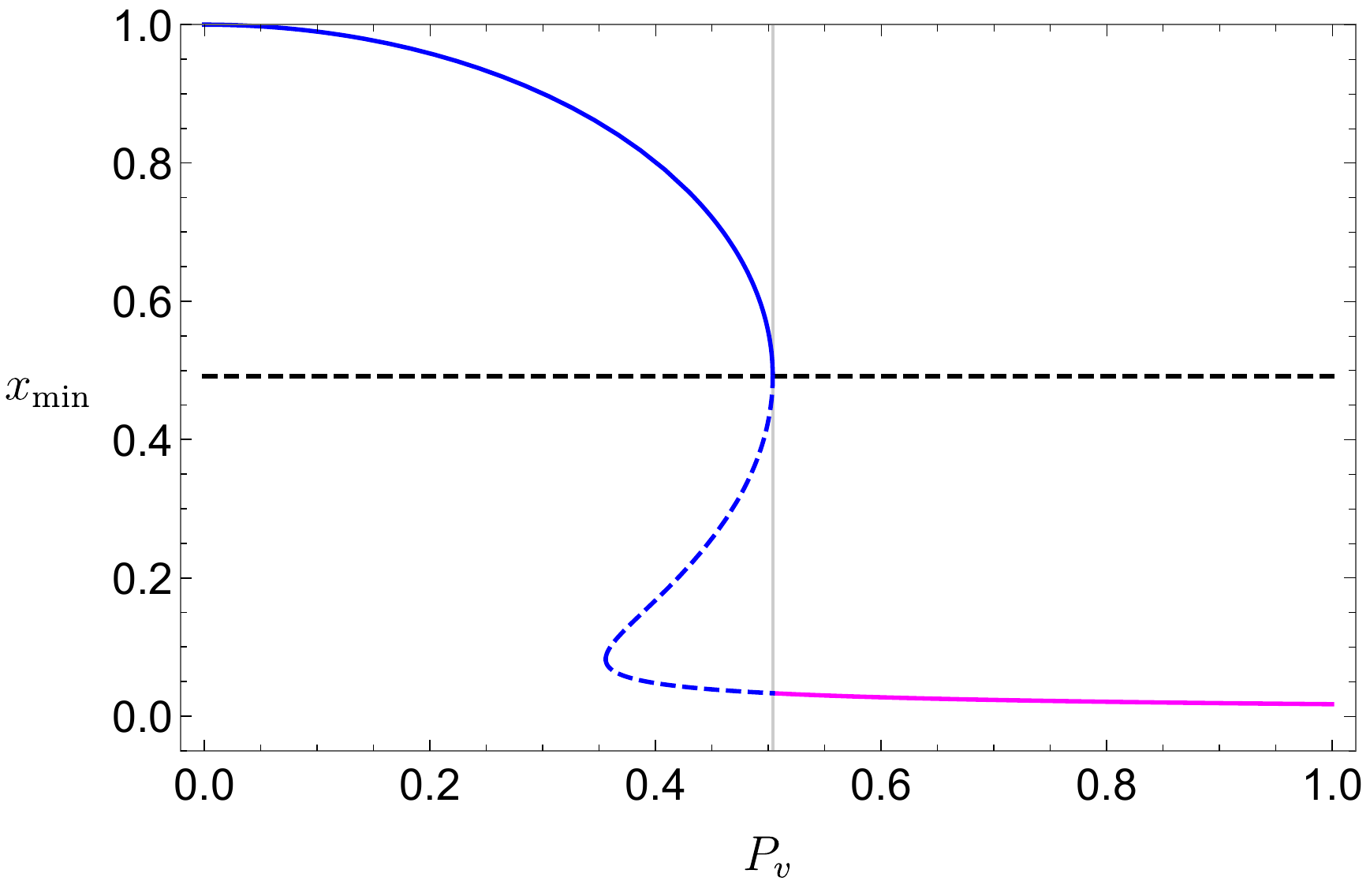}\,
	\includegraphics[width=2.9in]{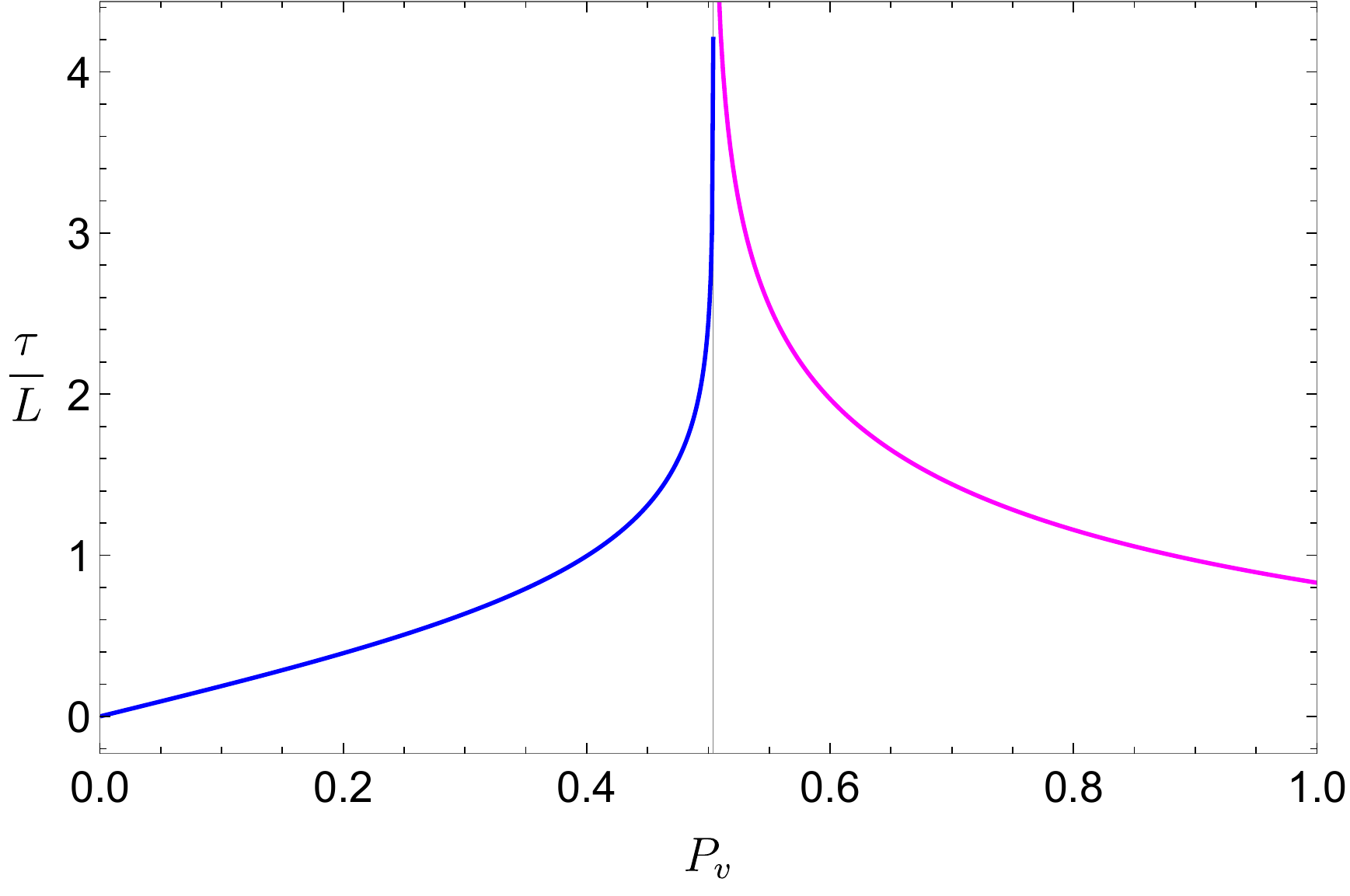}
	\caption{Left: The minimal radius as a function of the conserved momentum $P_v$. Here we choose the dimensionless parameter $x_{\rm{min}}$. The solid blue curve and red curve correspond to the extremal hypersurfaces with $P_v \le P_{\infty}$ and $P_v \ge P_{\infty}$, respectively. The middle dashed line presents the position of the final slice at $r=r_f$. Right: The relation between the boundary time $\tau$ and the conserved momentum $P_v$. The middle transparent line indicates the critical momentum $P_{\infty}$. For both plots, we choose $\tilde{\lambda} = \frac{1}{500}, d=3$.}	
	\label{fig:tvP}
\end{figure}

The basic intuitive description of the extremal surfaces remains the same as for the extremal volume case that is discussed in \cite{Stanford:2014jda,Chapman:2018dem,Belin:2021bga}. The relevant trajectories begin at $r=\infty$, approach the positive ridge in $U_0(r)$ reaching a turning point at $r=r_{\rm{min}}$, and then return out to $r=\infty$. Explicitly, the turning point is given by the conserved momentum $P_v$ via
\begin{equation}
	P_v^2=U_0(r_{\rm{min}}) \equiv -f(r_{\rm{min}}) a^2(r_{\rm{min}}) \(\frac{r_{\rm{min}}}{L}\)^{2(d-1)} \,,  \qquad \text{at} \qquad t=0\,.
\end{equation} 
See Figure \ref{fig:tvP} for an example of the relation between $r_{\rm{min}}$ and $P_v$.
Note that the turning point at $r=r_{\rm{min}}$ is behind the horizon,\ie $r_{\rm{min}} \le r_h$, and this is the minimum radius on the extremal surface, \ie the point of closest approach to the singularity. Compared with the potential from the volume, one new feature will be that since curvature invariants will generically diverge at $r=0$, we can expect that $a^2(r)$ and the effective potential will diverge to positive infinity there. Hence there will be no trajectories that crash into the singularity. Rather for arbitrarily large energy $P_v^2$, the incoming trajectories will be reflected out to infinity. While these trajectories can probe arbitrarily close to the singularity, we will show in the following that they do not correspond to the extremal surfaces with a maximal generalized volume. Rather those will correspond to trajectories that are tuned to reflect off of a maximum in the potential, \ie with $r_{\rm{min}}\simeq r_{\rm{max}}$.
When the conserved momentum $P_v$ is tuned to near the critical value $P_{\infty}$ that is defined by 
\begin{equation}
P_{\infty}^2 = U_0(r_f) \,, 
\end{equation}
the trajectory takes a large amount of effective time approaching the turning point such that $r\sim r_{\rm{min}} \simeq r_f$. This behaviour corresponds to late boundary times (\ie $\tau \to \infty$) when the extremal surface is very close to the final slice at $r=r_f$ over a large section inside the horizon.
Certainly, these are the trajectories that will yield linear growth at late times. It is natural to think that in this situation that there will be multiple trajectories for a given boundary time, \eg one reflecting off of the barrier near $r=0$ and another reflecting off the maximum, but that the second of these will yield the maximal `volume' -- as we will verify below. Of course, when the parameter $\tilde{\lambda} <-1$ or $\tilde{\lambda} > \tilde{\lambda}_{\rm{crt}1}$, the positive maximum in the effective potential of the volume (\ie $a^2(r)=1$) is eliminated because of the $a^2(r)$ profile, \ie there is no local maximum of the effective potential inside the horizon (see the right plot in Fig \ref{fig:potentialC2} for illustration). For these cases, we will show that there is no late-time limit. In other words, there is no extremal surface when the boundary time is larger than the critical value. 

Similar to what we have shown in eq.~\eqref{eq:dVdt01}, the time derivative of the extremal generalized volume $\mathcal{C}_{\rm gen} $ is also controlled by the momentum evaluated on the boundary, namely 
\begin{equation}\label{eq:dVdtau}
 \frac{d 	\mathcal{C}_{\rm gen} }{d\tau} = \frac{V_x}{\GN L}P_v(\tau)\,.
\end{equation}
By using the extremality equation related to the coordinate time $t$ in eq.~\eqref{eq:dtdr} and fixing the boundary condition of a specific extremal hypersurface, \ie the boundary time $\tau$, one can determine the conserved momentum $P_v(\tau)$ as follows 
\begin{equation}\label{eq:tauPv}
	\frac{\tau}{2}=t_{\mt{R}}=t_{\mt{L}}= \int^{r_{\rm{max}}}_{r_{\rm{min}}(P_v)} dr\, \frac{-P_v}{f(r)\sqrt{P_v^2+f(r)\,a^2(r)\,(r/L)^{2(d-1)}}}  \,.
\end{equation}
It is also obvious that the low bound of the conserved momentum $P_v^2=0$ corresponds to the extremal hypersurface anchored at $\tau =0$.  In Fig \ref{fig:tvP}, we show the numerical results of this relations with a fixed parameter $\tilde{\lambda}$.  In the following, we will analyze the late-time limit for different choices of the parameter $\tilde{\lambda}$.

\subsection{Extremal Surfaces at Late Times }
Our first goal is to show the linear growth of the generalized volume at late times, \ie 
\begin{equation}
	\text{Linear Growth:}\qquad \lim\limits_{\tau\to\infty} 	\mathcal{C}_{\rm gen}(\tau) \sim P_{\infty}\, \tau  \,,
\end{equation}
with a growth rate 
\begin{equation}
 P_{\infty} \equiv \lim_{\tau \to \infty}  P_v(\tau)\,. 
\end{equation}
 It is worth noting that the linear growth of the generalized volume $\mathcal{C}_{\rm gen}$ in the late-time limit is equivalent with the condition that its time derivative approaches a constant, \ie
\begin{equation}
	\lim_{\tau \to \infty}  \(    \frac{d	\mC_{\rm gen}}{d\tau} \)  = \frac{V_x}{\GN L} P_{\infty} \,.
\end{equation}
due to the equivalence $\lim\limits_{\tau\to\infty}   \(  \frac{\mC_{\rm gen}}{ P_v \tau}\) = \lim\limits_{\tau\to\infty} \(  \frac{d\mC_{\rm gen}/d\tau}{ P_v }\)$. Since we have shown the generic time derivative is given by the conserved momentum $P_v(\tau)$ in eq.~\eqref{eq:dVdtau}, the linear growth at late times then reduces to the following conditions:
\begin{itemize}
	\item the existence of the extremal surface in the late-time limit $\tau \to \infty$;
	\item the late-time limit of the conserved momentum $P_v(\tau)$ exists and is finite.
\end{itemize}

In the following, we will show that the generalized volume presents the expected linear growth if its effective potential $U_0(r)$ contains a local maximum inside the horizon, \ie when $-1\le \tilde{\lambda} \le \tilde{\lambda}_{\rm{crt}1}$ for the generalized volume with a Weyl square term. From the relation between $P_v$ and $\tau$ in eq.~\eqref{eq:tauPv}, one sees that the existence of the extremal hypersurface at $\tau \to \infty$ implies that there is a divergence in the integral in eq.~\eqref{eq:tauPv}. Regardless of the value of $\tilde{\lambda}$, the integrand contains multiple potential singularities at $r \to r_{\rm{min}}$ with ${P_v^2+f(r_{\rm{min}})\,a^2(r_{\rm{min}})\,(r_{\rm{min}}/L)^{2(d-1)}}=0$, and $r \to  r_h$ with $f(r_h)=0$. However, it is easy to see that the integral near these two points is still finite. More explicitly, we can take the leading-order expansion as follows
\begin{equation}
	\lim_{r\to r_{\rm{\min}}}  \frac{1}{\sqrt{ P_v^2 +f(r)\,a^2(r)\,(r/L)^{2(d-1)}}} = \lim_{r\to r_{\rm{\min}}} \frac{1}{\sqrt{  U_0(r_{\rm{min}})-U(r) }} \sim \frac{1}{\sqrt{ U_0'(r_{\rm{min}})  \( r_{\rm{\min}}-r \)}} \,,
\end{equation}
which implies the integral in eq.~\eqref{eq:tauPv} is regular around the minimal radius. Furthermore, although the integral near the horizon at $r_h$ presents the logarithmic divergence due to $f(r) \sim f'(r_h) (r-r_h)$, we remark that the extremal hypersurface crosses the horizon, and this kind of divergence is canceled between the two sides of the horizon. In other words, the {\it Cauchy principal value} of this integral for the singularity at the horizon $r_h$ is still finite, \ie 
\begin{equation}
	\begin{split}
			&\lim_{\epsilon \to 0} \[  \int^{r_h -\epsilon}_{r_{\rm{min}}} dr\, \frac{-P_v}{f(r)\sqrt{P_v^2+f(r)\,a^2(r)\,(r/L)^{2(d-1)}}} + \int_{r_h +\epsilon}^{r_{\rm{max}}} dr\, \frac{-P_v}{f(r)\sqrt{P_v^2+f(r)\,a^2(r)\,(r/L)^{2(d-1)}}} \] \\
			&\sim \mathcal{O}(1)\,. 
	\end{split}
\end{equation}
Finally, two other types of singularities may also appear, depending on the value of $\tilde{\lambda}$. The following shows that these different cases correspond to distinct fates for the extremal surfaces at late times. 

\subsubsection{No extremal hypersurface at late times} 
First of all, let us illustrate why there is no extremal hypersurface at late times for the generalized volume with $ \tilde{\lambda} < -1$ or $\tilde{\lambda} > \tilde{\lambda}_{\rm{crt}1}$. Except for the two potential divergences discussed above, there is one more singularity at $r=0$ for these regimes of $\tilde{\lambda}$. The extremal surface can arbitrarily approach the singularity by pushing the momentum to infinity, \ie $P_v \to \infty$ with $r_{\rm{min}} \to 0, f(r_{\rm{min}}) \to - \infty$. In this limit, one obtains the corresponding value of the boundary time 
\begin{equation}
	\lim_{r_{\rm{min}} \to 0} \tau = \lim_{P_v \to \infty} \tau = \int^{\infty}_{0} \frac{-dr}{f(r)} =\frac{L^2}{r_h}  \frac{\pi}{d}\cot \( \frac{\pi}{d} \) \,,
\end{equation}
where we have taken the limit $r_{\rm{min}} \to 0$ first and then applied the limit $r_{\rm{max}} \to \infty$. As a summary for these regimes of $\tilde{\lambda}$, we conclude that the boundary time $\tau$ is always finite. Correspondingly, there is no extremal hypersurface at late times.

\subsubsection{Linear growth at late times with $ -1\le \tilde{\lambda} \le \tilde{\lambda}_{\rm{crt}1}$}
As we illustrate before, the effective potential contains saddle points inside the horizon when $ -1\le \tilde{\lambda} \le \tilde{\lambda}_{\rm{crt}1}$. Importantly, there is a local maximum given by the final slice located at $r=r_f$. As a consequence, the integrand in eq.~\eqref{eq:tauPv} presents a higher order divergence, since $U'(r_{f})=0$. Correspondingly, we can push the extremal hypersurface near the final slice by taking $P_v \to P_{\infty} \equiv \sqrt{U_0(r_f)}$. Writing the expansion near the final slice, \ie 
\begin{equation}
	\lim_{r\to r_f}  \frac{1}{\sqrt{ P_v^2 +f(r)\,a^2(r)\,(r/L)^{2(d-1)}}} = \lim_{r\to r_f} \frac{1}{\sqrt{  U_0(r_f)-U_0(r) }} \sim \sqrt{\frac{-2}{ U_0''(r_f)}} \frac{1}{ \( r-r_f \)} \,,
\end{equation}
we find that the boundary time is approaching infinity
\begin{equation}
	\lim_{P_v \to P_{\infty}} \tau \sim \lim_{r_{\rm{min}} \to r_f} \log \( \frac{1}{ r_{\rm{min}}-r_f } \) \to \infty \,.
\end{equation}
This thus implies the existence of the extremal hypersurface at late times. As shown in Fig \ref{fig:tvP}, the boundary time is approaching infinity when $P_v$ approaches the critical value $P_{\infty}$. In the meantime, we arrive at the expected linear growth at late times, \viz
\begin{equation}
 \lim_{\tau \to \infty}   \frac{d \mC_{\rm gen}}{d\tau}  =  \frac{V_x P_{\infty}} {\GN L}  =\frac{16\pi M}{d-1}\sqrt{\frac{16 (1-x_f)^3 x_f}{(3-2 x_f)^2}} \,,
\end{equation}
where the explicit solution of the final slice for the generalized volume with the Weyl square term reads
\begin{equation}\label{eq:finalSlice}
\begin{split}
w_f= &\frac{1}{6} \left(1+\frac{1+12 \tilde{\lambda} }{\({1-144 \tilde{\lambda} +6  \sqrt{-3\tilde{\lambda}  (4 \tilde{\lambda}  (4 \tilde{\lambda} -47)+3)}} \)^{1/3}}+  \right. \\
& \qquad \left. + \({1-144 \tilde{\lambda} +6 \sqrt{-3\tilde{\lambda}  (4 \tilde{\lambda}  (4\tilde{\lambda} -47)+3)}} \)^{1/3}  \right) \,.
\end{split}
\end{equation}

\subsubsection{Local maxima among multiple extremal surfaces}\label{sec:localmaxima}
Due to the facts that
\begin{equation}
	\dot{r} \big|_{r=r_f} = 0 =\ddot{r}\big|_{r=r_f} \,, \qquad \text{with} \qquad  \(\partial_rU_0(r) \)\big|_{r=r_f} =0\,,
\end{equation} 
the extremal hypersurface anchored at infinite time $\tau =\infty$ is given by a constant-$r$ slice, \ie the final slice at $r=r_f$. This branch of extremal hypersurfaces is related to $P_v\le P_{\infty}$. As shown in Fig \ref{fig:tvP}, there is also another branch of solutions with $P_{\mt{S}} \ge P_{\infty}$, which also have an infinite time limit and the corresponding linear growth of the generalized volume. At late times, these surfaces hug the final slice $r=r_f$ most of the time, but have a region where they dip towards the singularity. Let us call this the dipping branch. We will now show that the generalized volume with $P_{\mt{L}} \le P_{\infty}$ (denoted by $\mC_{\mt{L}}$) is larger than the one (denoted by $\mC_{\mt{S}}$) with $P_{\mt{S}} \ge P_{\infty}$, \ie the dipping branch has smaller generalized volume. Using the extremality condition in eq.~\eqref{eq:dtdr},  the generalized volume $\mC_{\rm gen}$ (both for $\mC_{\rm L}$ and $\mC_{\rm S}$) can be recast as  
\begin{equation}
	\begin{split}
	\mC_{\rm gen }(\tau)  & =\frac{V_x}{\GN L}\int^{\tau}_{0} d\tau \( \frac{U_0(r(t))}{P_v}  \)\,.
	\end{split}
\end{equation}
We note again that the momentum $P_v$ is a constant along any extremal surface.
In the infinite time limit $\tau \to \infty$, the conserved momentum $P_v$ reaches the critical value $P_{\infty}$, and the entire extremal surface moves inside the horizon. We can then evaluate the late limit of the generalized volume $\mC_{\rm gen }$ of the non-dipping branch by 
\begin{equation}
	\begin{split}
	\lim\limits_{\tau\to\infty} \mC_{\mt{L}}(\tau) &=\frac{V_x}{\GN L} \int^{\tau \to \infty}_{\tau=0} P_{\infty} \, d\tau  \,,
	\end{split}
\end{equation}
where the extremal surface is the constant $r=r_f$ slice. For the dipping branch of extremal surfaces with the generalized volume $\mC_{\mt{S}}$, the late time surface is still entirely inside the horizon. It starts from the minimal radius $r_{\rm{min}}$ that is very near to the singularity and evolves towards the final slice until $\tau \to \infty$. Irrespective of the details of these two solutions, we find
\begin{equation}
	\begin{split}
	\lim\limits_{\tau\to\infty} \(	{\mC_{\mt{L}}(\tau)-\mC_{\mt{S}}} (\tau) \) &= \frac{V_x}{\GN L}\int^{\tau \to \infty}_{\tau=0} d \tau \,\( P_{\infty} -\frac{U_0(r)}{P_{\infty}}   \)  >0\,, 
	\end{split}
\end{equation}
because the effective potential along the extremal surface with $P_v^2 \ge P_{\infty}^2$ satisfies $U(r) \le P_{\infty}^2$. So far, we have shown that the final slice is the maximal extremal surface (with the larger generalized volume) anchored at infinite boundary time. Then we can use the time derivative of the generalized volume as shown in eq.~\eqref{eq:dVdtau} to obtain the generalized volume at finite boundary time $\tau$, \ie 
\begin{equation}
	\begin{split}
			\mC_{\mt{L}}(\tau)&=\mC_{\mt{L}} \(\infty\)  - \frac{V_x}{\GN L} \int^{\infty}_{\tau} P_{\mt{L}} \( \tau \) d \tau \,, \\
			 \mC_{\mt{S}}(\tau)&=\mC_{\mt{S}} \(\infty\)  - \frac{V_x}{\GN L}\int^{\infty}_{\tau} P_{\mt{S}} \( \tau \) d \tau \,.
	\end{split}
\end{equation}
Combing the above results, we arrive at 
\begin{equation}
	\begin{split}
		\mC_{\mt{L}}(\tau)- \mC_\mt{S} \( \tau \) &= \mC_{\mt{L}} \(\infty\)  -  \mC_{\mt{S}} \(\infty\)  + \frac{V_x}{\GN L}\int^{\infty}_{\tau} \(P_{\mt{S}}\( \tau \)- P_{\mt{L}}\( \tau \) \)  d \tau  >0 \,,
	\end{split}
\end{equation}
which is obviously positive because of $P_{\mt{L}}\( \tau \) \le P_{\infty} \le P_{\mt{S}}\( \tau \)$. Therefore, we conclude that for a fixed boundary time $\tau$ the generalized volume of the non-dipping surfaces $P_v \le P_{\infty}$ is larger than the dipping surfaces $P_v \ge P_{\infty}$.  In other words, we can find that the extremal surfaces from the non-dipping branch and dipping branch correspond to the local maximal surface and locally minimal surface, respectively. Finally, let us remark that the above proof for the local maximal could be easily generalized to the general case with an arbitrary potential $U_0(r)$ with multiple peaks, \ie multiple extremal surfaces at the late-time limit. In the same spirit, one can conclude that the maximal generalized volume $\mC_{\rm gen}$ is always given by the one with the largest momentum $P_\infty$, \ie 
\begin{equation}
 \max \( \mC_{\rm{gen}}(\tau \to \infty) \) \qquad  \longleftrightarrow  \qquad  \max \,(P_{\infty}) \,.
\end{equation}
In other words, the maximal extremal surfaces at late times are given by the branch whose growth rate $P_v(\tau)$ approaches the highest peak of the effective potential $U_0(r)$ from below.

\section{Adding Extrinsic Curvature Terms}\label{sec:appB}
In section \ref{zero} and appendix \ref{revone}, we explore several explicit example for {\it complexity equals anything}  proposal. In particular, these previous examples consider specific functional constructed from geometric tensor of the bulk spacetime and hypersurfaces. Focusing on the codimension-one case, we can define the holographic complexity measure as
\begin{equation}\label{eq:dCdt}
	\mC_{\rm{gen}}(\tau) =\max_{\partial\Sigma(\tau)=\Sigma_\tau} \[ \frac{1}{G_{\mt{N}} \,L}  \int_{\Sigma} \! d^d\sigma \,\sqrt{h} \,F_1(g_{\mu\nu};X^\mu(\sigma))  \].
\end{equation}
with $X^\mu(\sigma)$ as the embedding function for the hypersurface $\Sigma(\tau)$ at a boundary time $\tau$. In the most general case, we should point out that the codimension-zero functional would also depend on the extrinsic geometric quantities of the hypersurface. Inspired by the ADM Hamiltonian formalism  or canonical quantization of general relativity where the phase space is given by  $(h_{ij}, \pi^{ij})$ with $\pi^{ij} \equiv \sqrt{h}\( K^{ij} - K h^{ij} \)$ as the conjugate momentum, we would like to explore the role of the extrinsic curvature $K_{ij}$ of the hypersurface for the measure of holographic complexity $\mC_{\rm{gen}}(\tau)$ \footnote{\label{footnoteKterm} The extrinsic curvature terms also naturally appear in the generalized volume for holographic complexity of higher curvature gravity theory \cite{Hernandez:2020nem}.} in this section.  

For example, we can consider the generalized volume with including the extrinsic curvature terms, \eg 
\begin{equation}
	\mC_{\rm{gen}}(\tau) = \max_{\partial\Sigma(\tau)=\Sigma_\tau} \[ \frac{1}{G_{\mt{N}} \,L}  \int_{\Sigma} \! d^d\sigma \,\sqrt{h} \left(1 + g(h_{ij}, K^{ij}) \right)   + \text{Boundary terms} \big|_{\Sigma_\tau} \].
\end{equation}
where $g(K_{ij})$ is referred to as an arbitrary function of the induced metric and extrinsic curvature. Noting that the extrinsic curvature terms include higher derivatives, we need to have an appropriate boundary term in order to have a well-posed extremization for the functional $\mC_{\rm{gen}}(\tau)$. Because of the appearance of higher derivative terms
we also need to generalize the results in section \ref{zero} to the most general case with an arbitrary function $g(h_{ij}, K^{ij})$. 

Similar to the analysis we discussed before in section \ref{zero} and appendix \ref{revone}, we can recast the problem as a classical mechanics problem which is defined by an effective action $S = \int \mathcal{L}\, d\lambda$. Supposing the effective Lagrangian for the dynamical fields $y(\lambda)$ includes second derivatives, \eg $\mathcal{L} (y(\lambda), y'(\lambda), y''(\lambda))$,  the action with the well-posed variational problem leads to the equation of motion, namely
\begin{equation}\label{eq:EOM}
	\text { EoM: } \quad \frac{\partial \mathcal{L}}{\partial y}-\frac{d}{d  \lambda} \frac{\partial \mathcal{L}}{\partial y^{\prime}}+\frac{d^{2}}{d \lambda^{2}}\left(\frac{\partial \mathcal{L}}{\partial y^{\prime \prime}}\right)=0 \,.
\end{equation}
If we start from the on-shell solutions with satisfying the equation of motion, the on-shell condition simply implies that only boundary terms contribute to the perturbation with changing the endpoints, \ie $\Delta S= P_y \delta y \big|_{\rm bdy}$. The conjugate momentum $P_y $ is defined by 
	\begin{equation}\label{eq:Py}
		P_{y} \equiv \frac{\partial \mathcal{L}}{\partial y^{\prime}}-\frac{d}{d \lambda}\left(\frac{\partial \mathcal{L}}{\partial y^{\prime \prime}}\right)\,,
	\end{equation}
which is decoded in the equations of motion as follows: 
\begin{equation}\label{eq:Pv02}
	\frac{\partial \mathcal{L}}{\partial y}-\frac{d P_y}{d  \lambda}=0 \,.
\end{equation}
In the following, we take two simple examples by including the extrinsic curvature terms and focusing on the evolution of the extremal surface and the growth rate of holographic complexity measure $\mC_{\rm{gen}}$.  In particular, we can find that the extremal surface is determined by the equations of motion in eq.~\eqref{eq:EOM} and the growth rate is still controlled by the conjugate momentum defined in eq.~\eqref{eq:Py}.

\subsection{Linear $K$ term}

\begin{figure}[!]
	\centering
	\includegraphics[width=5.5in]{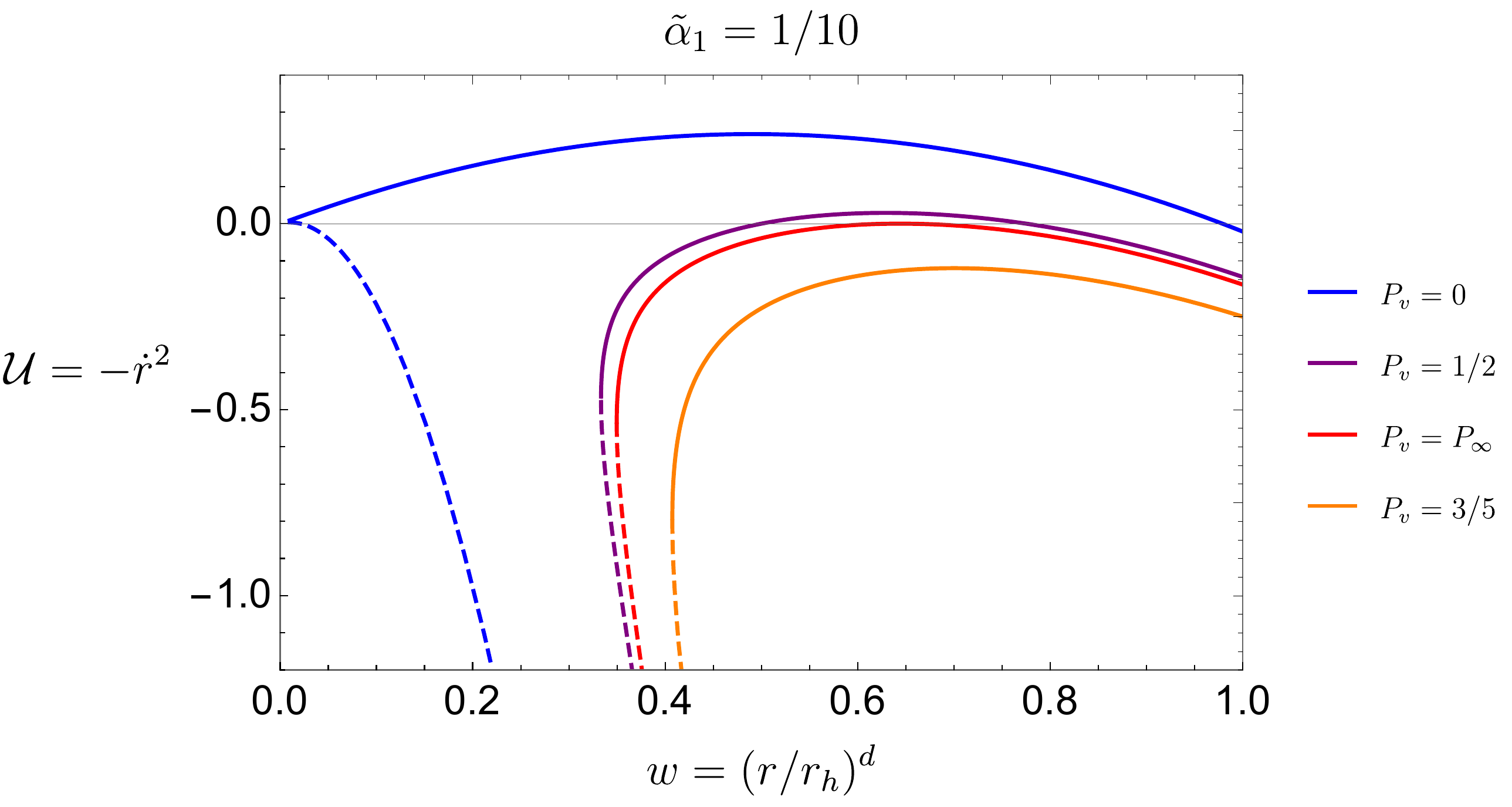}
	\caption{The effective potentials $\mathcal{U}(P_v, r)$ defined in eq.~\eqref{eq:UKtwo} (for a planar black hole solution) with various fixed conserved momentum $P_v$. The solid/dashed curves denote the positive/negative branch derived in eq.~\eqref{eq:UKtwo}, respectively. We have fixed $r_h/L=1, d=3$ for this plot.}	
	\label{fig:tvPK}
\end{figure}

The simplest codimension-one functional with containing extrinsic curvature is taken as
\begin{equation}\label{eq:linearK}
	\mC_{\rm{gen}}(\tau) =  \max_{\partial\Sigma(\tau)=\Sigma_\tau} \[ \frac{1}{G_{\mt{N}} \,L}  \int_{\Sigma} \! d^d\sigma \,\sqrt{h}  (1 + \alpha_1 K)   \]\,,
\end{equation}
where we include the trace of the extrinsic curvature besides of the volume term. Taking the bulk spacetime as the planar black hole whose metric is defined in eq.~\eqref{infall}, one can consider any hypersurface parameterized by $v(r)$ and derive its trace of extrinsic curvature as 
\begin{equation}
K =\frac{ 4(d-1)v'(r)  - \(2(d-1) f(r)  - r f'(r)\) (3- f(r)v'(r)) (v'(r))^2    - 2r v''(r)}{ 2r \( 2 v'(r) - f(r) v'(r)^2) \)^{3/2}} \,,
\end{equation}
where the second derivative term $v''(r)$ arises by definition.  Similar to previous ansatz, we parameterize the hypersurface $\Sigma(\tau)$ by introducing a free parameter $\lambda$, \ie $\Sigma (v(\lambda), r(\lambda))$.  More explicitly,  we rewrite the codimension-one observable as 
\begin{equation}\label{eq:CgenK}
	\begin{split}
&	\mC_{\rm{gen}}=\frac{V_x}{G_N L } \int_{\Sigma} d\sigma\, \(\frac{r}{L}\)^{d-1} \left[ \sqrt{-f(r\,){\dot v}^2+2\dot v\,\dot r} +\frac{\alpha_1}{2r(-f \dot{v}^2+2\dot{v}\dot{r})} \right.  \\ 
	&  \left.   \times \(  4(d-1)\dot{v} \dot{r}^2 +  f(2(d-1)f + rf') \dot{v}^3-3 \dot{v}^2 \dot{r} (rf'(r)+2(d-1)f)+2r\dot{v}\ddot{r}  -2r \dot{r}\ddot{v}  \)  \right]
	\,,\\
	\end{split}
\end{equation}
Noting the corresponding functional $\mC_{\rm{gen}}$ still has the reparametrization invariance for the gauge parameter $\lambda$, one can freely choose the gauge condition as before, \viz 
\begin{equation}\label{eq:gauge01}
	\sqrt{-f(r)\dot v^2+2  \dot v \dot r}=\(\frac{r}{L}\)^{d-1} \,.
\end{equation}
Using the definition for the conjugate momentum in eq.~\eqref{eq:Py}, we can obtain that the momentum conjugate to infalling time $v$ (\ie, also the coordinate time on the conformal boundary):
\begin{equation}\label{eq:PvK}
	\begin{split}
		P_v &\equiv \frac{\partial \mathcal{L}}{\partial \dot{v}} - \frac{d}{d\lambda} \(  \frac{\partial \mL}{\partial \ddot{v}}   \)\\
		&=  \sqrt{\dot{r}^2- f(r)\(\frac{r}{L}\)^{d-1} } +\tilde{\alpha}_1 \[   \(\frac{r}{L}\)^{d-2}  \(  (d-1) f(r)+ \frac{rf'(r)}{2}\)   - (d-1)\(\frac{L}{r}\)^{d} \dot{r}^2   \] \,, 
	\end{split}
\end{equation} 
with the dimensionless parameter $\tilde{\alpha}_1 = \alpha_1/L$. Thanks to the independence of $v(\lambda)$, the equations of motion from eq.~\eqref{eq:Pv02} implies that $P_v$ is also conserved along the extremal surfaces.  For a given conserved momentum $P_v$ that would be determined by the boundary conditions, the extremal surface is derived by solving the two independent equations in eqs.~\eqref{eq:gauge01} and \eqref{eq:PvK}. One can recast the radial equations in terms of the effective potential, \ie 
\begin{equation}\label{eq:UKtwo}
	\begin{split}
			\dot{r}^2 = - \mathcal{U}(P_v, r) &= \frac{1}{2(d-1)^2 \tilde{\alpha}_1^2}\left[  \( \frac{r}{L}\)^{2d}  - 2\tilde{\alpha}_1 P_v(d-1)  \(\frac{r}{L}\)^{d}    \right. \\
			&  +\tilde{\alpha}_1^2 (d-1) \( \frac{r}{L}\)^{2(d-1)}  (2(d-1)f(r) +r f'(r))    \\
			& \left.    \pm  \( \frac{r}{L}\)^{2d} \sqrt{1 - 4 \tilde{\alpha}_1 P_v (d-1) \(\frac{L}{r} \)^d  + 2\tilde{\alpha}_1^2 (d-1) \frac{L^2 f'(r)}{r}  }   \right]  \,,
	\end{split}
\end{equation}
where the minus branch reduces to the extremal volume case in the limit $\tilde{\alpha}_1 \to 0$. Similarly, one can find that the growth rate of the codimension-one observable $\mC_{\rm{gen}}$ along the symmetric extremal surface ($t_{\mt{R}}=\frac{\tau}{2}=t_{\mt{L}}$) is still derived as 
\begin{equation}
   \frac{d	\mC_{\rm gen}}{d\tau}   = \frac{V_x}{\GN L} P_{v}(\tau) \,.
\end{equation}

In the following, let us take the planar black hole given in eq.~\eqref{infall} as the bulk spacetime for more detailed analysis. Due to the double branches for the solutions of equations of motion, one can find that there are more types of extremal surfaces. First of all, it is straightforward to find from eq.~\eqref{eq:UKtwo} that the value of the conserved momentum is constrained by (the square root part should always be positive) 
\begin{equation}
	\tilde{\alpha}_1 P_v \le \frac{1}{4} \( \frac{r_h}{L}   \)^d \(  \frac{w}{d-1}  + 2\tilde{\alpha}_1^2 (2w +d -2) \) \,,
 \end{equation}
with using the dimensionless radius coordinate $w= (r/r_h)^d$. Supposing $\alpha_1 >0$, it is then clear that the extremal surface with $P_v \ge P_{\rm crt, 1}= \frac{1}{4 \tilde{\alpha}_1} \( \frac{r_h}{L}   \)^d \(  \frac{1}{d-1}  + 2d \tilde{\alpha}_1^2 \)$ would not extent to the black hole interior, which means this type of extremal surface can not connect both sides of CFT.  Sitting in the regime with $P_v \ge P_{\rm crt, 2}= \frac{\tilde{\alpha}_1(d-2)}{2 }  \( \frac{r_h}{L}   \)^d$, the extremal surfaces can extent from the left/right boundary to the singularity located at $x=0$. We are more interested in the middle regime with 
\begin{equation}
	P_{\rm crt, 1} > P_v > P_{\rm crt, 2} \,, 
\end{equation}
where the extremal surface cross the $t=0$ slice and joint the left and right boundaries. Similarly, we can define the minimal radius of this type of extremal surface by solving $ \mathcal{U}(P_v, \rmin)=0$, \ie 
\begin{equation}
	w_{\rm min } \equiv  \(\frac{\rmin}{r_h}\)^d = \frac{1}{2(1+ \tilde{\alpha}_1^2d^2)}  \(  1+ \tilde{\alpha}_1^2d^2 + 2 d \tilde{\alpha}_1  P_v \( \frac{L}{r_h}\)^d     \pm  \sqrt{1+ \tilde{\alpha}_1^2d^2 - 4 P_v^2 \(\frac{L}{r_h}\)^{2d}}   \)\,.
\end{equation}
The critical case is given by the conserved momentum $P_{\infty} $, \viz
\begin{equation}
 P_{\infty} =   \frac{1}{2} \(\frac{r_h}{L}\)^d \sqrt{1+ \tilde{\alpha}^2_1 d^2 } \,. 
\end{equation}
This also corresponds to the solutions of $\partial_r \mathcal{U}(P_v, \rmin)=0$ with the constant$-r$ slice located at 
\begin{equation}
	w_f = \frac{1}{2}  \(  1+ \frac{d \tilde{\alpha}_1}{\sqrt{1+ \tilde{\alpha}^2_1 d^2}} \) \,. 
\end{equation} 
Obviously, this extremal surface is inside the black hole interior and located at the infinity boundary time $\tau \to \infty$. The late-time linear growth thus reads 
\begin{equation}
\lim_{\tau \to \infty}  \(    \frac{d	\mC_{\rm gen}}{d\tau} \)  = \frac{V_x P_{\infty}}{\GN L}  =  \sqrt{1+ \tilde{\alpha}^2_1 d^2} \,  \frac{8 \pi M}{d-1} \,.
\end{equation}

\subsection{Higher Orders}
Besides the simple linear $K$ term, it is also natural to include higher orders. It also introduces more derivatives and results in more solutions for the extremal surfaces. We can not generalize the above analysis to the most general case. However, it is easy to find that the extremal surface corresponds to the infinity time limit $\tau \to \infty$ since the spacetime we are studying is time-translation invariant.  

\begin{figure}[h]
	\centering
	\includegraphics[width=5in]{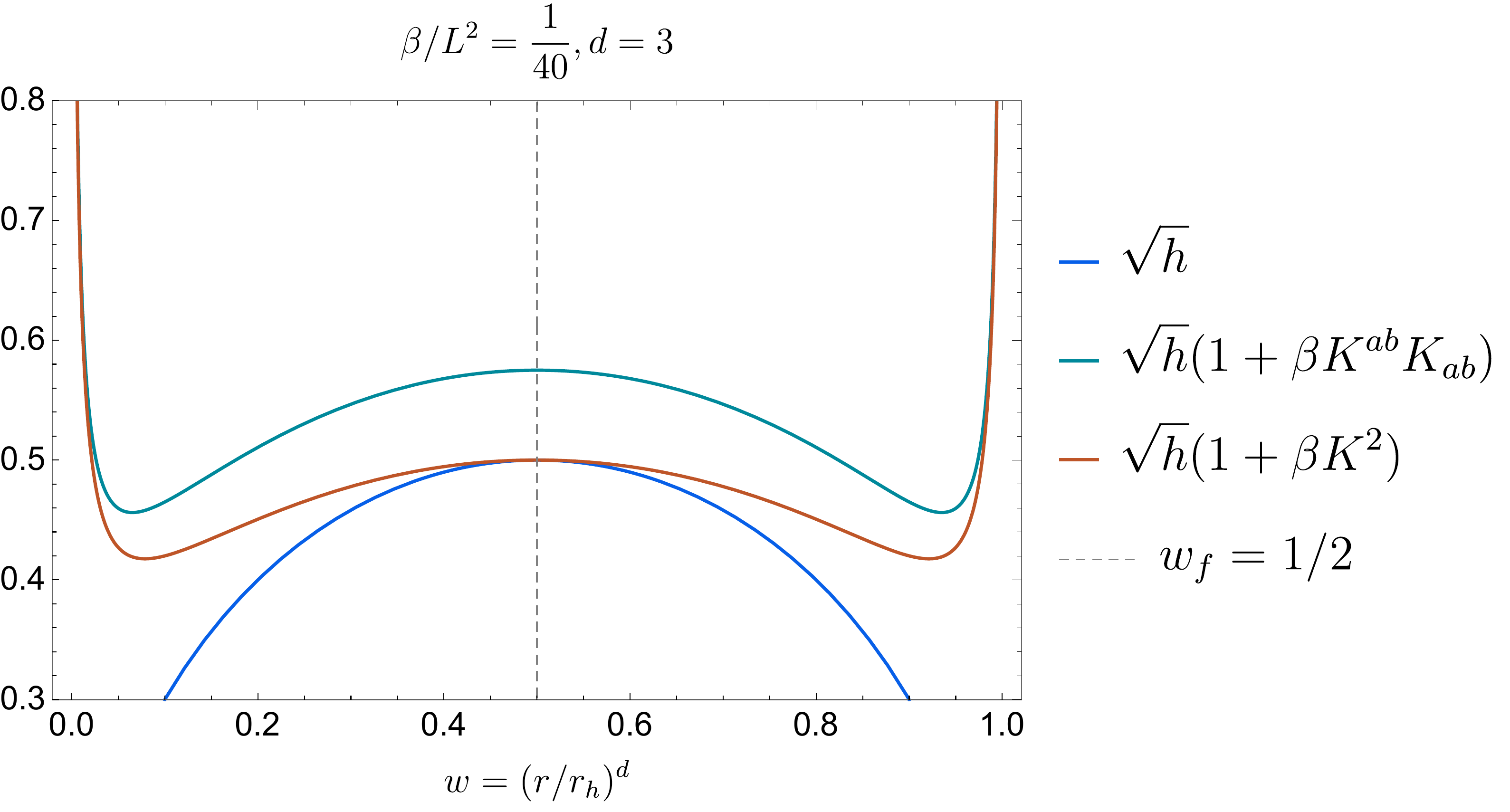}
	\caption{The volume measures on the constant$-r$ slices 
(rescaled by $L^d/r_h^d$) for different $	\mC_{\rm{gen}}$ with distinct extrinsic curvature terms. We have fixed $\frac{\beta}{L^2}=\frac{1}{50}, d=3$ for this plot.}	
	\label{fig:Volumes}
\end{figure}

A simple example at the quadratic order is given by 
\begin{equation}\label{eq:Ksquare}
	\mC_{\rm{gen}}(\tau) =  \max_{\partial\Sigma(\tau)=\Sigma_\tau} \[ \frac{1}{G_{\mt{N}} \,L}  \int_{\Sigma} \! d^d\sigma \,\sqrt{h}  (1+ \alpha_2 (K^{\mu\nu} K_{\mu\nu} -K^2))   \] \,,
\end{equation}
If we the bulk spacetime is given by the vacuum solution ($ \mR_{\mu\nu} =- \frac{D-1}{L^2} g_{\mu\nu}$), 
we can find that the above functional is completely described by the intrinsic curvature, \ie 
\begin{equation}
	\mC_{\rm{gen}}(\tau)=\frac{1}{G_{\mt{N}} \,L}\int d^d\sigma\,\sqrt{h}\left[ 1 + \alpha_2 \(R[h_{ij}]  + \frac{d(d-1)}{L^2}\)  \right]\,,
\end{equation}
by using Gauss-Codazzi equation:
\begin{equation}\label{eq:Gauss}
	\mR - 2 \varepsilon n^{\mu} n^{\nu}  \mR_{\mu\nu}= R+ \varepsilon \( K^{\mu\nu} K_{\mu\nu} - K^2 \)\,.
\end{equation}
As a result, we do not need to worry about the boundary term, and the surface variations are well defined. Now, one can process the analysis with the same method before by recasting it as a classical mechanics problem. We will not repeat similar calculations. Instead, let us focus on the time evolution of the codimension-one observable in eq.~\eqref{eq:Ksquare}. The corresponding conserved momentum is then derived as 
\begin{equation}
	P_v =   \(   1+ \tilde{\alpha}_2 \(   d(d-1) - 2(d-2) \dot{r}^2 \(\frac{L}{r} \)^{2(d-1)} \)     \) \sqrt{\dot{r}^2- f(r)\(\frac{r}{L}\)^{d-1} } 
\end{equation}
where we have used the same gauge choice to simplify the expression and redefine a dimensionless coupling constant $\tilde{\alpha}_2= \alpha_2/L^2$. The final slice locates at the constant-r slice inside the black hole interior, \ie $w_f=1/2$, which is the same for the extremal-volume hypersurface at $\tau \to \infty$. We find that the linear growth rate is modified in terms of 
\begin{equation}
\lim_{\tau \to \infty}  \(    \frac{d	\mC_{\rm gen}}{d\tau} \)  = \frac{V_x P_{\infty}}{\GN L}  = \(  1+ \tilde{\alpha}_2 d (d-1) \)\,  \frac{8 \pi M}{d-1} \,.
\end{equation}
We want to remark that the local maximal surface at $\tau \to \infty$ arrives at the same one as that for the volume case. It has a simple reason for this fact. In order to see this, let us restrict to the constant-$r$ slices. One can then find that the extrinsic curvature term, appearing in the codimension-one observable defined in eq.~\eqref{eq:Ksquare} reduces to 
\begin{equation}
	\begin{split}
		 \(K^{\mu\nu}K_{\mu\nu}- K^2\)_{r=\rm{con}} &=   \frac{-f(r)}{4}  \left((d-1) \left(\frac{2}{r}\right)^2+\left(\frac{f'(r)}{f(r)}\right)^2\right)  - \( \sqrt{-f} \(  \frac{f'(r)}{2f(r)} + \frac{d-1}{r}\)  \)^2  \\ 
		 &= \frac{(d-1) \left((d-2) f(r)+r f'(r)\right)}{r^2} \\
		 &= \frac{(d-1) d}{L^2} \,,
	\end{split}
\end{equation}
which is a constant with using the planar black hole solution. From another perspective, the simple truth of the final slice at $w=\frac{1}{2}$ is that it is an extremal surface with not only $K=0$ but also a vanishing extrinsic curvature $K_{\mu\nu}=0$.

\begin{figure}[!]
	\centering
	\includegraphics[width=3in]{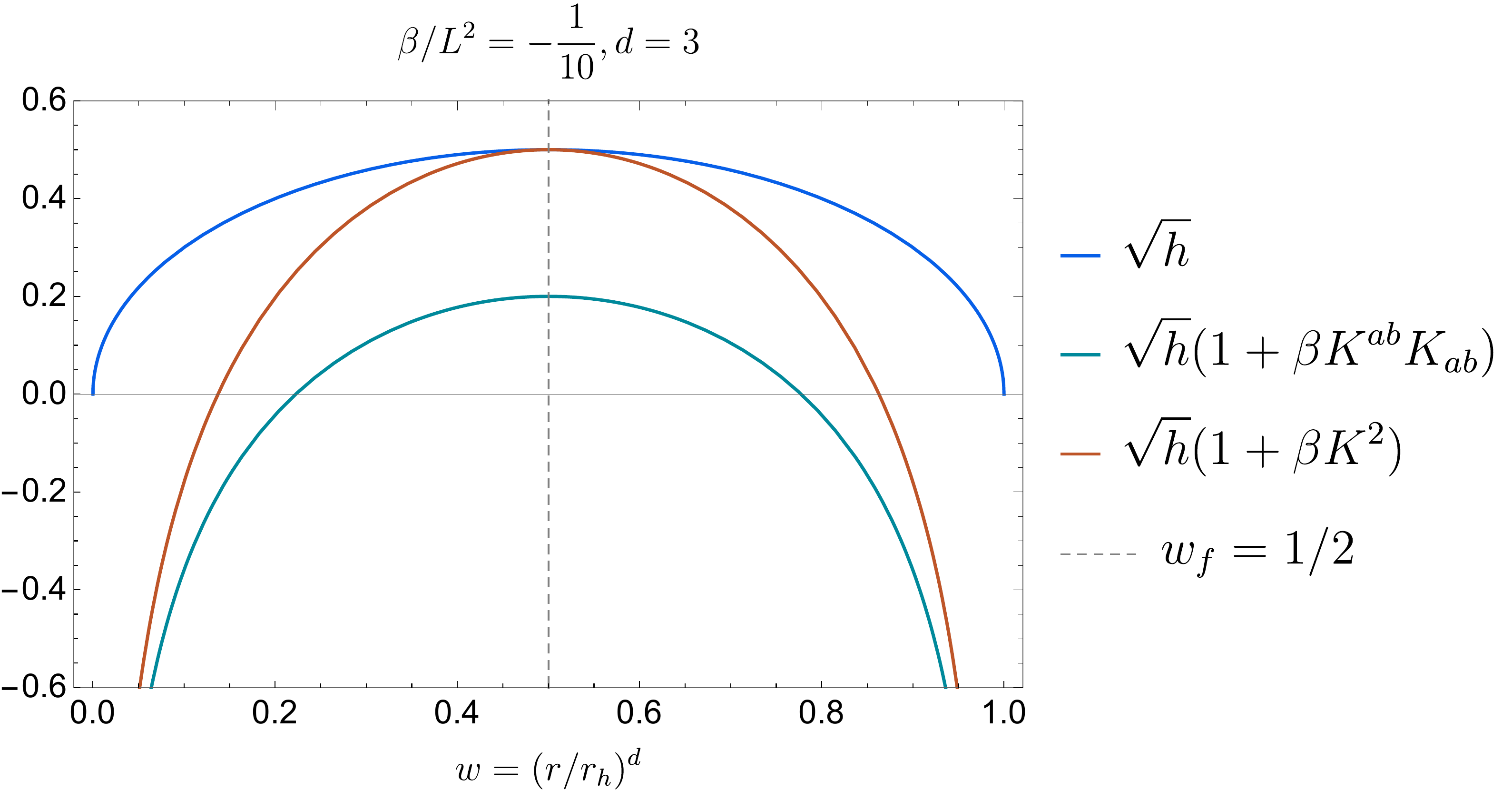}
		\includegraphics[width=3in]{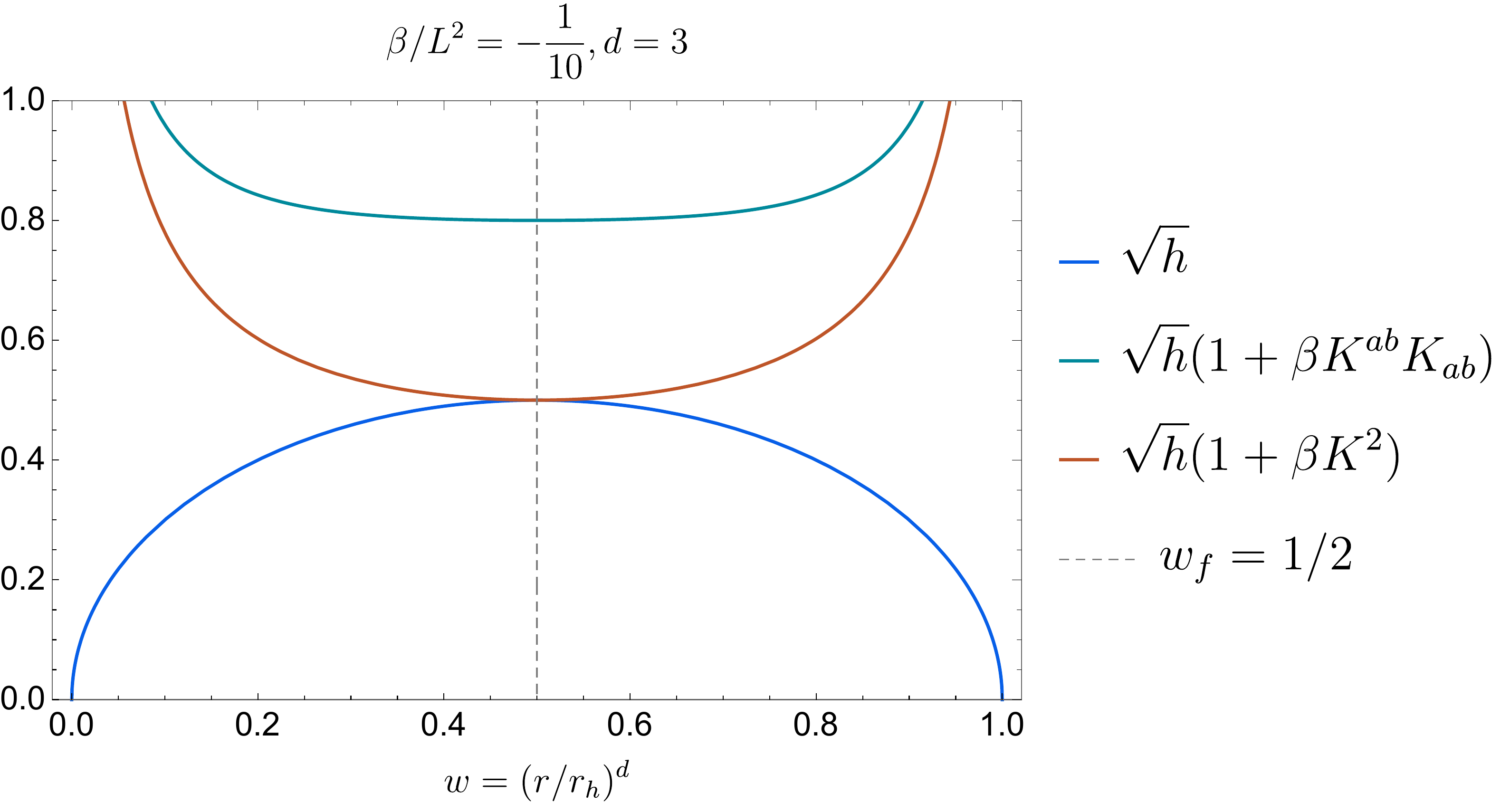}
	\caption{The effective volume measure for different $	\mC_{\rm{gen}}$ with different extrinsic curvature terms. We have fixed $\frac{\beta}{L^2}= \pm \frac{1}{10}, d=3$ for this plot.}	
	\label{fig:Volumes02}
\end{figure}

Finally, let us comment on the multiple extremal surfaces anchoring on the same boundary time. Especially, we are interested in the late-time limit. Different from what have shown before, the final slices at a constant$-r$ surface could be even more than one candidate for the most general functionals. For example, we show the effective volume measures with $K^2, K^{\mu\nu}K_{\mu\nu}$ in Figure \ref{fig:Volumes} for illustration. More explicitly, one can find that the functional defined by $\int \sqrt{h}\( 1+ \beta K^{\mu\nu}K_{\mu\nu} \) $ would contain at most three final slices, \ie 
\begin{equation}
 w_{f_1}= \frac{1}{2}\,, \quad w_{f_{2,3}} = \frac{1}{2} \left(1
\pm \sqrt{\frac{L^2-\beta d (d+1)}{L^2-\beta  d}}\right)\,,
\end{equation}
where the last two exist for the regime $ \beta <\frac{L^2}{d (d+1)}$ (as shown in Figure \ref{fig:Volumes}). As a summary, one can conclude that there is one final slice at $ w_{f_1}$ as a local maximum for $\beta <0$, one final slice as a local minimum for $\beta> \frac{L^2}{d (d+1)} $ . At the middle regime with $0<\beta< \frac{L^2}{d (d+1)}$ (see Figure \ref{fig:Volumes02}), one can find three final slices of which two at $ w_{f_{2,3}}$ are local minimum but one at $ w_{f_{1}}$ is the local maximum. Similarly, we can obtain the linear growth at the late-time limit when the local maximum exists. In addition to those simple final slices located on the constant$-r$ slice, there are other types of extremal surfaces anchoring at infinity boundary time. However, those potential competing extremal surfaces require more detailed analysis for the time evolution, which we leave for future exploration. 

\section{Existence and uniqueness of CMC slices}
\label{newapp}

We return here to the observable \reef{eq:CMCfunc} examined in section \ref{sec:spacetimevolume}. Using previous analysis by \cite{marsden1980maximal,witten2017}, we argued that the extremal boundaries $\Sigma_\pm$ are constant mean curvature (CMC) slices, \eg see eq.~\reef{eq:CMCextr}. Given this elegant geometric framework, we now show that these slices $\Sigma_{\pm}$ always exist and are unique, as long as the cosmological constant is negative and no matter fields are excited.\footnote{Or rather as long as the timelike convergence condition is satisfied -- see eq.~\reef{TCC}.} For example, this would consider the eternal black hole solution \reef{eq:BHmetric} considered in section \ref{zero}. Our analysis closely follows the arguments in \cite{witten2017}.  

Uniqueness is proved by showing that every extremum is a maximum.  If there were two distinct maxima, the functional $\mC_{\rm gen}$ would have to possess another critical point that is either a saddle point or a minimum, or otherwise become singular somewhere.\footnote{Technically, this conclusion requires that the functional $\mC_{\rm gen}$ satisfies the Palais-Smale condition, see e.g.\
\cite{Bisgard2015}.  As discussed below, since we expect the space of embedded surfaces to be compact, this condition should be automatically satisfied.}  However, $\mC_{\rm gen}$ is manifestly positive and hence bounded below by $0$, and also bounded above by 
$\frac{1}{G_{\mt{N}} L}\left(\frac{\alB}{L} V_{\mt{WDW}} +(\alpha_++\alpha_-)V_{\text{max}}\right)$, where $V_{\mt{WDW}}$ is the volume of the Wheeler-de Witt patch and $V_{\mt{max}}$ is the volume of the maximal volume surface $\Sigma$.  Hence, no singularity in the functional can occur, intuitively because adding irregularities to the surfaces $\Sigma_{\pm}$ tends to only decrease the functional $\mC_{\rm gen}$.  The fact that all extrema are maxima therefore rules out the possibility of having more than one extremum.  


We need only show  now that all extrema are maxima. We compute second variations in a gauge adapted to the variation, \ie $N$ characterizes the arbitrary variation of a surface that we put at ADM time $t=0$ (without loss of generality) and the shape variation is then equivalent to an ADM time derivative 
In this adapted gauge, the shift vector is set to zero, $N^i=0$,
and  the lapse is chosen to be spatially constant, $D_i N =0$, which
simplifies the ADM equations of motion used below.  
For the contributions on the future and past boundaries in eq.~\eqref{eq:CMCfunc}, we have
\begin{equation}\label{eq:surfacevariation}
\frac{1}{\sqrt{h}}\frac{d^2}{dt^2}\sqrt{h} =N^2 K^2+\dot{N} K+N \dot K =N^2\(\frac{2}{d-1}\Lambda+K^2-K_{ij}K^{ij}\) + \dot{N} K\,,
\end{equation}
where we obtain $\dot{h}_{ij}, \dot{K}_{ij}$ from the reduced ADM equations, \ie
\begin{equation}
    \begin{split}
  \dot{h}_{ij} &\equiv \frac{d h_{ij} }{dt}  = 2N K_{ij} \,, \\
    \dot{K}_{ij} &\equiv\frac{d K_{ij}}{dt}  = N\( \tensor{h}{^\mu_i} \tensor{h}{^\nu_j} \mathcal{R}_{\mu\nu} - R_{ij}- KK_{ij}+2 K_{ia}\tensor{K}{^a_j} \) \,, \\
    \end{split}
\end{equation}
and use the Hamiltonian constraint (\ie Gauss-Codazzi equation) to eliminate the intrinsic Ricci scalar (see \eg \cite{Gourgoulhon:2007ue,Poisson:2009pwt}). For simplicity in the later analysis, we have also used the vacuum Einstein equation to replace the bulk curvature term with the cosmological constant in eq.~\eqref{eq:surfacevariation}. If we also allow for matter contributions, we need to assume the following energy condition
\begin{equation}
\mathcal{R}_{\mu \nu} n^\mu n^\nu =-\frac{2 \Lambda}{d-1}+8 \pi G_{\mathrm{N}}\left(T_{\mu \nu}-\frac{T g_{\mu \nu}}{d-1}\right) n^\mu n^\nu \geq 0 \,, \label{TCC}
\end{equation}
which is the so-called timelike convergence condition associated with any timelike vector \eg the normal vector $n^\mu$. For the bulk term $\int_{\mathcal{M}}\sqrt{g}$, we have
\begin{equation}
\frac{d^2}{dt^2} \int dt N \sqrt{h} =\frac{d}{dt}\(N \sqrt{h}\)\Big|_{t_{-}}^{t_{+}}
=\[\dot N \sqrt{h}+N^2 \sqrt{h}K\]_{t_{-}}^{t_{+}}\,,\label{bound4}
\end{equation}
where the notation $t_\pm$ indicates that the final expression is evaluated on the boundary surfaces $\Sigma_\pm$.
So the contribution to the second variation of the functional \eqref{eq:CMCfunc} coming from the future boundary $\Sigma_+$ is proportional to\footnote{Note that there are no cross terms in the second variation between the top and the bottom surface in the case where they have the same boundary condition.}
\beq
\sqrt{h}\left[\alpha_+\(N^2 \(\frac{2}{d-1}\Lambda+K^2-K_{ij}K^{ij}\)+\dot N K\) +\frac{\alB}L\( \dot N +N^2 K\) \right]\,.
\eeq
We now substitute $K=-\alB/(\alpha_+ L)$, the equation of motion \reef{eq:CMCextr} for $\Sigma_+$, and we find that the terms involving $K$ and $\dot N$ cancel, leaving us with
\beq
\alpha_+\,\sqrt{h}\,N^2\(\frac{2}{d-1}\Lambda - K_{ij}K^{ij}\)\,.
\eeq
For a negative cosmological constant, this quantity is negative definite provided $\alpha_+>0$. 

For the past boundary $\Sigma_-$, there are two changes to the above: First, the second variation of the bulk term gets an extra minus sign since now the contribution comes from the lower end of the $t$ integral. Second, the equation of motion \reef{eq:CMCextr} for the past boundary is $K=+\alB/(\alpha_-L)$. This way, the cancellation found about still happens, the variation is proportional to
\beq
\alpha_-\,\sqrt{h}\,N^2\(\frac{2}{d-1}\Lambda - K_{ij}K^{ij}\)\,, \label{extra1}
\eeq
which is again negative as long as $\alpha_->0$ (and $\Lambda<0$). 
Hence we find a unique maximum in solving for the boundary slices provided that $\alpha_{\pm}>0$. 

Implicitly, we are also assuming that $\alB>0$, which, in addition to the requirements $\alpha_\pm>0$, ensures that the $\mC_{\rm gen}$
functional (\ref{eq:CMCfunc}) is positive, as is necessary for its interpretation as the holographic dual of complexity.  This further implies that
$K_{\Sigma_+}<0<K_{\Sigma_-}$, guaranteeing that $\Sigma_+$ lies to the future of $\Sigma_-$ and also that the region bounded by $\Sigma_+$ and $\Sigma_-$ is convex.  Note that the requirement that $\Sigma_+$ be to the future of $\Sigma_-$ allows for either $\alpha_+$ or $\alpha_-$ to be negative, as long as $\alpha_+<-\alpha_-$ when they have opposite signs (again assuming $\alB>0$).  In this case, the region between the surfaces will be concave.  The case where both $\alpha_\pm<0$ always produces $\Sigma_+$ to the past of $\Sigma_-$, and hence we exclude it from consideration.  In the concave case, the functional $\mC_{\rm gen}$ achieves a local minimum when extremizing whichever surface $\Sigma_+$ or $\Sigma_-$ corresponds to a negative value of $\alpha_+$ or $\alpha_-$, respectively.  However, because the surface with negative $\alpha_\pm$ is always extremized at a minimum, and additionally the extremization conditions for $\Sigma_+$ and $\Sigma_-$ are independent, it is still true that this extremal surface will be unique.  We therefore conclude that the uniqueness result holds independently of the signs of $\alpha_\pm$, although it may no longer be the case that $\mC_{\rm gen}$ is positive if one of $\alpha_\pm$ is negative.


Now let us discuss the existence of solutions. Here again, our argument is similar to that for the case of the codimension-one volume \cite{witten2017}. First note that if we vary the boundary surfaces, the variation of the observable \reef{eq:CMCfunc} is given by
\begin{equation}\label{varied}  
\begin{split}
  \delta \mC_{\rm gen}=-\frac{1}{\GN L } &\Big[ \int_{\Sigma_+}\!\!\!d^d\sigma\,\sqrt{h}\(\alpha_+\,K_{\Sigma_+} +\frac\alB{L}\)n\cdot\delta X_+\\
&+\int_{\Sigma_-}\!\!\!d^d\sigma\,\sqrt{h}\(\alpha_-\,K_{\Sigma_-} -\frac\alB{L}\)n\cdot\delta X_-\Big]\,.
\end{split}
\end{equation}
Note that the sign of the bulk volume contribution is the opposite between the two lines because pushing $\Sigma_+$ to the future (\ie $n\cdot\delta X_+<0$) increases the volume, while pushing $\Sigma_-$ to the future (\ie $n\cdot\delta X_-<0$) decreases the volume. Further, as discussed above, we will assume that $\alB$ and $\alpha_\pm$ are all positive throughout the following.

Now let us start with both $\Sigma_\pm$ on the past light cone of the WDW patch. Of course, this configuration yields $\mC_{\rm gen}=0$. Now, the extremization of the two surfaces is independent and so we start by varying $\Sigma_+$.  Pushing $\Sigma_+$ forward from the past light cone while holding $\Sigma_-$ fixed, we find that $\mC_{\rm gen}$ is increasing from $0$, and hence $\delta\mC_{\rm gen} >0$ in this region. Now we can push $\Sigma_+$ all the way forward to the future light cone. At this point, we have $\mC_{\rm gen}>0$ (\ie it is given by the spacetime volume of the WDW patch) but we want to show that $\mC_{\rm gen}$ passed through a local maximum before reaching this boundary of the configuration space. If we look at eq.~\reef{varied} as $\Sigma_+$ approaches the future lightcone, $K_{\Sigma_+}$ is very large and negative. Therefore $\delta \mC_{\rm gen}<0$ approaching the other end configuration. Since $\mC_{\rm gen}$ is increasing as one moves away from the two bounding configurations of $\Sigma_+$ at the past or future lightcones, we would like to conclude that there is a configuration in between in which the functional must have achieved a maximum. This conclusion could be evaded if the observable \reef{eq:CMCfunc} met a singularity or a discontinuity. However, as mentioned above, $\mC_{\rm gen}$ is bounded above, and since wiggles in $\Sigma_+$ do not increase the value of the functional, there cannot be any such divergences or discontinuities. A rigorous proof that the functional achieves the maximum for some configuration could be made along similar lines as those employed in \cite[Section 9.4]{Wald:1984rg} for the existence of geodesics. This would require showing that the (infinite-dimensional) space of embeddings for the surface $\Sigma_+$ is 
compact in a natural topology for the problem,\footnote{Roughly, this should follow from the compactness of the causal domain in which the surfaces $\Sigma$ are being embedded. Since we are often dealing with Wheeler-de Witt patches that extend out to infinity, making the compactness argument rigorous likely involves a careful treatment of boundary conditions.} and furthermore that the functional $\mC_{\rm gen}$ is upper semicontinuous in this topology. We expect both of these properties to hold. Since an upper semicontinuous, bounded function on a compact space achieves a maximum, we conclude that there exists surface $\Sigma_+$ that maximizes $\mC_{\rm gen}$.  By the previous discussion, the maximizing surface is unique.

Next let us assume that $\Sigma_+$ sits on the maximal configuration identified above. We can now vary $\Sigma_-$ from the past light cone up to this surface. At both end configurations, we have $\mC_{\rm gen}>0$ but as above, we want to show these boundary values are minima for the functional. First, consider $\Sigma_-$ on the past lightcone and pushing it forward. Analogously to the reasoning above, the variation \reef{varied} gives $\delta\mC_{\rm gen}>0$ (\ie $K_{\Sigma_-}$ is very large and positive near the past lightcone). Now consider starting with $\Sigma_-$ at the extremal $\Sigma_+$ and pushing it to the past. On this surface, we know that $K<0$ (\ie $K_{\Sigma_+}=-\alB/(\alpha_+ L)$). Hence, both terms in the brackets are negative, but $n\cdot\delta X_->0$ because we are pushing $\Sigma_-$ to the past. Hence we find $\delta\mC_{\rm gen}>0$ again. Therefore following the reasoning above, there must have been a local maximum in the variations of $\Sigma_-$ between these two configurations. Hence both extremal surfaces are given by local maxima somewhere in the interior of the WDW patch, rather than at its boundaries.  

Note this argument can be further simplified by only considering surfaces of constant mean curvature when moving $\Sigma_{\pm}$ between the bounding configurations.  In this case, there is simply a one-parameter family of surfaces, and hence the existence of maxima for $\mC_{\rm gen}$ within this family follows directly from the intermediate value theorem applied to the derivative of $\mC_{\rm gen}$ within this family.  Since we know extrema of the functional must have constant mean curvature, the maximum found by this procedure is in fact a local maximum among all possible choices of embeddings. Note that this argument somewhat begs the question by assuming the existence of the constant mean curvature foliation, but intuitively illustrates why we expect a maximal configuration to exist.  


The above proof should also be extended for black hole backgrounds where the WDW patch may run into the (spacelike) singularities behind the event horizon. However, this is straightforward, \eg for the planar AdS black hole in eq.~\reef{eq:BHmetric} because the local volume measure of the boundary surfaces shrinks to zero as they approach the singularities, and the trace of the extrinsic curvature diverges and is negative (positive) as $\Sigma_\pm$ approaches the future (past) singularity. Hence the above arguments proceed without any essential changes.  

\section{Null limit of the gravitational action} \label{app:null}

Here we will show that the gravitational action associated with the subregion between two spacelike slices $\Sigma_\pm$ (see figure \ref{fig:spacetimevolume}) approaches a quantity closely related to the action of the Wheeler-de Witt patch in the limit that the surfaces $\Sigma_{\pm}$ are taken to be null. The resulting null limit action agrees with the action for a subregion with null boundaries explored in detail in, \eg \cite{Parattu:2015gga, Lehner:2016vdi,Hopfmuller:2016scf, Chandrasekaran:2020wwn}, up to corrections related to the surface gravity of the null surfaces. In particular, we will show that the null limit of the usual Gibbons-Hawking-York (GHY) boundary term is finite, as has recently been noted in \cite{Chandrasekaran:2021hxc}. We also examine the behaviour of the Hayward term, included for codimension-two corners of the subregion, and find that it diverges in the null limit. However, the divergent piece of the Hayward term can be shown to be constant in the null limit, which can be removed by a counterterm without affecting the variational principle. The resulting finite null limit action agrees with standard expressions for the action for subregions with null boundaries associated with a Dirichlet variational principle, again up to finite corrections involving contributions from the surface gravity quantities.  

The key to obtaining the null limit action is to work with manifestly finite quantities as the surfaces $\Sigma_\pm$ are taken to be null. The unit normal $n_\mu$  is not such a quantity since it is constructed by dividing by the norm of a generic normal, which goes to zero in the null limit. Instead, we work with an unnormalized normal $\un{n}_\mu$, which is related to the unit normal according to
\begin{equation}
n_\mu = B \un{n}_\mu,\qquad B = (-\un{n}\cdot\un{n})^{-\frac12} \,.
\end{equation}
It will be convenient throughout this section to consider an entire foliation of surfaces $\Sigma$ limiting to the null surface $\ns$, and to take $\un{n}_\mu$ to be the normal to the foliation, allowing it to be defined throughout the entire subregion.  The projector onto the tangent space of the foliation is then given by
\begin{equation}
h\indices{^\mu_\nu} = \delta\indices{^\mu_\nu} - \frac{\un{n}^\mu\un{n}_\nu}{(\un{n}\cdot\un{n})}\,,
\end{equation}
which is well-defined away from $\ns$. Using this projector, we can define the shape operator $\un{K}\indices{^\mu_\nu}$ associated with the foliation by the equation
\begin{equation}\label{eqn:unK}
   \un{K}\indices{^\mu_\nu} = h\indices{^\mu_\alpha} h\indices{^\beta_\nu}
\nabla_\beta \un{n}^\alpha = h\indices{^\beta_\nu}
\left(\nabla_\beta \un{n}^\mu - \frac{\un{n}^\mu \nabla_\beta(\un{n}\cdot\un{n})}{2(\un{n}\cdot\un{n})}\right) \,. 
\end{equation}
One can straightforwardly derive that $\un{K}\indices{^\mu_\nu}$ is related to the standard mixed index extrinsic curvature tensor $K\indices{^\mu_\nu}$ by a simple rescaling \cite{Mars:1993mj},
\begin{equation}
\label{eqn:KB}
K\indices{^\mu_\nu} = B \un{K}\indices{^\mu_\nu} \,. 
\end{equation}
Additionally, the natural volume form $\un{\eta}$
induced on the surfaces of the foliation is defined by the
relation 
\begin{equation}
\epsilon = -\un{n}\wedge \un{\eta}\,, 
\end{equation}
where $\epsilon$ is the spacetime volume form. The volume form $\eta$ associated with the unit normal $n$ is similarly defined by the equation $\epsilon = - n\wedge \eta$, hence we immediately derive
\begin{equation}\label{eqn:etaB}
\eta = \frac{1}{B} \un{\eta}\,.
\end{equation}
Given the relations (\ref{eqn:KB}) and (\ref{eqn:etaB}),
we find 
\begin{equation}\label{eqn:KKtilde}
\un{K}\indices{^\mu_\nu}\un{\eta} = K\indices{^\mu_\nu}\eta\,.
\end{equation}
Although $\un{K}\indices{^\mu_\nu}$ diverges in the null limit due to the factors of $\frac{1}{\un{n}\cdot \un{n}}$ appearing in its expression, we will see that after pulling back the $\nu$ index to the foliation surfaces, the divergent terms are set to zero, giving a finite null limit for the quantity $\un{K}\indices{^i_j}\un{\eta}$, where the indices $i,j$ are intrinsic indices on the surfaces. Since the right-hand side of (\ref{eqn:KKtilde}) is used to construct the GHY term, we see that we should expect it to have a finite null limit upon pulling back the $\nu$ index, given that it is related to an expression in terms of quantities that are manifestly finite in the null limit.  

In order to determine the explicit expression for the null limit of the GHY term,  we first need to introduce several notions of surface gravity associated with the null surface $\ns$. All of the definitions given here agree on Killing horizons, but, as noted in \cite{Jacobson:1993pf}, they all disagree for more general null surfaces. Using the notation of \cite{Jacobson:1993pf}, the three surface gravity quantities $\kappa_i$ we will need are defined by
\begin{align}
\nabla_\alpha(\un{n}\cdot\un{n}) &\overset{\ns}{=}
-2\kappa_1 \un{n}_\alpha \label{eqn:kappa1} \\
\un{n}^\alpha\nabla_\alpha\un{n}^\beta &\overset{\ns}{=}
\kappa_2 \un{n}^\beta \\
\nabla_{[\alpha} \un{n}_{\beta]} \nabla^{\alpha}\un{n}^\beta
&\overset{\ns}{=} -2\kappa_3^2\,, \label{eqn:kappa3}
\end{align}
where the equalities all hold only on $\ns$. For any null surface, the surface gravities are not fully independent, but instead satisfy the relation\footnote{A fourth definition for surface gravity was considered in \cite{Jacobson:1993pf}, which for generic null surfaces can be defined by
\begin{equation}
\lim_{\Sigma\rightarrow \ns}
(a\cdot a) (\un{n}\cdot \un{n}) = -\kappa_4^2\,,
\end{equation}
where $a^\mu = n^\nu\nabla_\nu n^\mu$ is the acceleration of the unit normalized normal, which diverges in the null limit. This definition is most closely related to the notion of surface gravity as the force per unit mass applied at infinity to hold a test particle near the horizon stationary.  With some effort, one can show that this surface gravity is related to others by the equation
\begin{equation}
\kappa_4 = \kappa_3 - \frac12 n^\alpha \nabla_\alpha \log \kappa_1\,.
\end{equation}}

\begin{equation}
\kappa_3 = \frac{\kappa_1 +\kappa_2}{2}\,.
\end{equation}

The null limit can now be computed by introducing a function $\Phi$ whose level sets the label of the leaves of the foliation, with the null surface $\ns$ lying at $\Phi = 0$.  The normal $\un{n}_\alpha$ is proportional to the gradient of $\Phi$,
\beq \label{eqn:AdPhi}
\un{n}_\alpha = A\, \nabla_\alpha \Phi\,,
\eeq
with $A$ some scalar function. Since $\un{n}\cdot \un{n}$ 
vanishes in the null limit $\Phi\rightarrow 0$, consistency with the definition (\ref{eqn:kappa1}) of $\kappa_1$ requires that 
\beq \label{eqn:ndotn}
\un{n}\cdot \un{n} = -2\kappa_1 A\Phi + C\Phi^2\,,
\eeq
where $C$ is some function that is regular as $\Phi\rightarrow 0$. Then we find that 
\begin{align}
\nabla_\alpha \log(\un{n}\cdot \un{n}) 
&=
\nabla_\alpha \log \Phi +\nabla_\alpha \log(-2\kappa_1 A + \Phi C)
\nonumber \\
&= \frac{\un{n}_\alpha}{A\Phi} +\nabla_\alpha\log(-2\kappa_1 A)
+ \frac{C \nabla_\alpha\Phi}{-2\kappa_1 A} + \mathcal{O}(\Phi)
\nonumber \\
&=
\frac{-2\kappa_1 \un{n}_\alpha}{\un{n}\cdot \un{n}} +\nabla_\alpha \log
\kappa_1 + \nabla_\alpha\log A +\un{n}_\alpha
\left(\frac{C}{-2\kappa_1 A^2} \right) + \mathcal{O}(\Phi)\,. 
\end{align}
The projection of this expression onto constant $\Phi$ surfaces is then given by
\beq
h\indices{^\mu_\nu}\nabla_\mu \log(\un{n}\cdot\un{n})
= h\indices{^\mu_\nu}\left(\nabla_\mu\log\kappa_1 
+\nabla_\mu \log A\right)
+\mathcal{O}(\Phi)\,. 
\eeq
Examining equation (\ref{eqn:unK}), we now see that
$\un{K}\indices{^\mu_\nu}$ limits to
\beq\label{eqn:unKlim}
\un{K}\indices{^\mu_\nu} \overset{\Phi\rightarrow 0}{\rightarrow}
h\indices{^\alpha_\nu}\left(\nabla_\alpha \un{n}^\mu
-\frac12(\nabla_\alpha\log\kappa_1 +\nabla_\alpha\log A)\right)\,.
\eeq

When pulled back to the null surface, the first term in 
(\ref{eqn:unKlim}) becomes the shape operator $S\indices{^\mu_\nu}= h\indices{^\alpha_\nu}\nabla_\alpha \un{n}^\mu$ of the null surface, whose trace is $S = S\indices{^\mu_\mu} = \Theta + \kappa_2$, with $\Theta$ the expansion of the null surface (see, \eg \cite{Gourgoulhon:2005ng, Chandrasekaran:2020wwn} for 
a review of the geometrical quantities associated with
null surfaces). Finally, 
the definition (\ref{eqn:AdPhi}) implies that $d\un{n} = d\log A\wedge\un{n}$, which along with the definition (\ref{eqn:kappa3}) for $\kappa_3$ implies that 
\begin{equation}
\kappa_3 = \frac12 \un{n}^\alpha\nabla_\alpha \log A\big|_{\ns}\,.
\end{equation}
These relations finally determine the null limit of the GHY term as 
\beq \label{eqn:GHYNL}
\frac{1}{8\pi G} K\eta = \frac{1}{8\pi G}\un{K}\un{\eta}
\overset{\Phi\rightarrow 0}{\rightarrow} \frac{1}{8\pi G}\un{\eta}\left(\Theta + \kappa_2
 -\kappa_3 -\frac12 \un{n}^i D_i \log\kappa_1\right)\,,
\eeq
where $D_i$ denotes the intrinsic derivative on the null surface.  

The terms involving $\Theta +\kappa_2$ are the standard null boundary terms associated with the Dirichlet variational principle for a subregion with null boundaries, as explored in \cite{Parattu:2015gga, Lehner:2016vdi,Hopfmuller:2016scf, Chandrasekaran:2020wwn}. Hence, we see that the null limit of the GHY boundary terms involves the correction $\frac{\un{\eta}}{8\pi G}
\left(-\kappa_3-\frac12\un{n}^i D_i\log\kappa_1\right)$ to
the standard null boundary terms in the action of a Wheeler-de Witt patch. In terms of the variational principle, these additional terms require that the surface gravities $\kappa_3, \kappa_1$ be held fixed in addition to the intrinsic quantities defined on the null surface. Since these surface gravities relate to the behaviour of the metric away from the null surface, it seems likely that this modified variational principle can always be satisfied by an appropriate gauge fixing.

A particular case of interest for a null surface on which to evaluate (\ref{eqn:GHYNL}) is that of a Killing horizon. All of the surface gravities $\kappa_i$ coincide in such a situation. In addition, the expansion of such a horizon vanishes, and, by the zeroth law of black hole mechanics, the surface gravity $\kappa_1$ is constant over the horizon. These facts imply that the null limit boundary term (\ref{eqn:GHYNL}) vanishes on a Killing horizon.  

In addition to the GHY term, we would also like to examine the behaviour of the Hayward term
\cite{Hayward:1993my,1994PhRvD..50.4914B}
added to codimension-two corners of the subregion in the null limit. For two intersecting spacelike surfaces $\Sigma_\pm$, the Hayward term can be expressed as \cite{Lehner:2016vdi} 
\beq
\frac{1}{8\pi G} \mu\, \alpha, \qquad \alpha = \log\big((s_+ - n_+)\cdot n_-\big)\,,
\eeq
where $\mu$ is the volume form on the codimension-two corner 
$\partial\Sigma_\pm$, $n_\pm$ are the future-directed unit normals of the surfaces $\Sigma_{\pm}$, and $s_+$ is the 
the outward-directed spacelike unit normal to $\partial\Sigma_+$ that lies in the plane of $\Sigma_+$.  

As before, when examining the null limit, we should work instead with unnormalized normals that have finite null limits. Hence we define
\beq
n_\pm = B_\pm \un{n}_\pm, \qquad s_\pm = B_\pm \un{s}_\pm, \qquad
B_\pm = (-\un{n}_\pm \cdot\un{n}_\pm)^{-\frac12}\,,
\eeq
where we have normalized the spacelike normals so that 
$|\un{s}_\pm| = |\un{n}_\pm| = B_\pm^{-1}$. In terms of these, the expression for $\alpha$ becomes
\beq
\alpha = \log\big((\un{s}_+ - \un{n}_+)\cdot\un{n}_-\big) 
+\log B_+ + \log B_-\,,
\eeq
Since $B_\pm \rightarrow \infty$ as $\Sigma_\pm$ are taken to be null, we see that, unlike the GHY term, the Hayward term diverges in the null limit. This divergence in the null limit of the Hayward term has recently been noted in \cite{Chandra:2022pgl}.

In order to resolve this issue, we would like to subtract a divergent piece from $\alpha$ to obtain a finite quantity in the null limit. However, we should also ensure that this subtraction does not affect the Dirichlet variational principle, meaning the counterterm we subtract should be purely a function of intrinsic variables on the codimension-two surface. Subtracting off $\log B_+ + \log B_-$ would produce a finite quantity in the null limit, but these are not good counterterms since generically $\delta B_\pm \neq 0$ in the null limit. Hence, we must find additional terms to subtract to preserve the Dirichlet variational principle.  

Using the standard boundary condition for the unnormalized normals $\delta \un{n}^\pm_\mu = 0$, and parameterizing the normals as in (\ref{eqn:AdPhi}) $\un{n}^\pm_\mu = A_\pm \nabla_\mu \Phi^\pm$, we find that 
\beq
\delta A_\pm \nabla_\mu \Phi^\pm = -A_\pm \nabla_\mu \delta\Phi^\pm\,.
\eeq
Since the null boundaries $\ns_\pm$ lie at $\Phi^\pm = 0$, and this condition is preserved under variations, it must be that $\delta\Phi^\pm= \psi^\pm \Phi^\pm$, where $\psi^\pm$ are scalar functions that are regular as $\Phi^\pm \rightarrow 0$. Applying this to the above identity, we find that 
 \beq
 \delta\log A_\pm \un{n}_\mu^\pm = -\delta\log\Phi^\pm\, \un{n}^\pm_\mu
 -\Phi^\pm A_\pm \nabla_\mu \psi^\pm\,,
 \eeq
and hence $\delta \log A_\pm 
= -\delta\log\Phi_\pm +\mathcal{O}(\Phi^\pm)$. This further implies that $\delta(A_\pm\Phi^\pm) = \mathcal{O}(\Phi_\pm^2)$, which, after using the relation (\ref{eqn:ndotn}), can be used to show that 
\beq
\delta \log B_\pm = -\frac12\delta \log(-\un{n}_\pm\cdot\un{n}_\pm)
 = -\frac12\delta\log\kappa_1^\pm + \mathcal{O}(\Phi^\pm)\,.
\eeq
Hence, to leading order in  $\Phi^\pm$, the log of the surface gravities $\kappa_1^\pm$ have variations proportional to $\delta\log B_\pm$, meaning that 
\begin{equation}
\log B_\pm + \frac12\log\kappa_1^\pm
\end{equation}
has zero variation in the null limit. This, therefore, serves as a suitable counterterm to obtain a finite null limit of the Hayward term. The renormalized Hayward term is then given by
\beq
\alpha_\text{ren} = \alpha -\log B_+ -\frac12\log\kappa_1^+
-\log B_- -\frac12\log\kappa_1^-\,.
\eeq
Taking the null limit of this quantity results in
\beq
\alpha_\text{ren} \overset{\Phi^\pm\rightarrow 0}{\longrightarrow}
\log\big(-\un{n}_+\cdot\un{n}_-\big)
-\frac12\log\kappa_1^+ - \frac12\log\kappa_1^-+\log 2\,.
\eeq
The first term in this expression agrees with the standard null joint corner term considered in \cite{Lehner:2016vdi}, while the terms involving the surface gravities $\kappa_1^\pm$ and the constant $\log 2$ represent corrections to this expression coming from the null limit. The $\log 2 $ term is a constant that can be removed by an additional finite counterterm. Just as in the null limit of the GHY boundary term, the presence of the $\log \kappa_1^\pm$ terms in the null limit implies that 
the variational principle associated with this action involves fixing $\kappa_1^\pm$ in addition to the standard Dirichlet conditions on the codimension-two surface. As before, it appears likely that this additional boundary condition is accessible as a gauge choice.

\section{Conjugate Variations for $h^{ij}$ and $K_{ij}$}\label{app:variation}

Considering a functional $W(h_{ij}, \pi^{ij})$ defined as the integral along a Cauchy slice, \ie 
\begin{equation}
W (h_{ij}, \pi^{ij}):= \int_{\Sigma} w(h_{ij},\pi^{ij}) \, d^d\sigma\,,
\end{equation}
we start by assuming the variation will emerge from symplectic form,\ie 
\begin{equation}
\Omega(\delta, \delta_{w})= \int_{\Sigma} \, \delta w \( h_{ij}, \pi^{ij}  \).
\end{equation}
Combining the definition of the symplectic form 
\begin{equation}
 \Omega(\delta, \delta_{w}) \equiv  \int_{\Sigma} \(\delta \pi^{ij}\delta_{w} h_{ij}- \delta h_{ij}  \delta_{w} \pi^{ij}    \) \,, \\
\end{equation}
along with the general variation of the density $w( h_{ij}, \pi^{ij} )$
\begin{equation}
 \delta w \( h_{ij}, \pi^{ij}  \) =  \frac{\delta w}{\delta h_{ij}} \bigg|_{\pi}\, \delta h_{ij} + \frac{\delta w}{\delta \pi^{ij}}  \bigg|_{h}\, \delta \pi^{ij} \,,
\end{equation}
one can derive the following two relations for the conjugate transformation $\delta_{w}$:
\begin{equation}\label{eq:NewYorKVar}
\delta_{w} \pi^{ij}=-\frac{\delta w}{\delta h_{ij}}\bigg|_{\pi}, \quad \delta_{w} h_{ij} = \frac{\delta w}{\delta \pi^{ij}}\bigg|_{h}\,.
\end{equation}
Implicitly, we are considering $W=W \( h_{ij}, \pi^{ij}  \) $ as a functional of the intrinsic metric $h_{ij}$ and the conjugate momentum $\pi^{ij}$, as the two independent variables. Alternatively, it will be convenient/useful to consider  $W =W \( h^{ij}, K_{ij}  \) $, \ie take $W$ as a functional of the inverse metric $h^{ij}$ and the extrinsic curvature $K_{ij}$. 

In general relativity, the momentum density tensor can be written as
\begin{equation}
\pi^{ij}  = \sqrt{h} \(  K^{ij} - h^{ij} K  \) 
=  \frac{1}{\sqrt{h^{-1}}} \(  h^{ik} h^{j\ell} K_{k\ell} - h^{ij} 
h^{k\ell}K_{k\ell}  \)  \,.
\end{equation}
Instead, we can also rewrite $K_{ij}$ as a functional of $h_{ij}$ and $\pi^{ij}$, namely
\begin{equation}\label{eq:KGR}
K_{ab} = \frac{1}{\sqrt{h}} \(  h_{ia}h_{jb} \pi^{ij}   -\frac1{d-1} h_{cd}\pi^{cd} h_{ab}   \) \,,
\end{equation}
where we are working in $d+1$ dimensions, \ie the Cauchy surface is $d$-dimensional. Hence, it is straightforward to rewrite the variations in \eqref{eq:NewYorKVar} in terms of the conjugate transformations, $\delta_{w} K_{ab}$ and $\delta_{w} h_{ab}$. 

In order to simplify the final expressions below, it is convenient to introduce two new tensors
\begin{equation}
\label{waltz}
A^{ab}=\frac{1}{\sqrt{h}}\frac{\delta w}{\delta K_{ab}}\bigg|_{h}  \quad 	 {\rm and}\quad
B_{ab}=\frac{1}{\sqrt{h}}\frac{\delta w}{\delta h^{ab}} \bigg|_{K}  \,,\
\end{equation}
where we assume $W=W \( h^{ij}, K_{ij}  \)$.\footnote{Note that we do not distinguish holding fixed $h_{ij}$ or $h^{ij}$ in the functional derivatives here and in the following.}
First of all,, we note that 
\begin{equation}
	B^{ab}=h^{ac}h^{bd}B_{cd}=-\frac{1}{\sqrt{h}}\frac{\delta w}{\delta h_{ab}} \bigg|_{K} \,.
\end{equation}
It is then straightforward to evaluate $\delta_{w} h^{ij}$ by using
\beq
h^{ac} h_{cb} = \delta^a{}_b
\quad\implies\quad
\delta_{w} h^{ab}=-h^{ac}h^{bd}\delta_{w} h_{cd}\,.
\label{wallf1}
\eeq
Furthermore, one can obtain
\begin{equation}
\begin{split}
\delta_{w} h_{ij}&= \frac{\delta w}{\delta \pi^{ij}} \bigg|_{h} = \frac{\delta w}{\delta K_{ab}} \bigg|_{h} \frac{\delta K_{ab}}{\delta \pi^{ij}}\bigg|_{h} \,.
\end{split}
\end{equation}
The explicit variation of \reef{eq:KGR} for GR yields
\begin{equation}
\frac{\delta K_{ab}}{\delta \pi^{ij}}  = \frac{1}{\sqrt{h}} \(  h_{a(i}h_{j)b}  - \frac{h_{ab}h_{ij}}{d-1} \) \,.
\end{equation}
Hence we arrive at the conjugate transformation of the inverse metric
\begin{equation}\label{eq:deltah}
\delta_{w} h^{ij}= -A^{ij}  + \frac{A }{d-1}\,h^{ij} \,,
\end{equation}
where $A= h_{ab} A^{ab}$. This is the expression given in the first line of eq.~\eqref{eq:genYorkK}.

Let us now turn to evaluating $\delta_{w} K_{ab}$. 
Here we begin with the expression for  $K_{ab}=K_{ab}\(h_{ij},
\pi^{ij}\)$, shown in \eqref{eq:KGR}, to write
\begin{equation}\label{dyk}
    \begin{split}
       \delta_{w} K_{ab} &= -\frac{1}{2\sqrt{h}} h^{ij} \delta_{w} h_{ij} \(    \pi_{ab} -  \frac{\pi h_{ab}}{d-1}   \)   \\
&+\frac{1}{\sqrt{h}}  \(  \delta_{w} h_{ia} \pi^i_b+\delta_{w} h_{ib} \pi^i_a  + h_{ia}h_{jb}\delta_{w} \pi^{ij} -\frac{h_{cd}\pi^{cd}}{d-1}\delta_{w}h_{ab}  - \frac{h_{ab}}{d-1} \(  \delta_{w} h_{cd} \pi^{cd} + h_{cd} \delta_{w} \pi^{cd}  \) \) \,. \\
    \end{split}
\end{equation}
Using the above analysis, we can get
\begin{equation}\label{walf2}
\delta_{w} h_{ij}= A_{ij}  - \frac{A }{d-1}\,h_{ij} \,,
\end{equation}
\ie combining eqs.~\reef{wallf1} and \reef{eq:deltah}.
Next, we would like to express $\delta_{w} \pi^{ij}$ in terms of $A^{ab}$ and $B_{ab}$ in \reef{waltz}. 
Hence considering $W\(h^{ij},K_{ij}(h,\pi)\)$, we write\footnote{An alternate approach writes the variation $\delta_{w} \pi^{ij}$ as 
\begin{equation}
\delta_{w} \pi^{ij}= -\frac{\delta w}{\delta h_{ij}} \bigg|_{\pi} = -\frac{\delta w}{\delta h_{ij}} \bigg|_{K}+  \frac{\delta w}{\delta K_{ab}} \bigg|_{h} \frac{\delta \pi^{cd}}{\delta h_{ij}}  \( \frac{\delta \pi^{cd}}{\delta K_{ab}} \bigg|_{h}  \)^{-1}  \,. \\
\end{equation}
The two expressions are equivalent by noting the identify 
\begin{equation}
\frac{\delta \pi^{ab}}{\delta h_{cd}}\bigg|_{h} \frac{\delta h_{cd}}{\delta h_{ij}}+ \frac{\delta \pi^{ab}}{\delta K_{cd}}\bigg|_{h} \frac{\delta K_{cd}}{\delta h_{ij}}\bigg|_{\pi} =0\,.
\end{equation}}
\begin{equation}\label{eq:deltapi}
\begin{split}
\delta_{w} \pi^{ij}&= -\frac{\delta w}{\delta h_{ij}} \bigg|_{\pi} = -\frac{\delta w}{\delta h_{ij}} \bigg|_{K}- \frac{\delta w}{\delta K_{ab}} \bigg|_{h} \frac{\delta K_{ab}}{\delta h_{ij}}\bigg|_{\pi} \\
&=\sqrt{h}\[ B^{ij}  +\frac{1}{2} A^{ab}K_{ab} h^{ij} + KA^{ij} + \frac{A}{d-1} \( K^{ij}-h^{ij}K  \)  -2 A^{b(i}K^{j)}{}_b \] \,.\\
\end{split}
\end{equation}
Here the second line is derived by using \reef{eq:KGR} to find
\begin{equation}
\frac{\delta K_{ab}}{\delta h_{ij}}\bigg|_{\pi}= -\frac{1}{2} h^{ij} K_{ab} + K \delta_a^{(i}\delta^{j)}_b 
- \frac{h_{ab}}{d-1} \( K^{ij} -h^{ij} K \) + \frac{1}{\sqrt{h}} \delta^{(i}_a\pi^{j)}_b + \frac{1}{\sqrt{h}} \delta^{(i}_b\pi^{j)}_a\,.
\end{equation}
Substituting eqs.~\eqref{walf2} and \eqref{eq:deltapi} into \reef{dyk} then yields 
\begin{equation}
\delta_{w} K_{ab} = B_{ab} - \frac{A\, K_{ab}}{2(d-1)}  - \frac{h_{ab}}{d-1}  \(  B - \frac{1}{2} A^{ij}K_{ij}  \) \,,
\end{equation}
which is the result presented in eq.~\eqref{eq:genYorkK}. 

\section{Complexity=Anything for Rotating BTZ} \label{sec:approtate}

In the main text, we focused on examining our new gravitational observables in the eternal planar black hole background \reef{eq:BHmetric}. Here we take a first step in investigating our complexity=anything proposal in  rotating black hole backgrounds. In particular, we consider the three-dimensional rotating BTZ black hole. But let us first discuss our expectations for the late-time growth rate in these backgrounds. Following the discussion in \cite{Susskind:2014moa}, one should find that the rate is proportional to the product of the black hole's entropy and temperature. First, since the complexity should be proportional to the number of active degrees of freedom in the system, both the complexity and it's growth rate are expected to be proportional the entropy. Second, the growth rate has units of energy and so the natural scale to introduce is the temperature so that it remains extensive. Of course, for the static background in eq.~\reef{eq:BHmetric}, this product gives the mass of the black hole and we already confirmed that the growth rate was generally proportional to the mass for our new gravitational observables. For the rotating BTZ background, the suggested result was confirmed for the CA and CV proposals in \cite{Brown:2015lvg} and \cite{Couch:2018phr}, respectively. The growth rate was also studied for rotating black holes in higher dimensions with the standard proposals for holographic complexity, and the desired result was recovered for the limit of large black holes \cite{Couch:2018phr,Cai:2016xho,Auzzi:2018zdu,Frassino:2019fgr,AlBalushi:2020rqe,AlBalushi:2020ely,Bernamonti:2021jyu}. Here we return to the rotating BTZ background and examine the late-time growth rate associated with gravitational observable in eq.~\reef{eq:CMCfunc} with general coefficients. We will find that the rate is indeed proportional to the product of the entropy times the temperature.

Let us begin with the three-dimensional rotating black hole, \ie BTZ black hole \cite{Banados:1992wn}, whose metric is given by  
\begin{equation}
ds^2 = -f(r) dt^2 + \frac{dr^2}{f(r)} + r^2 \(d\phi - \frac{r_+r_-}{L^2 r^2} dt \)^2 \,,
\end{equation}
where $f(r)=\frac{(r^2-r^2_+)(r^2-r_-^2)}{L^2 r^2}$ and $r_\pm$ denote radius of the outer/inner black hole horizon. Recall that the physical quantities of BTZ black hole are recast in terms of $r_\pm$ as 
\begin{equation}
T= \frac{r^2_+ -r_-^2}{2\pi L^2\, r_+}, \quad M= \frac{r_+^2+r_-^2}{8\GN L^2}\,, \quad S=\frac{\pi r_+}{2\GN}\,, \quad J= \frac{r_+r_-}{4\GN L}\,, \quad \Omega=\frac{r_-}{\GN r_+}\,.
\end{equation}
Note that implicitly with these expressions, we have set the spatial volume of the boundary time slices to be $V_\Phi=2\pi L$.
With using the ingoing coordinates, we can also rewrite the BTZ metric as \cite{Bernamonti:2021jyu}
\begin{equation}
 ds^2 = -f(r) dv^2 + 2 dv dr + r^2 \(d \Phi  - \frac{r_+r_-}{L^2 r^2} dv  \)^2 \,,  
\end{equation}
where the angular coordinate $\Phi = \phi + \int \frac{r_+r_-}{L^2 r^2}\frac{dr}{f(r)}$. (Note that $\Phi  \in [0, 2\pi]$.)
From the blackening factor $f(r)$ of the rotating BTZ black hole, it is obvious that we have two independent length scales $r_+, r_-$ in general. If the late-time growth rate was proportional to M, we would have $\frac{d}{d\tau}\mC_{\rm{gen}}\propto (r_+^2+r_-^2)$,
whereas if it is proportional to $T\,S$, we would find $\frac{d}{d\tau}\mC_{\rm{gen}}\propto  (r_+^2-r_-^2)$. Below, we will show that the latter case applies.

To begin, let us consider the hypersurface parametrized by $(v(\sigma),r(\sigma), \Phi)$ and evaluate the volume of a hypersurface, \ie CV proposal (see \cite{Bernamonti:2021jyu} for more detailed analysis). The effective potential captures the time evolution of the extremal surface, \ie
\begin{equation}
    U_0(r):=-f(r) \(\frac{r}{L} \)^2 =- \frac{(r^2-r^2_+)(r^2-r^2_-)}{L^4}\,,
\end{equation}
as we have shown in the general analysis in section \ref{zero} and appendix \ref{revone}. By noting that the final slice is located at 
\begin{equation}
r_f^{(0)} = \sqrt{\frac{r^2_+ + r^2_-}{2}} \quad \in \quad (r_-, r_+) \,, 
\end{equation}
the corresponding critical momentum can be derived as $P_{\infty} = \frac{r_+^2-r_-^2}{2L^2} $. It is then straightforward to obtain the late-time growth rate as \cite{Couch:2018phr,Bernamonti:2021jyu}
\begin{equation}\label{eq:CVBTZ}
 \lim_{\tau \to \infty} \frac{d\mC_{\rm V}}{d\tau}= \frac{V_{\Phi} P_{\infty}}{\GN L}= \frac{\pi}{\GN L^2}\,(r_+^2-r_-^2)= 4 \pi\, T\,S \,,
\end{equation}
with the spatial volume taken as $V_\Phi = 2\pi L$. Hence we see that the CV proposal yields the desired result for the late-time growth rate,  $TS$ which is proportional to $r^2_+ - r^2_-$. 

Instead, let us move to the infinite codimension-zero functionals defined in eq.~\eqref{eq:CMCfunc} with taking the bulk background as BTZ black hole. The corresponding effective potential for $\Sigma_\pm$ is then derived as 
\begin{equation}
 \mathcal{U}(r;P_v^\varepsilon) = U_0(r) - \( P_v^\varepsilon + \frac{\varepsilon \alB}{2\alpha_\varepsilon}\frac{r^2}{L^2}\)^2\,.
 \end{equation}
In terms of the minimal radius $\rmin$, the conserved momentum can be rewritten as 
\begin{equation}
P_{v}^{\varepsilon} = \frac{2\sqrt{(r^2_+ - r_{\rm min}^2)(r_{\rm min}^2 - r_-^2)} - \varepsilon \, r_{\rm min}^2 
}{2 \alpha_{\varepsilon} L^2} \,. 
\end{equation}
For each hypersurface $\Sigma_\pm$, it is straightforward to obtain the position of the final slice from the effective potential $\mathcal{U}(r;P_v)$. It is worth noting that the solutions depend on the sign of $\alpha_\pm$. Supposing $\alpha_\pm>0$, the corresponding two final slices (with $\tau \to +\infty$) are derived as 
\begin{equation}
\begin{split}
   r_f(\Sigma_+)=  r_f^{+} &= \frac{\sqrt{\left(\sqrt{4 \alpha_+ ^2+\alB^2}-\alB\right) r_+^2+\left(\sqrt{4 \alpha_+ ^2+\alB^2}+\alB\right)r_-^2}}{\sqrt{2 \left(4 \alpha_+ ^2+\alB^2\right)}} <  r_f^{(0)}  \,, \\
    r_f(\Sigma_-)=   r_f^{-} &= \frac{\sqrt{\left(\sqrt{4 \alpha_- ^2+\alB^2}+\alB\right) r_+^2+\left(\sqrt{4 \alpha_- ^2+\alB^2}-\alB\right)r_-^2}}{\sqrt{2 \left(4 \alpha_- ^2+\alB^2\right)}} > r_f^{(0)}   \,.\\
\end{split}
\end{equation}
The associated critical momentum $P_{\infty}$ then reads 
\begin{equation}
    \begin{split}
    P_{\infty}^{+} (\alpha_+>0) &= \frac{r^2_+( \sqrt{4\alpha_+^2+\alB^2}-\alB)-r^2_-( \sqrt{4\alpha_+^2+\alB^2}+\alB) }{4\alpha_+ L^2} \,, \\
     P_{\infty}^{-}(\alpha_->0)&= \frac{r^2_+( \sqrt{4\alpha_-^2+\alB^2}+\alB)-r^2_-( \sqrt{4\alpha_-^2+\alB^2}-\alB) }{4\alpha_- L^2} \,,\\
    \end{split}
\end{equation}
which both depend on the two independent scales $r_+, r_-$. However, combining the contributions from both surfaces $\Sigma_\pm$, the late-time growth rate of the codimension-zero functional \eqref{eq:CMCfunc} for the rotating BTZ black hole reduces to 
\begin{equation}\label{eq:CMCBTZ}
\begin{split}
 \lim_{\tau \to \infty} \frac{d\mC_{\rm{gen}}}{d\tau} = \frac{V_{\Phi}}{\GN L}\( \alpha_+ \, P_{\infty}^{+} +  \alpha_- P_{\infty}^{-}\) =2\pi \(\sqrt{4\alpha_+^2 +\alB^2}+\sqrt{4\alpha_-^2 +\alB^2}\)T\,S\,.\\
\end{split}
\end{equation}
Taking the limit $\alpha_\pm \to 0$ as in eq.~\reef{limit2}, we  recover the result of CV2.0 proposal for  rotating BTZ \cite{Couch:2018phr,Bernamonti:2021jyu}, \viz  
\begin{equation}
 \lim_{\tau \to \infty} \frac{d \cvv}{d\tau}= \lim_{\alpha_\pm \to 0}\lim_{\tau \to \infty} \frac{1}{\alB}\frac{d\mC_{\rm{gen}}}{d\tau}= \frac{\pi (r_+^2 -r_-^2)}{\GN L^2}= 4\pi TS  \,.
\end{equation}
We should note that this simple result depends on the choice of $\alpha_\pm >0$, \ie taking the codimension-zero functional defined in eq.~\eqref{eq:CMCfunc} as a local maximum rather than a saddle point. For the branches with $\alpha_+$ or $\alpha_-$ negative, one would instead have $r_f(\Sigma_+)= r_f^- $ and $ r_f(\Sigma_-)= r_f^+$, respectively. The associated critical momentum $P_{\infty}$ becomes
\begin{equation}
    \begin{split}
    P_{\infty}^{+} (\alpha_+<0) &=\frac{r^2_-( \sqrt{4\alpha_+^2+\alB^2}-\alB)-r^2_+( \sqrt{4\alpha_+^2+\alB^2}+\alB) }{4\alpha_+ L^2}  \,, \\
     P_{\infty}^{-}(\alpha_-<0)&= \frac{r^2_-( \sqrt{4\alpha_-^2+\alB^2}+\alB)-r^2_+( \sqrt{4\alpha_-^2+\alB^2}-\alB) }{4\alpha_- L^2} \,.\\
    \end{split}
\end{equation}
However, the time evolution of these two hypersurfaces $\Sigma_\pm$ is problematic since the relation $r_f(\Sigma_+)> r_f^{(0)}> r_f(\Sigma_-)$ implies that the upper boundary $\Sigma_\pm$ would intersect with the lower boundary $\Sigma_-$ and finally become even below the bottom boundary $\Sigma_-$. For other two configurations with $\alpha_+ \alpha_- <0$, we finally obtain a 
growth rate proportional to $\pm(r_+^2 -r_-^2) \(\sqrt{4\alpha_+^2+\alB^2}-\sqrt{4\alpha_-^2+\alB^2}  \) $. It would result in a vanishing growth rate when $|\alpha_+ |= |\alpha_-|$. This is expected since the bulk contribution vanishes when the codimension-zero subregion with $r_f(\Sigma_+)= r_f(\Sigma_-)$ collapses to nothing and the two boundary terms are exactly canceled after taking $\alpha_+ = -\alpha_-$.

In this appendix, we only considered three-dimensional BTZ black holes. It would also be interesting to the complexity=anything proposal for higher dimensional rotating black holes. Although we do not explore this here, it is expected that one would {\it not} find the growth rate in eq.~\eqref{eq:CMCBTZ} for a generic higher dimensional Kerr black hole, as was found for the CV and CA proposals \cite{Couch:2018phr,Cai:2016xho,Auzzi:2018zdu,Frassino:2019fgr,AlBalushi:2020rqe,AlBalushi:2020ely,Bernamonti:2021jyu}. The simple expression $\frac{d}{d\tau}\mC_{\rm{gen}}\propto T\,S$ is only expected to arise for large black holes with $r_+ \to \infty$. That is, this result arises in the limit that $L/r_+ \to 0$ and $r_-/r_+ \to 0$.

\bibliographystyle{jhep}
\bibliography{references}
\end{document}